\documentclass[usenatbib]{mnras}
\usepackage{newtxtext,newtxmath}
\usepackage[T1]{fontenc}

\usepackage{amsmath}
\usepackage{graphicx}
\usepackage[usenames,dvipsnames]{xcolor}
\usepackage{mathtools}
\usepackage{gensymb}
\usepackage{tabularx}
\usepackage{subfigure}
\usepackage{physics}
\usepackage{subfiles}

\newcommand{\dobib}{\bibliography{references}}

\def\gsim{\;\rlap{\lower 2.5pt \hbox{$\sim$}}\raise 1.5pt\hbox{$>$}\;}
\def\lsim{\;\rlap{\lower 2.5pt \hbox{$\sim$}}\raise 1.5pt\hbox{$<$}\;}

\newcommand{\cm}{ \,\ifmmode{{\mathrm{cm}}}\else cm\fi}
\newcommand{\ccm}{\,\ifmmode{\mathrm{cm}^{-3}}\else cm$^{-3}$ \fi}
\newcommand{\kms}{\,\ifmmode{\mathrm{km}\,\mathrm{s}^{-1}}\else km\,s$^{-1}$\fi}

\title{Cloudy with A Chance of Rain: Accretion Braking of Cold Clouds}

\author[B. Tan, S.P. Oh and M. Gronke] {
  Brent Tan$^1$\thanks{E-mail: zunyibrent@physics.ucsb.edu},
  S. Peng Oh$^1$,
  and Max Gronke$^{2}$\\
    $^1$University of California - Santa Barbara,
    Department of Physics, CA 93106-9530, USA\\
    $^2$Max Planck Institut f\"{u}r Astrophysik, Karl-Schwarzschild-Straße 1, D-85748 Garching bei M\"{u}nchen, Germany
}
\date{Accepted XXX. Received YYY; in original form ZZZ}
\pubyear{2022}

\begin{document}
\renewcommand{\dobib}{}
\label{firstpage}
\pagerange{\pageref{firstpage}--\pageref{lastpage}}
\maketitle

\begin{abstract}
Understanding the survival, growth and dynamics of cold gas is fundamental to galaxy formation. While there has been a plethora of work on `wind tunnel' simulations that study such cold gas in winds, the infall of this gas under gravity is at least equally important, and fundamentally different since cold gas can never entrain. Instead, velocity shear increases and remains unrelenting. If these clouds are growing, they can experience a drag force due to the accretion of low momentum gas, which dominates over ram pressure drag. This leads to sub-virial terminal velocities, in line with observations. We develop simple analytic theory and predictions based on turbulent radiative mixing layers. We test these scalings in 3D hydrodynamic simulations, both for an artificial constant background, as well as a more realistic stratified background. We find that the survival criterion for infalling gas is more stringent than in a wind, requiring that clouds grow faster than they are destroyed ($t_{\rm grow} < 4\,t_{\rm cc} $). This can be translated to a critical pressure, which for Milky Way like conditions is $P \sim 3000  \,{\rm k}_B {\rm K}\,{\rm cm}^{-3}$ . Cold gas which forms via linear thermal instability ($t_{\rm cool}/t_{\rm ff} < 1$) in planar geometry meets the survival threshold. In stratified environments, larger clouds need only survive infall until cooling becomes effective. We discuss applications to high velocity clouds and filaments in galaxy clusters.
\end{abstract}



\section{Introduction}  \label{sect:intro}

The cycle of baryons -- particularly that of cold gas, the fuel for star formation -- is absolutely fundamental to galaxy formation and a crucial link between galactic and cosmological scales \citep{2020ARA&A..58..363P}. This cycle can take various forms: (i) Outflows due to feedback processes \citep{thompson16,schneider18}. Observationally, cold gas is frequently seen outflowing at velocities comparable to virial/escape velocities \citep{veilleux05,steidel10,rubin14,heckman17}. (ii) Inflow of cold gas which forms via thermal instability in the halo \citep{joung12,sharma12,fraternali15,voit19,tripp22}, or is supplied by direct cosmology accretion (cold streams; \citealp{keres05,dekel06}), and falls under gravity. (iii) Fountain recycling, which is a combination of these two processes. A useful analogy is the terrestrial water cycle, where evaporation, condensation and precipitation both play crucial roles. 

All of these motions involve velocity shear between cold gas clouds and background hot gas. A long-standing problem has been to understand why clouds are not shredded by hydrodynamic instabilities, particularly the Kelvin-Helmholtz instability. The hydrodynamic acceleration time for a cloud of radius $r$, overdensity $\chi$ embedded in a wind of velocity $v_{\rm w}$ is $t_{\rm acc} \sim \chi r/v_{\rm w}$, the timescale for the cloud to sweep up its own column density. By contrast, the cloud destruction (`cloud crushing') time is $t_{\rm cc} \sim \sqrt{\chi} r/v_{\rm w}$, i.e. of order the Kelvin-Helmholtz time, implying that $t_{\rm acc}/t_{\rm cc} \sim \sqrt{\chi}$, i.e. clouds should be destroyed before they can be accelerated \citep{klein94,zhang17}. Numerous simulation studies, including those with radiative cooling, concluded that cold clouds get destroyed before they can become entrained with the wind \citep[e.g.][]{cooper09,scannapieco15,schneider16}; magnetic fields can ameliorate but do not solve the problem \citep{mccourt15,gronke20-cloud}.  

In recent years, it was realized that there are regions of parameter space where the cooling efficiency of the mixed, `warm' gas is sufficiently large to contribute new comoving cold gas which can significantly exceed the original cold gas mass, enabling the cloud to survive. Cloud growth is thus mediated by these turbulent mixing layers \citep{begelman90,ji18,fielding20,tan21}. The criteria for this to happen is $t_{\mathrm{cool, mix}}/t_{\mathrm{cc}} < 1$, where $t_{\mathrm{cool,mix}}$ is the cooling time of the mixed warm gas (with $T_{\rm mix} \sim (T_{\rm hot} T_{\rm cold})^{1/2}$) and $t_{\mathrm{cc}}$ is the cloud crushing time \citep{gronke18}. This criterion is always satisfied for a large enough cloud $r > c_{\rm s, cold} t_{\rm cool,mix}$ (where $c_{\rm s, cold}$ is the sound speed of the cold gas),  which grows and entrains by gaining mass and momentum from cooling mixed hot gas. Thus, the cloud eventually comoves with the wind, with a cold gas mass which can be many times the original cloud mass. These conclusions have been borne out in many subsequent studies \citep[e.g.,][]{sparre20,li20,abruzzo21,girichidis2021,farber22}. 

However, cold gas survival and growth has only been understood for part of the baryon cycle, galactic outflows. To date, there have only been a handful of studies studying cold gas survival and growth during \textit{infall}, which is arguably even more fundamental to processes such as star formation. 

An important outstanding problem in galaxy evolution is that the observed star formation rates (SFRs) in galaxies at a range of redshift are unsustainable - they would rapidly deplete current existing gas reservoirs - and hence these galaxies require some form of continuous accretion to supply the necessary fuel \citep{erb08,hopkins08,putman09}. For example, our Milky Way has a SFR of $\sim 2$ M$_{\odot}$\,yr$^{-1}$ but only $\sim 5 \times 10^9$\,M$_{\odot}$ of existing fuel, and would thus burn through this supply in just 2-3\,Gyrs \citep{chomiuk11,putman12}. Supplementary inflow must come in the form of low-metallicity ($Z < 0.1 \, Z_{\odot}$) gas, so as to satisfy constraints from disk stellar metallicities and  chemical evolution models \citep{schonrich09,kubryk13}. 

At the same time, we see infall in the form of `high-velocity' and `intermediate-velocity'  clouds (HVCs and IVCs; \citealt{putman12}) with relatively low metallicities, as well as a galactic fountain with  continuous circulation of material  between the  disk and corona \citep{shapiro76, fraternali08}. Fountain-driven accretion could supply the disk with gas for star formation, and explain the observed kinematics of extra-planar gas \citep{armillotta16,fraternali17}. It is tempting to speculate from the results of wind tunnel simulations that star formation in the disk exerts a form of positive feedback: cold gas thrown  up into the  halo `comes back with interest', by mixing with low metallicity halo gas which cools and increases the cold gas mass. 

HVCs are also good candidates and could provide a significant amount of the necessary fuel for star formation, provided they survive their journey to the disk \citep{vanwoerden,putman12,fox19}. First detected in HI\,21\,cm emission by \citet{muller63}, HVCs are gas clouds observed moving at high velocities relative to the local standard of rest. The traditional definition for HVCs is thus those clouds with velocities in the Local Standard of Rest frame $|v_{\rm LSR}| \geq 90$\,km/s \citep{wakker91} (although similar clouds whose velocities significantly overlap that of the disk may be missed; \citealp{zheng15}). They have been observed in all regions of the sky, and come in a range of sizes \citep{putman12}. Clouds are grouped into various complexes based on spatial and kinematic clustering but because of their proximity, precise distances to HVCs are difficult to measure. The main method of doing so is to use halo stars of known distances in the same sky region to bracket the cloud distance by looking for absorption lines (or lack thereof) in the stellar spectra. By determining if a HVC is in front of or behind each star, the HVC's distance can thus be effectively constrained. Most HVCs with distances measured as such are found between 2-15\,kpc, with most heights above the disk  $<10$\,kpc \citep{wakker08,thom08}. The head-tail morphology observed in many HVCs \citep{putman11}, along with observations that the majority of high velocity absorbers kinematically and spatially lie in the vicinity of HVCs \citep{putman12}, strongly suggest that the HVCs are mixing as they travel through the ambient medium. There is a wealth of literature on observations of HVCs -- we refer the reader to reviews such as \citet{putman12} for a more comprehensive account. 

As we have discussed, the survival of HVCs is inherently problematic, since they are vulnerable to hydrodynamic instabilities while travelling through the hot background \citep{klein94,zhang17}. Early theoretical efforts to model HVCs initially focused on predicting their velocity trajectories, without taking into consideration their mass evolution. These early models assumed that these HVCs fell ballistically \citep{bregman80} or reached a terminal velocity when eventually slowed by hydrodynamic drag forces \citep{benjamin&daly97}, and were used in evaluating the contributions of HVCs in larger feedback models \citep{maller04}. However, the decoupling of the velocity and mass evolution implied by this approach has been shown to be untenable for HVCs with the advent of high resolution hydrodynamical simulations, many of which show that the mass and morphology of the clouds evolve significantly \citep[e.g.,][]{kwak11,armillotta17,gritton17,gronke20-cloud}. While wind-tunnel setups are numerous, the number of 3D simulations of clouds falling under the influence of gravity and including radiative cooling is more limited \citep{heitsch09,heitsch22,gronnow22}. The survival criterion for infalling clouds has not been quantified, and analytic models for mass and velocity evolution which match simulations do not yet exist. We will tackle these challenges in this paper. 

Presumably, similar considerations apply, with a minimum cloud size $r_{\rm crit} \sim c_{\rm s, cold} t_{\rm cool,mix}$ required for survival and growth. However, this ignores a crucial distinction between outflowing and infalling cold gas clouds. Outflowing gas clouds gradually entrain, so destruction processes become weaker as the velocity shear is reduced. The cloud only has to survive until it becomes comoving with the hot gas, at which point hydrodynamic instabilities are quenched (and mass growth peaks). Indeed, wind tunnel simulations (particularly for clouds with sizes just above $r_{\rm crit}$) often show clouds which initially break up into small fragments, with a significant amount mixed into the hot medium, but eventually survive as the fragments entrain and grow. The cold fragments then coalesce -- the cloud `rises from the dead' to a peaceful environment. In contrast, infalling clouds accelerate under the action of gravity, with continually {\it increasing} velocity shear, and consequently increasing cloud destruction rate, which is maximized at the cloud terminal velocity. Thus, the cloud instead is exposed to continually worsening conditions, and somehow has to survive an unrelenting hot wind. Moreover, the properties of the wind change with time, as the cloud falls through a background stratified hot medium. 

The survival and growth of a cold cloud under such conditions is the focus of this paper. We develop simple analytic scalings which we test in 3D hydrodynamic simulations. Unsurprisingly, several important aspects, such as cloud survival criteria, are quite different from the wind tunnel case. 

What is at stake? As previously mentioned, if clouds can survive and grow, the ultimate fuel supply for star formation could simply be coronal gas, whose condensation is triggered by star formation feedback and Galactic fountain recycling. During this process, cold gas also exchanges angular momentum with coronal gas, which links fountain circulation to the observable kinematics of coronal gas. More broadly, the physics of radiative turbulent mixing layers is complex, and theoretical studies demand empirical tests. Unlike clouds embedded in galactic winds, which lie at extra-galactic distances and are difficult to resolve, there is a plethora of spatially and kinematically resolved observational data for intermediate and high velocity clouds in the Milky Way. There is also ample similar data for infalling filaments in galaxy clusters \citep[e.g.][]{russell19}. Such systems can be used as laboratories for the interaction between multiphase gas, mixing, and radiative cooling, which is also critical to galactic winds but difficult to test there. We shall see that we predict sub-virial terminal velocities at odds with standard predictions (which balance hydrodynamic drag with gravity) but in much better agreement with observations. Moreover, the predicted terminal velocity from the model is an observable that can be tested, at least on a statistical basis (given observational uncertainties and degeneracies). Such empirical tests have thus far been sorely lacking in cloud physics models. 

The outline of this paper is as follows. In Section~\ref{sect:analytics}, we outline analytic theory and predictions for the dynamics, growth and survival of infalling cold clouds. In Section~\ref{sect:methods}, we describe our simulation setup. In Sections~\ref{sect:results} and \ref{sect:results_stratified}, we describe simulation results, both for an artificial constant background (which allows us to test analytic scalings), as well as a more realistic stratified background. In Section~\ref{sect:discussion}, we discuss applications to the Milky Way (HVCs) and galaxy clusters (infalling filaments). Lastly, we summarize and conclude in Section~\ref{sect:conclusions}. 

\dobib

\section{Dynamics of Infalling Clouds}  \label{sect:analytics}

\subsection{Cloud Evolution and Terminal Velocities} \label{sec:terminal_velocity}
A falling cloud growing via accretion can be described by the following set of differential equations:
\begin{align} 
    \frac{\dd z}{\dd t}    &= v \label{eq:diff_eq1} \\
    \frac{\dd (mv)}{\dd t} &= mg - \frac{1}{2}\rho_{\rm hot} v^2 C_0 A_{\rm cross} \label{eq:diff_eq2} \\
    \frac{\dd m}{\dd t}    &= \frac{m}{t_{\rm grow}} 
    \label{eq:diff_eq3}
\end{align}
where $z$, $v$ and $m$ represent the distance fallen, velocity, and mass of the cloud respectively, $t_{\rm grow} \equiv m/\Dot{m}$ is the growth timescale (which we discuss in Section~\ref{sec:growth}), $g$ is the gravitational acceleration, $C_0$ is the drag coefficient (geometry dependent; of order unity here), $\rho_{\rm hot}$ is the density of the background medium, and $A_{\rm cross}$ is the cross-sectional area which the cloud presents to the background flow. We shall see that it is important to distinguish $A_{\rm cross}$ from $A_{\rm cloud}$, the overall surface area of the cloud.  We shall also see that $t_{\rm grow}$ is roughly independent of mass growth, so that from equation~\eqref{eq:diff_eq3}, mass growth is nearly exponential. Note that equation~\eqref{eq:diff_eq3} assumes steady growth and omits terms which contribute to cloud destruction. Thus, it does not apply to clouds which are losing rather than gaining mass. In this paper, we focus on scenarios where clouds survive and grow, which is the novel feature in our new model (previous works, e.g. \citealp{afruni19}, have looked at scenarios with significant mass loss). In Section~\ref{sec:survival} we will quantify the criterion for cloud survival. In this work, we only consider the hydrodynamic case and leave investigation of other factors such as magnetic fields, externally driven turbulence, and cosmic rays to future work.

The terms on the right hand side in the momentum equation (equation \eqref{eq:diff_eq2}) represent the gravitational and hydrodynamic drag forces. In standard models, these two terms are assumed to balance one another in steady-state, giving the hydrodynamic drag terminal velocity
\begin{align}
    v_{\rm T,drag} = \sqrt{\frac{2mg}{\rho_{\rm hot} C_0 A_{\rm cross}}} \simeq \sqrt{\frac{2 \chi L g}{C_0}}
    \label{eq:vtdrag}
\end{align}
for a falling cloud with volume $\sim A_{\rm cross} L$ and $\chi = \rho_{\rm cloud}/\rho_{\rm hot}$. The hydrodynamic drag time (momentum divided by the drag force) is given by $t_{\rm drag} \sim \chi L/v$. In fact, {\it this gives the terminal velocity only if the left hand side of equation \eqref{eq:diff_eq2} vanishes, $\dot{p} = m \dot{v} + \dot{m} v = 0 \Rightarrow \dot{v} =0$, which is correct only if cloud mass does not evolve so $\dot{m}=0$.}  If $\Dot{m} > 0$, i.e. the cloud grows by accreting mass from the background, then from momentum conservation, since the background gas is at rest and has zero initial momentum, this will slow down the cloud. In the limit that the hydrodynamic drag term is small compared to $\Dot{m}v$,
\begin{align}
    v = \frac{m}{\Dot{m}}(g - \Dot{v}).
    \label{eq:nonconstantv}
\end{align}
Thus, if $\dot{v} \ll g$, there is a second terminal velocity
\begin{align}
    v_{\rm T,grow} = \frac{mg}{\Dot{m}} = g t_{\rm grow}.
    \label{eq:vtgrow}
\end{align}
The two terminal velocities in equations~\eqref{eq:vtdrag} and \eqref{eq:vtgrow} represent regimes where the cloud acceleration under gravity is predominantly balanced by either hydrodynamic drag or the momentum transfer from background accretion respectively. We can separate them by considering the ratio $t_{\rm grow}/t_{\rm drag}$. When this ratio is large, gravity is balanced by drag. Conversely, when this ratio is small, gravity is balanced by accretion. The transition between the two is marked by where $t_{\rm grow} \sim t_{\rm drag}$. 
\begin{figure}
    \centering
	\includegraphics[width=\columnwidth]{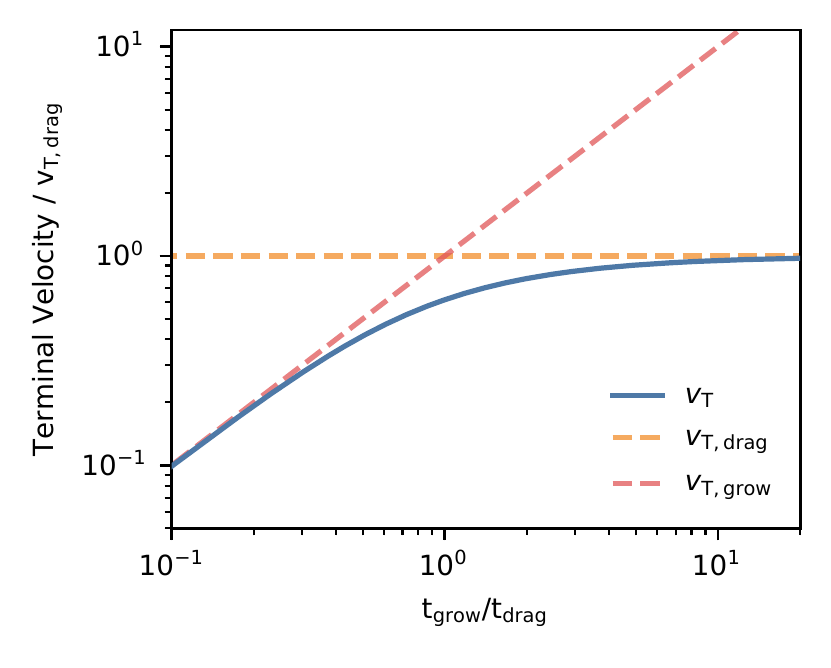}
	\caption{The terminal velocity (normalized by the drag terminal velocity) as a function of the ratio of the growth time $t_{\rm grow}$ and the drag time $t_{\rm drag}$. The dashed lines show the corresponding values for the terminal velocities when assumed to be set by either drag or growth. As the ratio increases, there is a smooth transition from $v_{\rm T,grow}$ to $v_{\rm T,drag}$.}
	\label{fig:fig_tv}
\end{figure}
We can illustrate this by solving equations~\eqref{eq:diff_eq1} -- \eqref{eq:diff_eq3} numerically for a constant $t_{\rm grow}$. The result is shown in Fig.~\ref{fig:fig_tv}, where we plot the terminal velocity as a function of $t_{\rm grow}/t_{\rm drag}$. We can see that when $t_{\rm grow} \ll t_{\rm drag}$, the terminal velocity follows $v_{\rm T,grow}$, and when $t_{\rm grow} \gg t_{\rm drag}$, the terminal velocity follows $v_{\rm T,drag}$ as expected. 

Which regime is more realistic? Let us first explore this for the idealized case of spherical clouds. For a spherical cloud, the growth time is given by \citep{gronke20-cloud}: 
\begin{equation}
    t_{\rm grow} \equiv \frac{m}{\dot m}\sim \frac{\rho_{\rm cold}r^3}{\rho_{\rm hot}A_{\rm cloud} v_{\rm mix}} \sim \chi \frac{r}{v_{\rm mix}}. 
    \label{eq:tgrow-approx} 
\end{equation}
This seems long: if $v_{\rm mix} \sim c_{\rm s,cold}$ (a reasonable estimate; see Section~\ref{sec:growth} of \citealt{gronke20-cloud} and Sections 4.6 \& 5.3.3 of \citealt{tan21}), then $t_{\rm grow} \sim \chi t_{\rm sc}$, where $t_{\rm sc}$ is the sound crossing time across the cloud. By contrast, the hydrodynamic drag time for a spherical cloud (as mentioned previously) is: 
\begin{equation} 
    t_{\rm drag} \sim \chi \frac{r}{v},
\end{equation} 
which is much shorter, since $t_{\rm drag}/t_{\rm grow} \sim v_{\rm mix}/v \sim c_{\rm s,cold}/c_{\rm s,hot} \sim \chi^{-1/2} \ll 1$, if we assume the the virial velocity to be a characteristic infall speed, $v \sim v_{\rm vir} \sim c_{\rm s,hot}$. The fact that $t_{\rm drag} \ll t_{\rm grow}$ makes physical sense. The hydrodynamic drag time is also the timescale for a cloud to sweep up its own mass in hot gas ($\rho_{\rm hot}A_{\rm cross}vt_{\rm drag} \sim \rho_{\rm hot}r^3\chi \sim m$). Even if all this mass is incorporated into the cloud, then at best $t_{\rm grow} \sim t_{\rm drag}$. In fact, only a small fraction of this gas is actually incorporated into the cloud, so that $t_{\rm grow} \gg t_{\rm drag}$. This suggests that hydrodynamic drag is the main drag force, which results in a terminal velocity given by equation~\eqref{eq:vtdrag}.

However, as previously mentioned, clouds in a shearing wind do {\it not} remain spherical; they develop extended cometary tails (as seen both in simulations and observations). This change in geometry -- and in particular the large increase in surface area -- is {\it crucial} for enabling momentum transfer via mass growth. In hydrodynamic drag, $F_{\rm drag} \sim \rho_{\rm hot} v^2 A_{\rm cross}$, the area $A_{\rm cross}\approx \pi r^2$ is the cross-sectional area the cloud presents to the wind. Thus, $F_{\rm drag}$ remains roughly {\it constant} during cloud evolution. By contrast, in $\dot{m} \sim \rho_{\rm hot} A_{\rm cloud} v$, the area $A_{\rm cloud}$ is the surface area of the cloud available for mixing. In a cometary structure, this is dominated by the sides of the cylinder, so that $A_{\rm cloud} \sim 2 \pi r L$, where $L$ is the length of the tail. Thus, $\dot{m} \propto A_{\rm cloud} \propto L$ {\it increases} as a cloud develops a cometary tail. It is this increase in $\dot{m}$, and thus the effective momentum transfer rate $F_{\rm grow} \sim \dot{p}_{\rm grow} \sim \dot{m} v$, compared to a constant $F_{\rm drag}$, which causes mass growth to dominate momentum transfer: $t_{\rm grow} \sim \chi r/v_{\rm mix}$ is roughly constant, while $t_{\rm drag} \sim m v/F_{\rm drag} \sim \rho_{\rm cloud} A_{\rm cross} L v/(\rho_{\rm hot} A_{\rm cross} v^{2}) \sim \chi L/v$ increases as the mass of the cloud increases. In particular, 
\begin{equation}
    \frac{t_{\rm grow}}{t_{\rm drag}} \sim \frac{r}{L} \frac{v}{v_{\rm mix}} \sim \frac{r}{L} {\chi^{1/2}}.  
    \label{eq:tgrow_tdrag}
\end{equation}
In cloud crushing simulations, the tail grows during the process of entrainment to a length $L \sim vt_{\rm drag} \sim \chi r$ during the `tail formation' phase \citep{gronke20-cloud}, so that $t_{\rm grow}/t_{\rm drag} \sim \chi^{-1/2} \ll 1$. The continuous shear for infalling clouds can lead to even more extended tails since the cloud does not entrain, so $t_{\rm grow}/t_{\rm drag} \ll 1$ is easily satisfied\footnote[1]{Shorter entrainment times than $t_{\rm drag}$ have been observed in cloud crushing simulations \citep[e.g.,][]{gronke20-cloud,farber22}.}.

Finally, it is important to realize that there is a third timescale in the problem, the free-fall time $t_{\rm ff}\sim v_{\rm vir}/g$. This sets the evolutionary lifetime available to clouds, before they fall to the halo center. Clouds will not grow significantly (and reach the terminal velocity $v_{\rm T,grow}$ given by equation~\eqref{eq:vtgrow}), unless $t_{\rm grow} < t_{\rm ff}$.
Indeed, $t_{\rm grow} < t_{\rm ff}$ is required for a subvirial terminal velocity. We can show this by recalling that $F_{\rm grav} \sim m g \sim m v_{\rm vir}/t_{\rm ff}$, while the drag force from mass growth is $F_{\rm grow} \sim \dot{m}v \sim m v/t_{\rm grow}$. At the terminal velocity $v_T$, we have $F_{\rm grav} \sim F_{\rm grow}$, so that: 
\begin{equation} 
    f_{\rm sub-vir} \equiv \frac{v_{\rm T,grow}}{v_{\rm vir}} \sim \frac{v_{\rm T}}{c_{\rm s,hot}} \sim \frac{t_{\rm grow}}{t_{\rm ff}}.
    \label{eqn:alpha}
\end{equation} 
This is useful because $f_{\rm sub-vir}$ -- infall velocities, normalized to the virial velocity -- can be measured observationally. Indeed, $f_{\rm sub-vir} < 1$, sub-virial infall velocities, is commonly observed in LRGs \citep{huang16,zahedy19} and galaxy clusters \citep{russell16}, much lower than predicted terminal velocities from hydrodynamic drag models \citep{lim08}. Our models can explain these puzzling observations, as we describe in Section~\ref{sect:clusters}. It also allows for testable predictions. Since $f_{\rm sub-vir}$ is measured and $t_{\rm ff}$ is known from the density profile, we can predict $t_{\rm grow} \approx f_{\rm sub-vir} t_{\rm ff}$ from kinematic observations, assuming that clouds have reached terminal velocity. This can be compared with predictions for $t_{\rm grow}$ from equations~\eqref{eq:tgrow_scaling} and \eqref{eq:tgrow_scaling2}, given measured or inferred cloud and background hot gas properties. Lastly, the mass growth that a cloud experiences is $m/m_0 \sim \exp(t_{\rm ff}/t_{\rm grow}) \sim {\rm exp} (f_{\rm sub-vir}^{-1})$. Thus, a measurement of sub-virial velocities directly constrains the degree to which mixing and cooling enhances cool gas infall to the central galaxy. Significantly sub-virial infall implies that cold clouds grow considerably before reaching the halo center. These analytical estimates can be compared to measurements of the mass infall rate \citep[e.g.,][]{2006MNRAS.366..449F,Fox2019}. In Section \ref{sect:results_stratified}, we will also show the rather remarkable result that in an isothermal atmosphere with constant gravity, $f_{\rm sub-vir}$ is {\it fixed} by geometry, specifically the scaling between cloud mass and area ($\alpha$ in equation \eqref{eq:acloud}), {\it independent} of all other properties of the system. For our infalling clouds, we find $f_{\rm sub-vir} \approx 0.6$. 

\subsection{Cloud Growth}
\label{sec:growth}
Previous models of infalling clouds have considered the interplay between gravity and hydrodynamic drag forces, assuming a fixed cloud mass \citep{benjamin&daly97}. However, a fixed cloud mass is unrealistic due to various processes that trigger mixing with the hot background gas or shred the cloud. Mass evolution therefore cannot remain static; clouds should either be destroyed ($\dot{m} < 0$), or grow ($\dot{m} > 0$) over time. 

In the absence of cooling, clouds moving relative to a background medium are destroyed by hydrodynamic instabilities on the cloud crushing timescale \citep{klein94,scannapieco15}
\begin{align}
   t_{\rm cc} \sim \sqrt{\chi} \frac{r}{v},
\end{align}
where $\chi$ is the ratio of the cloud density to the background density, $r$ is the cloud radius, and $v$ is the magnitude of the relative velocity between the cloud and the background. This cloud crushing timescale reflects the destruction of the cloud via internal shocks induced inside the cloud due to its velocity with respect to the medium it is moving through (assuming that this velocity is supersonic with respect to the sound speed within the cloud), and is roughly the same timescale on which surface instabilities such as the Kelvin-Helmholtz and Rayleigh-Taylor instabilities grow to the cloud scale \citep{klein94}. This destructive fate can however be counteracted by mass growth due to cooling. In wind tunnel simulations of `cloud crushing', \citet{gronke18} found that in order for cold gas to survive, cooling needs to be strong enough to satisfy the criterion
\begin{align}
    t_{\rm cool,mix} < t_{\rm cc},
    \label{eq:t_cool_mix}
\end{align}
where $t_{\rm cool,mix}$ is the cooling time of the {\it mixed} gas, defined as $T_{\rm mix} \sim \sqrt{T_{\rm cloud} T_{\rm hot}}$ (in the spirit of \citealp{begelman90}, see also \citealp{2019ApJ...885..101H} for an alternative derivation). That is, if the cooling time of the mixed gas is shorter than the {\it initial} cloud crushing time, then cold gas survives and is eventually entrained in the hot background wind. 

However, infalling clouds have an important aspect that differentiates them from clouds in a wind -- gravity. Clouds encountering a hot wind gradually entrain in the wind, so that shear eventually drops to zero if the cloud manages to survive until entrainment. The cloud thus encounters destructive forces for a limited period of time. By contrast, clouds in a gravitational field will always keep falling and shearing against the background gas. Thus, the survival criterion is different, and more stringent; we discuss this in Section~\ref{sec:survival}.

Assuming cloud survival, let us quantify the timescale on which clouds grow. We first derive some scaling relations, before deriving numerical expressions. For now, we ignore fudge factors (due to geometry, etc.) which can be up to an order of magnitude. As in equation~\eqref{eq:tgrow-approx}, the mass growth rate of a cloud can be written as 
\begin{align}
    \Dot{m} \sim \rho_{\rm hot} A_{\rm cloud} v_{\rm mix},
    \label{eq:mdot}
\end{align}
where $\rho_{\rm hot}$ is the density of the hot background medium, $A_{\rm cloud}$ is the effective surface area of the cloud\footnote[2]{The effective surface area corresponds to the (smoothed) enveloping area of the cloud and not the (non-convergent) surface area of the cold gas. See \citet{gronke20-cloud} for further discussion of this distinction.} and $v_{\rm mix}$ is the the velocity corresponding to the mass flux from the hot background onto this surface. As above, if we write $m \sim \rho_{\rm cold} A_{\rm cloud} r$, this gives: 
\begin{equation} 
    t_{\rm grow} \sim \chi \frac{r}{v_{\rm mix}}. 
    \label{eq:tgrow-simple}
\end{equation}
Plane parallel simulations of mixing layers \citep{tan21} show: 
\begin{equation}
    v_{\rm mix} \sim u^{\prime 3/4} \left( \frac{r}{t_{\rm cool}} \right)^{1/4}
    \sim v_{\rm shear}^{3/5} v_0^{3/20} \left( \frac{r}{t_{\rm cool}} \right)^{1/4}
    \label{eq:vmix-simple}
\end{equation}
where $t_{\rm cool}$ is the cooling time in {\it cold} gas (the minimum cooling time in the mixing layer, a convention we adopt henceforth) and $u^{\prime}$ is the peak turbulent velocity in the mixing layer (usually in intermediate temperature gas). Note that while the first step in equation~\eqref{eq:vmix-simple}, i.e., $v_{\rm mix}(u')$, is generally valid, we have used the scaling $u^{\prime} \propto v_{\rm shear}^{4/5}$ for relating $u'$ to the parameters of the setup. This scaling was found numerically in \citet{tan21} for plane-parallel mixing layers and we have written it here as $u^{\prime} \sim v_{\rm shear}^{4/5} v_0^{1/5}$ to preserve dimensionality ($v_0$ simply encodes normalization). If we set $v_{\rm shear} \sim v_{\rm T,grow} \sim g t_{\rm grow}$, this yields: 
\begin{equation} 
    t_{\rm grow} \sim \chi^{5/8} \frac{r^{15/32} t_{\rm cool}^{5/32}}{g^{3/8} v_0^{3/32}}; \ \ {v_{\rm T,grow} < c_{\rm s,hot}}.
    \label{eq:tgrow_nonumbers}
\end{equation}
While the above scalings focus on the subsonic and transonic cases, large enough clouds can reach velocities exceeding the sound speed of the hot gas. In such a case, the turbulent mixing velocity saturates and stops scaling with the cloud velocity \citep{yang22}, changing the above scalings. In this case, from equations \eqref{eq:tgrow-simple} and \eqref{eq:vmix-simple}, we obtain:
\begin{equation}
    t_{\rm grow} \propto \frac{\chi  r^{3/4} }{{\rm c_{\rm s,hot}^{3/5}} t_{\rm cool}^{1/4}}. 
\label{eq:tgrow-sat-simple}
\end{equation}

We now give numerical expressions, which are calibrated to simulations.
For cooling dominated regimes (defined below), \citet{tan21} found that $v_{\rm mix}$ in turbulent mixing layers follows
\begin{equation}
    v_{\rm mix} \approx 
    9.5 {\mathrm{~km}\,\mathrm{s}^{-1}}
    \left(\frac{u^{\prime}}{50 \, {\rm km \, s^{-1}}}\right)^{3/4} 
    \left(\frac{L_{\rm turb}}{100\,{\rm pc}}\right)^{1/4}
    \left(\frac{t_{\rm cool}}{0.03\,{\rm Myr}}\right)^{-1/4},
    \label{eq:vmix}
\end{equation}
where $L_{\rm turb}$ is the outer scale of the turbulence.
Note that equation~\eqref{eq:vmix} only applies in the `fast cooling' (${\rm Da}_{\rm mix} \equiv L_{\rm turb}/(u^{\prime} t_{\rm cool,mix}) > 1$, where ${\rm Da}_{\rm mix}$ is the Damk{\"o}hler number; \citealp{tan21}) regime, where the cooling time
is much smaller than the turbulent mixing time $L_{\rm turb}/u'$. As we will discuss below, however, this is always true for surviving clouds.

\citet{tan21} note that $u'$ is geometry dependent, but find for shearing layers that
\begin{equation}
    u' \approx 50{\mathrm{~km}\,\mathrm{s}^{-1}}
    \mathcal{M}^{4/5}
    \left(\frac{c_{\rm s,hot}}{150 {\mathrm{~km}\,\mathrm{s}^{-1}}} \right)^{4/5}
    \left(\frac{t_{\rm cool}}{0.03\,{\rm Myr}}\right)^{-0.1} ,
    \label{eq:uprime} 
\end{equation}
for $\chi\gtrsim 100$ and $\mathcal{M}\equiv v_{\rm shear} / c_{\rm s,hot}$. From equation~\eqref{eq:vmix}, we can approximate $v_{\rm mix} \sim c_{\rm s,cold}$ for quick estimates. While \citet{tan21} only considered mixing layers with subsonic to transonic velocity shears, \citet{yang22} found that beyond $\mathcal{M} = 1$, $u'$ in the mixing region stops scaling with $\mathcal{M}$ and saturates. We include this in our model by setting $\mathcal{M} \rightarrow \min(1,\mathcal{M})$. We find good evidence for this in our simulations.

Equations \eqref{eq:vmix} and \eqref{eq:uprime} assume fully developed turbulence. When a cloud falls from rest however, there is a transient period when turbulence is developing. We hence set a time dependent weight factor $w_{\rm kh}(t)$ to account for the initial onset of turbulence. Turbulence develops over the timescale for the development of the Kelvin-Helmholtz instability; on the scale of the cloud $t_{\rm kh} = f_{\rm kh} t_{\rm cc}$ where $f_{\rm kh}$ is some constant of proportionality \citep{klein94}. We use the simplest ansatz that 
\begin{equation} 
    {v}_{\rm mix} \rightarrow w_{\rm kh}(t) v_{\rm mix};  \ \ 
    w_{\rm kh}(t) = \min\left(1, \frac{t}{f_{\rm kh} t_{\rm cc}}\right),
\label{eq:w_kh} 
\end{equation} 
which amounts to $v_{\rm in}$ growing linearly with time over the instability growth time, until fully developed and capped at unity. We will justify this ansatz in our simulations. Since $t_{\rm cc}$ is changing over time, we note that $t/t_{\rm cc} \propto vt \sim ~z$, where $z$ is distance the cloud has fallen. We find in our simulations that $f_{\rm kh} \sim 5$ for a constant background and $\sim 1$ for a stratified background. In a more realistic setting with less idealized initial conditions, this time-dependent weight factor might not be necessary as the initial mixing can be already seeded from the outflowing section (assuming $v<v_{\rm esc}$), extrinsic turbulence, or cooling induced pulsations \citep{gronke20-mist,gronke22}.

What is an appropriate scaling relation for the effective cloud surface area $A_{\rm cloud}$? In cloud crushing simulations, areal growth follows two phases \citep{gronke20-cloud,abruzzo22}. In the `tail-formation' phase, surface area growth is dominated by the formation of a cometary tail, with $A_{\rm cloud} \propto L \propto m$, where $L$ is the length of the tail. The stretching of the cloud means that the area to mass ratio $A_{\rm cloud}/m \approx $ constant, rather than $A_{\rm cloud}/m \propto m^{-1/3}$, as for fixed geometry. Once the tail grows to a length $L \sim \chi r$ (the hydrodynamic drag length), the cloud becomes entrained in the wind from efficient momentum transfer, and due to lack of shear the tail no longer grows. The cloud surface area thereafter scales roughly as $A_{\rm cloud} \propto (m/\rho_{\rm cloud})^{2/3}$, as one would expect for a monolithic cloud.  

However, our falling clouds do not get entrained - rather the opposite in fact, as they start at rest and accelerate until reaching some terminal velocity. This means they start `entrained' and then begin to shear against background gas. They never leave the `tail-formation' phase, since there is a constant velocity difference between the cloud and background medium. The cloud sees a continuous headwind which drives turbulence, mixing, and lengthening. Instead of $A_{\rm cloud} \propto m/\rho_{\rm cloud}$ or $A_{\rm cloud} \propto (m/\rho_{\rm cloud})^{2/3}$, we assume that $A_{\rm cloud} \propto (m/\rho_{\rm cloud})^{\alpha}$, where $\alpha$ is a growth scaling exponent between 2/3 and 1. Physically, this is because both mass growth onto the surface of the cloud and a lengthening tail are concurrent processes. We will demonstrate that this is a good assumption for the mass growth of the falling clouds in our simulations, where we find $\alpha \approx$~5/6. The cloud surface area is thus
\begin{align}
    A_{\rm cloud} \approx A_{\rm cloud,0} \left( \frac{m}{m_0} \frac{\rho_{\rm cloud,0}}{\rho_{\rm cloud}} \right)^{\alpha},
    \label{eq:acloud}
\end{align}
where $A_{\rm cloud,0}$, $\rho_{\rm cloud,0}$ and $m_0$ are the initial cloud surface area, density and mass respectively. Note that since $\dot{m} \propto m^{\alpha}$ where $\alpha = 5/6$ is close to 1, the growth is close to exponential\footnote[3]{Similar scalings $\dot{m} \propto A \propto m^{\alpha}$, where $\alpha \approx 0.8$, are seen in simulations of cloud growth when clouds are embedded in a turbulent medium \citep{gronke21}. This super-Euclidean scaling can be understood as the outcome of the fractal nature of the mixing surface, where area $A \propto m^{D/3}$, where $D$ is the fractal dimension \citep{barenblatt83}. In their mixing layer simulations, \citet{fielding20} measure $D \approx 2.5$, which gives $\alpha \approx D/3 = 5/6$, consistent with the above.}. The cloud density $\rho_{\rm cloud}$ changes because the ambient pressure increases as the cloud falls in a stratified medium, compressing the cloud. 

Using equations~\eqref{eq:mdot}, \eqref{eq:vmix}, \eqref{eq:w_kh} and \eqref{eq:acloud}, we can write the growth time $t_{\rm grow} \sim m/\Dot{m}$ as
\begin{align}
    t_{\rm grow} = \frac{t_{\rm grow,0}}{w_{\rm kh}(t)}
                   \left( \frac{c_{\rm s,150}}{v} \right)^{3/5}
                   \left( \frac{t_{\rm cool}}{t_{\rm cool,0}} \right)^{1/4}
                   \left( \frac{m}{m_0} \frac{\rho_{\rm hot,0}}{\rho_{\rm hot}} \right)^{1-\alpha},
    \label{eq:tgrow_scaling}
\end{align}
where $c_{\rm s,150} = 150\,{\rm km\,s}^{-1}$ is the sound speed of gas at $10^6$~K and the initial growth time $t_{\rm grow,0}$ is given by
\begin{align}
    t_{\rm grow,0} \approx 35\,{\rm Myr }\,
                     \left( \frac{f_{\rm A}}{0.23} \right)
                     \left( \frac{\chi}{100} \right)
                     \left( \frac{r}{r_{100}} \right)
                     \left( \frac{L_{\rm turb}}{L_{100}} \right)^{-1/4}
                     \left( \frac{t_{\rm cool,0}}{0.03\,{\rm Myr}} \right)^{1/4}.
    \label{eq:tgrow_scaling2}
\end{align}
where $r_{100} = L_{100} = 100$\,pc and $r$ is the initial cloud size. We will assume generally that $L_{\rm turb} \sim r$ (since the hydrodynamic instabilities which drive turbulence and mixing have an outer scale set by cloud size). We have included an unknown normalization factor $f_{\rm A}$ to account for uncertainties arising from geometrical differences between the single mixing layers in \citet{tan21} and our cloud setup here, the use of the initial size of the sphere as a characteristic scale (see discussion at the end of Section~\ref{sect:discussion}), and any other simplifying assumptions we might have made. We find in our simulations that $f_{\rm A} \sim 0.23$.  
We can simplify equations \eqref{eq:tgrow_scaling} and \eqref{eq:tgrow_scaling2} by ignoring the weak mass and hot gas density dependence, and setting $L_{\rm turb} \sim r$, to obtain: 
\begin{align}
    t_{\rm grow}     = \frac{35\,{\rm Myr }}{w_{\rm kh}(t)} \,
                     \left(\frac{f_{\rm A}}{0.23}\right) \left( \frac{c_{\rm s,150}}{v} \right)^{3/5}
                     \left( \frac{\chi}{100} \right)
                     \left( \frac{r}{r_{100}} \right)^{3/4}
                     \left( \frac{t_{\rm cool}}{0.03\,{\rm Myr}}\right)^{1/4}.
    \label{eq:tgrow_terminal}
\end{align}

Equations (\ref{eq:tgrow_scaling}) and \eqref{eq:tgrow_scaling2} should be used when evaluating $t_{\rm grow}$ if the velocity $v(t)$ varies with time (i.e., when solving equations \eqref{eq:diff_eq1} -- \eqref{eq:diff_eq3}. However, a key quantity is the growth time at the terminal velocity $v=g t_{\rm grow}$, which we shall see determines whether the cloud can survive (Section \ref{sec:survival}). Inserting $v = g t_{\rm grow}$ into equation \eqref{eq:tgrow_scaling}, setting $w_{\rm kh}(t)=1$, and using $f_A=0.23$, we obtain the numerical version of equation~\eqref{eq:tgrow_nonumbers}: 
\begin{align}
    t_{\rm grow}     = 40\,{\rm Myr }\,
                     \left( \frac{g}{g_{\rm fid}}\right)^{-3/8}
                     \left( \frac{\chi}{100} \right)^{5/8}
                     \left( \frac{r}{r_{100}} \right)^{15/32}
                     \left( \frac{t_{\rm cool}}{0.03\,{\rm Myr}} \right)^{5/32},
    \label{eq:tgrow_vgt}
\end{align}
where $g_{\rm fid} = 10^{-8}\,{\rm cm\,s}^{-2}$. On the other hand, for supersonic speeds, as we have discussed, the turbulent mixing velocity saturates and stops scaling with the cloud velocity \citep{yang22}. Setting $v \sim gt_{\rm grow}$ to $ v \sim c_{\rm s,hot}$ instead in equation~\eqref{eq:tgrow_terminal}, we find the numerical version of equation~\eqref{eq:tgrow-sat-simple}: 
\begin{align}
    t_{\rm grow}     = 35\,{\rm Myr }\,
                     \left( \frac{c_{\rm s,hot}}{c_{s,150}} \right)^{-3/5}
                     \left( \frac{\chi}{100} \right)
                     \left( \frac{r}{r_{100}} \right)^{3/4}
                     \left( \frac{t_{\rm cool}}{0.03\,{\rm Myr}} \right)^{1/4}.
    \label{eq:tgrow_sat}
\end{align}

\subsection{Cloud Survival}
\label{sec:survival} 

The model we have presented only accounts for mass growth of the cloud and does not include processes that result in mass loss. In addition, the initial onset and development of turbulence is only very crudely incorporated. The absence of these refinements mean that we should expect differences between model predictions and simulations, certainly for clouds that are losing mass, and at early times even for clouds that do survive and grow. We leave the inclusion and refinement of these components for future work, as we find that the model as presented works well for surviving clouds. Since the key assumption of our model is that the cloud is growing, we now discuss when this is a valid assumption. 

As we previously discussed, clouds placed in a wind tunnel encountering a hot wind can survive if $t_{\rm cool, mix} < t_{\rm cc}$ (equation~\eqref{eq:t_cool_mix}). 
Physically, $t_{\rm cool,mix}$ can be understood as the time it takes gas to cool in the downstream tail region of the cloud. Even if the initial pristine cloud material does not survive, if {\it mixed} gas can cool and survive, then the cold gas mass will increase. Since this mixed gas in the tail is cooling from the background, it is much more entrained in the wind than the initial cloud and hence able to survive -- once the cold gas is entrained, it is no longer subject to destruction by shear. 

The `usual' survival criterion $t_{\rm cool,mix} < t_{\rm cc}$ above is certainly a necessary condition for survival. If no gas can cool before the cloud is completely disrupted, the cloud cannot survive. However, this criterion is not a sufficient one. This is because the physical process associated with $t_{\rm cc}$ is not simply surface evaporation. If this were so, then the above criterion would indeed be sufficient as any mixing would lead to a net increase in cloud mass. Instead, the entire cloud is disrupted \citep[i.e. the cloud is broken up into smaller fragments;][]{klein94,schneider17}. Hence, as we shall see, it is not enough that mixed gas can cool faster than the cloud crushing time.

Compared to a wind tunnel setup, the considerations for an infalling cloud are different. Since the cloud's velocity increases instead, and there is no entrainment, $t_{\rm cc}$ decreases over time.
The only way for cold gas to survive is if it is produced at a rate faster than it is destroyed: 
\begin{align}
    t_{\rm grow} < f_{\rm S} t_{\rm cc},
    \label{eq:survive}
\end{align}
where $f_{\rm S}$ is some constant\footnote[4]{Although we find that a constant factor is sufficient for our purposes, this coefficient has been found to vary in supersonic flows. For example, \citet{scannapieco15} found that in the cloud crushing setup with a supersonic wind, $f_{\rm S}$ scales as $\sqrt{1+\mathcal{M}_{\rm hot}}$ where $\mathcal{M}_{\rm hot}$ is the Mach number of the hot medium \citep[see also][for alternative scalings]{li20,bustard2021}. However, we mostly probe the subsonic to transonic regime.} factor of order unity, which we shall calibrate in simulations. It encodes the fact cloud destruction takes place over several cloud crushing times \citep{klein94,scannapieco15}. In evaluating $t_{\rm cc} \sim \chi^{1/2} r/v$, the cloud radius is evaluated at its initial value. As in wind tunnel experiments, this turns out to be a very good approximation, since the cloud grows mostly in the streamwise direction. If the velocity is evaluated at the terminal velocity $v_{\rm T} \sim g t_{\rm grow}$, then equation \eqref{eq:survive} is equivalent to: 
\begin{equation}
    \frac{g t_{\rm grow}^{2}}{\chi^{1/2} r} < f_{\rm S}. 
    \label{eq:survive-terminal} 
\end{equation}

As we have seen, there are two regimes for $t_{\rm grow}$, subsonic and supersonic infall. The criterion for subsonic infall is $t_{\rm grow} < t_{\rm ff}$ (equation~\eqref{eqn:alpha}). Using equation~\eqref{eq:tgrow_vgt}, and assuming $t_{\rm ff} \sim c_{\rm s,hot}/g$, this can be rewritten as $r< r_{\rm sonic}$, where
\begin{equation}
    r_{\rm sonic} \sim 150\,{\rm pc }\,
         \left(\frac{t_{\rm cool}}{0.03\,{\rm Myr}}\right)^{-1/3}
         \left( \frac{g}{g_{\rm fid}} \right)^{-4/3}
         \left( \frac{\chi}{100} \right)^{-4/3}
        \left( \frac{c_{\rm s,hot}}{c_{\rm s,150}} \right)^{32/15}.
    \label{eq:rsonic}
\end{equation}
Thus, clouds must be {\it smaller} than some critical radius to fall at sub-virial velocities. In this regime, ($v_{\rm T} \lsim c_{\rm s,hot}$), $t_{\rm grow}$ is given by equation~\eqref{eq:tgrow_vgt}, and the survival criterion, equation~\eqref{eq:survive-terminal}, becomes: 
\begin{eqnarray}
    t_{\rm cool} < 5 \times 10^{-3} \, {\rm Myr} 
    \left(\frac{f_{\rm S}} {2}\right)^{16/5}
    \left(\frac{r}{r_{100}}\right)^{1/5} 
    \left( \frac{g}{g_{\rm fid}} \right)^{-4/5} 
    \left( \frac{\chi}{100} \right)^{-12/5} .
    \label{eq:survival_criterion}
\end{eqnarray}
Note that equation \eqref{eq:survival_criterion} is almost independent of cloud size. Indeed, $t_{\rm grow}/t_{\rm cc} \propto g t_{\rm grow}^2/r \propto r^{-1/16}$, i.e., a very weak scaling. We shall verify this in Section~\ref{sec:results-survival}. 

Is it possible for clouds to survive in the supersonic regime ($r> r_{\rm sonic}$)? This requires $t_{\rm ff} < t_{\rm grow} < f_S t_{\rm cc}$. This in turn requires that clouds be smaller than some critical size $r_{\rm SS}$, since $t_{\rm grow}/t_{\rm cc} \propto gt_{\rm grow}^{2}/r \propto r^{1/2}$ in the supersonic regime (using $t_{\rm grow} \propto r^{3/4}$ from equation \eqref{eq:tgrow_sat}). Thus, supersonic infall and survival requires: 
\begin{equation}
    r_{\rm sonic} < r < r_{\rm SS} ,
    \label{eq:survive_supersonic}
\end{equation} 
where $r_{\rm SS}$ is given by: 
\begin{equation}
    r_{\rm SS} = 100\,{\rm pc }\,
         \left(\frac{t_{\rm cool}}{0.03\,{\rm Myr}}\right)^{-1}
         \left( \frac{g}{g_{\rm fid}} \right)^{-2}
         \left( \frac{\chi} {100}\right)^{-2}
         \left( \frac{c_{\rm s,hot}}{c_{\rm s,150}} \right)^{12/5}.
    \label{eq:rSS}
\end{equation}
Note that equation~\eqref{eq:survive_supersonic} can only be fulfilled if $r_{\rm sonic}/r_{\rm SS} < 1$, where: 
\begin{equation}
    \frac{r_{\rm sonic}}{r_{\rm SS}} \sim 1.5 
    \left(\frac{t_{\rm cool}}{0.03\,{\rm Myr}}\right)^{2/3} 
    \left( \frac{g}{g_{\rm fid}} \right)^{2/3} 
    \left( \frac{\chi} {100}\right)^{2/3} 
    \left( \frac{c_{\rm s,hot}}{c_{\rm s,150}} \right)^{-4/15}.
\end{equation}
\begin{figure}
    \centering
	\includegraphics[width=\columnwidth]{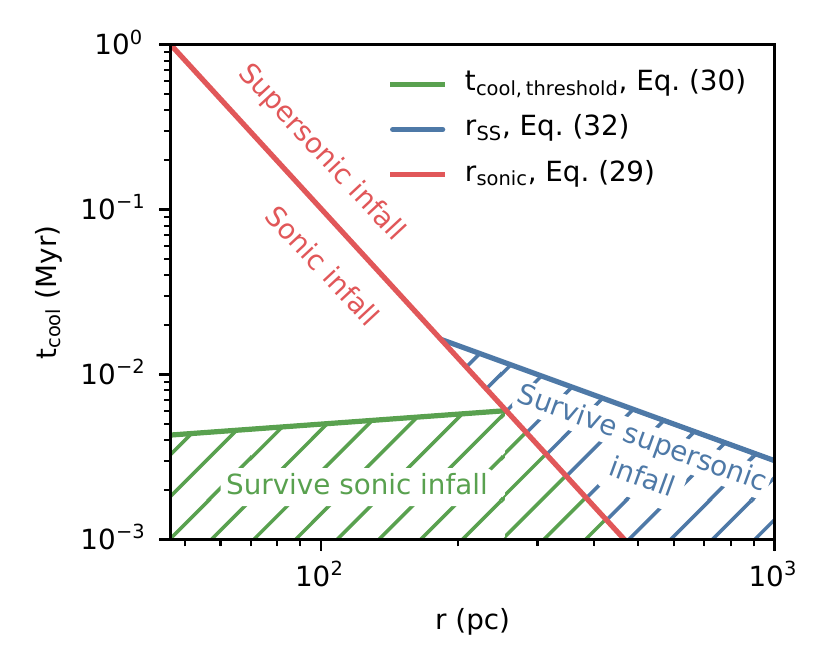}
	\caption{Cloud survival for subsonic and supersonic infall for different cloud          sizes and cooling times. Survival is mostly sensitive to the latter.}
	\label{fig:surv_analytic}
\end{figure}
Figure~\ref{fig:surv_analytic} shows the survival criteria above (equations \eqref{eq:survival_criterion} and \eqref{eq:rSS}) for $g = g_{\rm fid}$, $\chi=100$, $c_{\rm s,hot}=c_{\rm s,150}$, and $f_{\rm S}=2$. It is clear that survival is mostly independent of cloud size and depends instead on the cooling time. 

In practice, the subsonic case is of most interest. There, clouds must satisfy $t_{\rm grow} < {\rm min}(t_{\rm ff}, t_{\rm cc})$, which translates into a maximum allowed cloud size (equation~\eqref{eq:rsonic}) and a maximum allowed cooling time in cold gas (equation~\eqref{eq:survival_criterion}). The latter criterion is quite stringent. Since the dependence on size in equation (\ref{eq:survive_supersonic}) is weak, under isobaric conditions $t_{\rm cool} \propto 1/P$, we can translate equation~\eqref{eq:survival_criterion} into a critical pressure. For $f_{\rm S} = 2$, and ignoring the size dependence, equation~\eqref{eq:survival_criterion} is equivalent to:
\begin{align}
    P > 3000 \,{\rm k}_B\,{\rm K}\,{\rm cm}^{-3}
    \left(\frac{g}{g_{\rm fid}}\right)^{4/5} \left(\frac{\chi}{100}\right)^{12/5} ,
    \label{eq:safe_zone}
\end{align}
where the RHS is the critical pressure $P_{\rm crit}$ above which a falling cloud can survive. We can also write equation~\eqref{eq:survival_criterion} in terms of the cooling time of the hot gas $t_{\rm cool,hot} \sim \chi^2 t_{\rm cool} [\Lambda(T_{\rm cold})/\Lambda(T_{\rm hot})]$ and the free fall time $t_{\rm ff} \sim c_{\rm s,hot}/g$ to obtain: 
\begin{align}
    \frac{t_{\rm cool,hot}}{t_{\rm ff}} \lesssim 1 \left( \frac{\chi}{100} \right)^{-2/5} \left( \frac{\Lambda(T_{\rm cold})}{\Lambda(T_{\rm hot})} \right),
\label{eq:tcool_over_tff}
\end{align}
where we have ignored the weak dependence on $g$, $t_{\rm cool}/t_{\rm ff} \propto g^{1/5}$. This is similar to the criterion ($t_{\rm cool,hot}/t_{\rm ff} < 1$, where $t_{\rm cool}$ is evaluated at one scale height) for precipitation out of a thermally unstable background medium in a plane parallel atmosphere\footnote[5]{It is somewhat more stringent than the requirement for thermal instability in spherical systems ($t_{\rm cool,hot}/t_{\rm ff} < 10$; \citealt{sharma12}), where the gravitational acceleration $g$ and hence $t_{\rm ff}$ varies as a function of radius. However, it has been shown that there is no geometrical difference in cold gas condensation in plane parallel and spherical geometries; the apparent difference arises from definitional differences in where $t_{\rm cool}/t_{\rm ff} $ is evaluated and cold gas is located \citep{choudhury16}.} \citep{mccourt12}. Since all our analytics and simulations are in the framework of plane parallel systems, the numerical factor in equation~\eqref{eq:tcool_over_tff} will likely change in spherical systems. Equation~\eqref{eq:tcool_over_tff} has the very interesting implication that clouds which condense via thermal instability are able to survive subsequent infall, as long as they are below the critical size given by equation~\eqref{eq:rsonic}. Note that the physics of stratified thermal instability which leads to the $t_{\rm cool,hot}/t_{\rm ff} < 1$ criterion -- overstable gravity waves driven by cooling -- is quite different from what we have discussed here, so it is non-trivial (perhaps coincidental) that both thermal instability and falling cloud survival have similar criteria. 

\dobib

\section{Methods}  \label{sect:methods}
We carry out our simulations using the publicly available MHD code Athena\verb!++! \citep{stone20}. All simulations are run in 3D on regular Cartesian grids using the HLLC approximate Riemann solver and Piecewise Linear Method (PLM) applied to primitive variables for second order spatial reconstruction. By default, we use the second-order accurate van Leer predictor-corrector scheme for the time integrator, but switch to the third-order accurate Runge-Kutta method when the former is not stable enough, in particular for simulations with a constant background where the cooling time is extremely short throughout the entire simulation. 

Our simulation setups consist of rectangular boxes with identical $x, y$ dimensions and an extended vertical $z$ axis. They are filled with static hot $T_{\rm hot} = 10^6$\,K gas with initial density $n_0 = 10^{-4}\ccm$ at $z=0$. A cold $T_{\rm cold} = 10^4$\,K spherical cloud, initially at rest, is also inserted, usually a quarter box height from the bottom. This placement allows us to follow the development of a cometary tail behind the cloud as it falls. The initial cloud density is perturbed at the percent level randomly throughout the cloud to reduce numerical artifacts arising from the initial symmetry. We use outflowing boundary conditions, except at the bottom of the box (negative $z$) where the background profile is enforced in the ghost cells and the velocity is set to be that of the frame velocity. This is valid as long as cloud material does not interact with this bottom boundary. The frame velocity is based on a cloud-tracking scheme we implement where we continuously shift into the reference frame of the center of mass of the cold gas, defined as gas below a temperature of $T \sim 2 \times 10^4$\,K, an approach widely used in similar falling cloud simulations \citep{heitsch22} and wind tunnel simulations \citep{mccourt15,gronke18,gronke20-cloud}. This scheme allows our simulation box to `track' the cloud as it falls and hence reduces computational costs. The fiducial resolution of the boxes are $256^2\times2048$ (see Section~\ref{subsect:resolution} for a resolution test). The dimensions of the boxes are $10^2 \times 80$\,$r_{\rm cloud}$. This translates to $r_{\rm cloud}$ being resolved by $\sim 25$ cells.

During the simulations, the clouds are allowed to fall freely under gravity. We assume a constant gravitational acceleration $\boldsymbol{g} \equiv - g \hat{z} $, with $g = 10^{-8}$\,cm\,s$^{-2}$, as appropriate for the Milky Way, taken from the fit in \citet{benjamin&daly97} for distances between 1 and 10\,kpc. We discuss the impact of a more realistic gravitational profile and apply them within the scope of our model in Section \ref{sect:discussion}.

In our implementation of radiative cooling, we assume collisional ionization equilibrium (CIE) and solar metallicity ($X=0.7$, $Z=0.02$)\footnote[6]{We phrase our results in terms of cooling times, so they can easily be scaled for different cooling curves. We note however, that the minimum cooling time at $T \sim 1.5 \times 10^{4}$K, which is dominated by hydrogen cooling, is relatively insensitive to metallicity.}. We obtain our cooling curve by performing a piece-wise power law fit to the cooling table given in \cite{Gnat2007} over 40 logarithmically spaced temperature bins, starting from a temperature floor of $10^4$\,K, which we also enforce in the simulation. We then implement the fast and robust exact cooling algorithm described in \citet{townsend}. For this cooling curve, the cooling time in the cold gas is $t_{\rm cool} \sim 0.15$\,Myr. To emulate the effect of heating and to prevent the background medium from cooling over simulation timescales, we cut off any cooling above $5 \times 10^5$\,K. The particular choice of this value is unimportant \citep{gronke18,gronke20-cloud,abruzzo21}.

We run two different sets of simulations with different static background profiles. The first set has gravity acting on a cloud which is embedded in a constant background, i.e. constant hot gas temperature, density and pressure. This is obviously unphysical, since there are no pressure gradients in the background to counteract gravity. However, it is very useful for understanding the underlying physical mechanisms which affect the cloud, without the confounding effects of the varying background which a cloud falling through a stratified medium experiences. To prevent the background from falling under gravity, we introduce an artificial balancing force $\rho_{\rm hot}g$ upwards. The hot background thus feels a net zero force from gravity, while the cold cloud is negligibly affected. For this set of simulations, we also vary the cooling time by changing the normalization of the cooling function by a constant factor $\Lambda_0$. For example, $\Lambda_0 = 100$ would be a case where cooling is a hundred times stronger than the fiducial value, corresponding to cooling an environment where $n_{\rm hot} = \Lambda_0 n_0 = 10^{-2}\ccm$, or $nT = 10^4{\rm \,K} \ccm$, a relatively high pressure. For the constant background, we adopt $\Lambda_0 = 100$ as a default, so that cooling is extremely strong and cloud growth is guaranteed.
We emphasize that the constant background is simply used to provide a clean test of our analytic model, so that (for instance) the cooling time is not a function of position, as in a stratified atmosphere.  

The second, more realistic, setup is that of a hydrostatic isothermal halo. The density profile of the background is thus:
\begin{align}
    n(z) = n_0 \exp(-\frac{gm_H}{k_B T_{\rm hot}}z),
\end{align}
where $n_0$ is the midplane density, $z$ is the height above the disk and $H \equiv k_B T_{\rm hot}/g m_H = 2.8$\,kpc is the isothermal scale height (assuming the mean molecular weight $\mu = 1$). This is a simplified model that is likely to break down close to the disk below 2\,kpc, where it likely underestimates the background density, since the background gas is cooler. However, this simple model allows us to study the effects of both a changing background profile and the resultant decrease in cooling time as the cold gas falls inwards. Besides the initial setup of the background profile, since we are employing a cloud-tracking scheme, the boundary cells are set accordingly throughout the course of the simulation using this background profile and the current height of the cloud, which we also track.

Our cloud chambers are somewhat artificial in that they are arbitrarily long. Thus, for instance, in the stratified case, the cloud can fall through an unrealistically large number of scale heights (well beyond when the plane parallel approximation is valid). In practice, transition to a spherical gravitational potential with declining gravitational acceleration $g$ means that even if clouds fall ballistically, they will only accelerate to transonic velocities, rather than fall supersonically. However, our setup is a clean probe of the underlying physics. In all the cases we care about, where the cloud survives, infall is subsonic. 

In order to evaluate the cold gas mass $m$ as well as other related quantities such as the mass growth rate, we use a temperature threshold of $T \sim 2 \times 10^4$\,K below which we define the gas to be `cold'. No magnetic fields are included in our simulations. We leave the exploration of the MHD case to future work.

\dobib

\section{Results : Constant Background}  
\label{sect:results}

Our first objective is to test our semi-analytic model for falling clouds (equations~\eqref{eq:diff_eq1} -- \eqref{eq:diff_eq3}) against full 3D simulations. Hence, the first set of our simulations are set up with a constant background, where the properties of the background medium are held unchanged as the cloud falls. We use this setup as a simple way to explore and test our model in an environment where the cooling time is kept constant. This allows us to test the various components of our model by adjusting individual parameters, ceteris paribus.

\subsection{Time Evolution}
\begin{figure*}
    \centering
	\includegraphics[width=\textwidth]{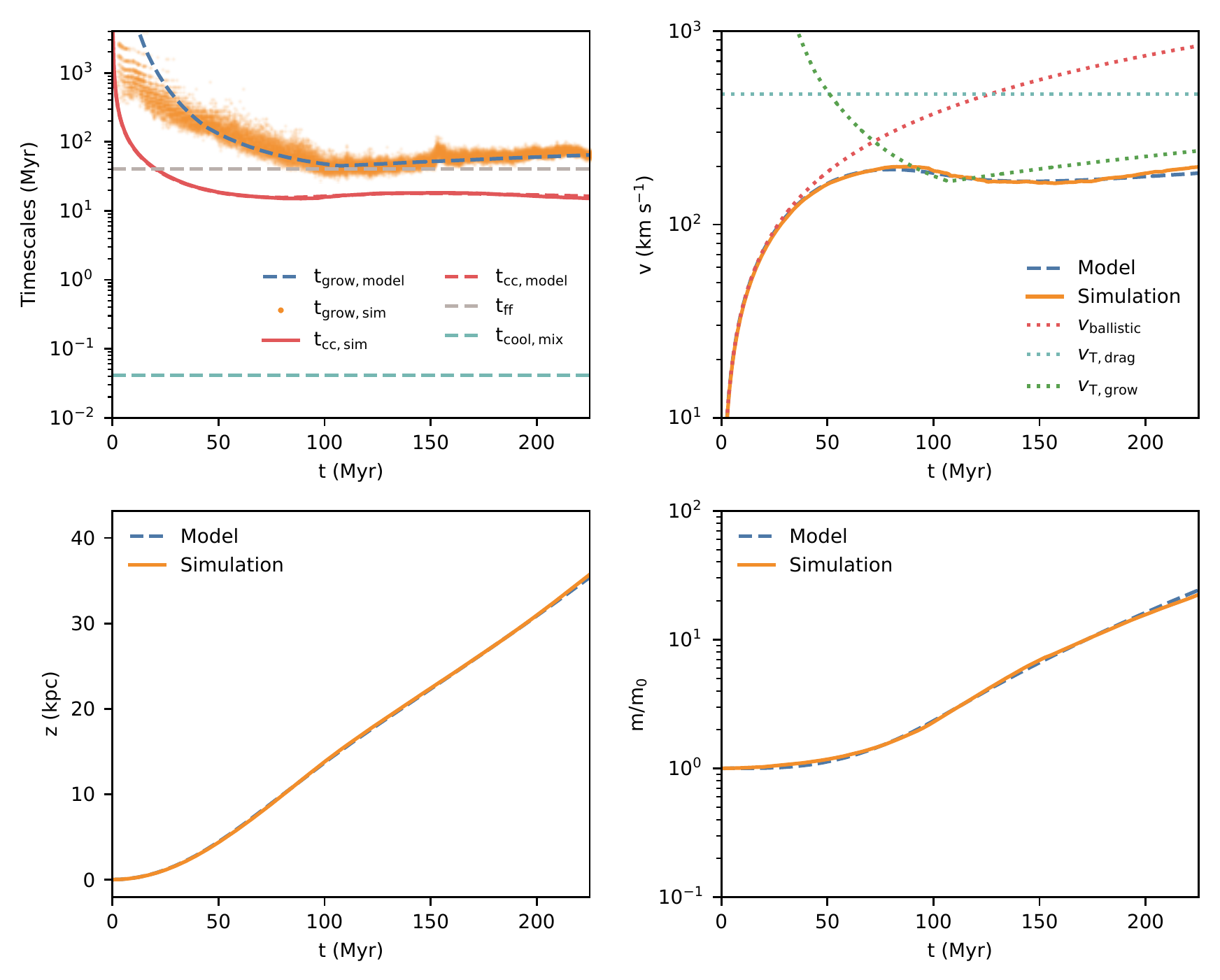}
	\caption{Time evolution of various quantities for a $r = 300$~pc cloud falling in a constant background. From left to right, top to bottom, the panels compare the growth time $t_{\rm grow}$, the velocity $v$, the distance fallen $z$, and the cold gas mass $m$ of the cloud in the simulation versus the model. The upper panels also include comparison with other quantities of interest. Model predictions are in good agreement with simulations results.}
	\label{fig:02}
\end{figure*}

In order to understand the dynamical evolution of a falling cloud, we first present the time history of various quantities of interest, both as predicted by the model and as seen in the simulations. Note that the model (equations \eqref{eq:diff_eq1} -- \eqref{eq:diff_eq3}) predicts $m(t)$, $v(t)$, and $z(t)$ independently, without using any input from the simulations. Figure~\ref{fig:02} shows the evolution of these quantities over the course of a simulation with an initial cloud radius $r = 300$\,pc. These are, from left to right and top to bottom - timescales, cloud velocity, distance fallen, and the total mass of cold gas. The simulation runs for over 200\,Myr, which is between 10 to 15 cloud crushing times.

The various timescales shown in the upper left panel of Fig.~\ref{fig:02} are as follows: The cooling time of the {\it mixed} gas $t_{\rm cool, mix}$, where mixed gas is defined as gas at $T_{\rm mix} \sim \sqrt{T_{\rm hot} T_{\rm cold}} \sim 10^5$\,K, the free-fall time $t_{\rm ff} = c_{\rm s,hot}/g$, the cloud crushing time $t_{\rm cc} = \sqrt{\chi}r/v$, which uses the {\it initial} cloud radius $r$ and the instantaneous cloud velocity, and the instantaneous cloud growth time $t_{\rm grow} = m/\Dot{m}$, computed using the mass of cold gas (defined as gas with $T<2\times10^4$\,K). For the latter two timescales ($t_{\rm cc}$ and $t_{\rm grow}$), both model and simulation results are shown for comparison. While wind tunnel setups define $t_{\rm cc}$ using the initial wind velocity, we use the instantaneous cloud velocity (defined as the center of mass velocity of cold gas) instead. This changes with time - it is initially infinitely long since the cloud starts at rest, but decreases as the cloud accelerates. Similarly, $t_{\rm grow}$ is initially infinite, since there is no turbulence at the start of the simulation (any mixing would be due to numerical diffusion, since we do not implement physical diffusion). Mass growth then begins with the initial onset of turbulence, which we have included in the model via the weight term $w_{\rm kh}(t)$. 
Our crude model for $w_{\rm kh}(t)$ means that our analytic model for $t_{\rm grow}$ is less accurate at these times. However, since $t_{\rm grow}$ is in any case long in these stages, with mass increasing very slowly, inaccuracy in modeling the growth of turbulence fortunately has little impact on $m(t)$ (and by extension $v(t)$ and $z(t)$). The model performs well at matching the simulation results for both $t_{\rm cc}$ and $t_{\rm grow}$. Since $t_{\rm grow} \sim t_{\rm ff}$, the terminal velocity of the cloud here is roughly the sound speed of the hot gas, as expected from equation~\eqref{eqn:alpha}. For all simulations, $t_{\rm cool,mix} \ll t_{\rm cc}$, as required to be in the fast cooling regime.

The upper right panel of Fig.~\ref{fig:02} shows the velocity evolution of the cloud, as measured by the center of mass velocity of the cold gas. We also show the velocity trajectory from the model, along with three other characteristic velocities. These are the ballistic velocity $v_{\rm ballistic} = gt$ and the `terminal' drag and growth velocities $v_{\rm T,drag}$ and $v_{\rm T,grow}$ respectively, as given by equations~\eqref{eq:vtdrag} and \eqref{eq:vtgrow}. The terminal velocities\footnote[8]{While we use the terminology of a `terminal' velocity, $v_{\rm T,grow} \approx g t_{\rm grow}$ is in fact time dependent here since $t_{\rm grow}$ has a mass dependence.} are computed using the size of the initial cloud, and we can see that $v_{\rm T, grow} < v_{\rm T,drag}$, as expected. 
The ram pressure drag experienced by the cloud is thus much weaker than the mixing-cooling induced drag due to momentum transfer as hot surrounding gas is accreted onto the cloud (as expected from the estimates presented in Section~\ref{sec:terminal_velocity}). The relative contribution of ram pressure drag can be seen in the small deviation of the model (which includes both effects) from $v_{\rm T,grow}$. The cloud initially accelerates ballistically, before reaching a high enough velocity where the cooling drag force kicks in and slows the cloud down. 
Since the cooling drag force operates on a timescale $t_{\rm grow}$, the cloud remains ballistic until $t \sim t_{\rm grow}$. This progression means that the cloud can experience a phase where its velocity is {\it decreasing} as it falls. While not strongly apparent in this setup, this effect can be pronounced when the background is not constant, which we discuss in the following section. The model does an excellent job at matching the evolution of the cloud velocity over time, and in particular the cloud reaches the asymptotic velocity $v_{\rm T,grow} \approx g t_{\rm grow}$ predicted by the mixing and cooling induced accretion of hot background gas.

The remaining two lower panels of Fig.~\ref{fig:02} show the distance the cloud has fallen and the total mass of cold gas. Of course, the two quantities are not independent from the upper panels: we expect to predict $z(t)$ accurately since we predict $v(t)$ accurately, and we expect to predict $m(t)$ accurately since we could predict $t_{\rm grow}$ accurately. Overall, it is remarkable how well our simple model of `accretion braking' matches the simulations. We now explore how it performs in different regions of parameter space.
 
\subsection{Area Growth Rate}
\begin{figure}
    \centering
	\includegraphics[width=\columnwidth]{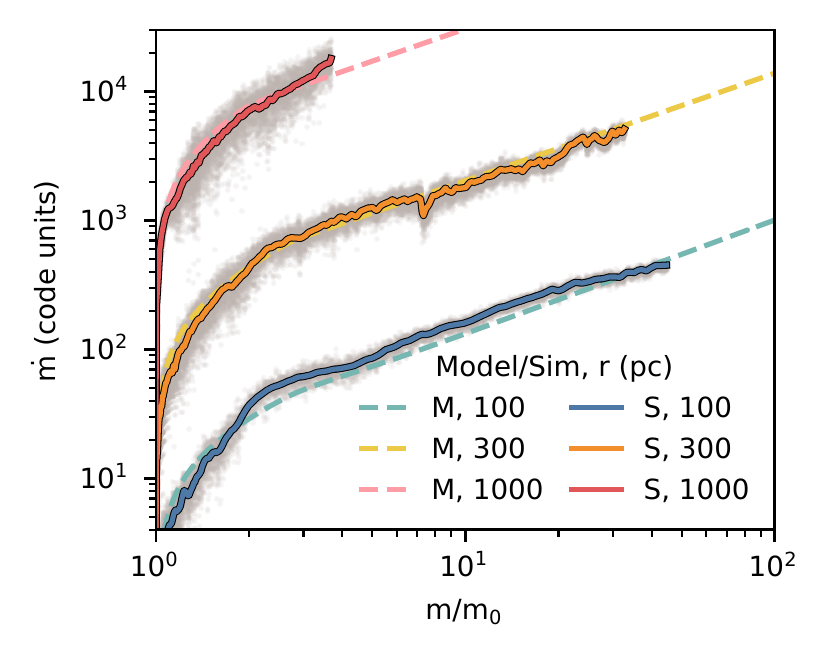}
	\caption{The mass growth rate as a function of cold gas mass for clouds of different initial sizes. Curves are labelled by the initial cloud radius and whether they represent model solutions (M) or simulations (S), which are shown as dashed and solid lines respectively. Using a scaling of $\alpha=5/6$ in the model matches the mass growth rate in the simulations well.}
	\label{fig:03}
\end{figure}

We first investigate the areal growth scaling in equation (\ref{eq:acloud}), where we stated that we expect the value of $\alpha$ to lie between 2/3 and 1. Equation~\eqref{eq:tgrow_scaling} can be rewritten as 
\begin{align}
    \Dot{m} = 
    \frac{m_0}{t_{\rm grow,0}} 
    \left( \frac{v}{c_{\rm s,6}} \right)^{3/5}
    \left( \frac{m}{m_0} \right)^\alpha.
\end{align}

Figure~\ref{fig:03} shows the mass growth rate of cold gas $\Dot{m}$ as a function of the cold gas mass $m$ normalized by the initial cloud mass $m_0$ in three simulations with $r = 100$, 300 and 1000\,pc. We expect from our model that past the turbulent onset and acceleration phases, the cloud should reach terminal velocity and its mass growth rate should thus follow lines with slope $\alpha$. The dashed lines in Fig.~\ref{fig:03} show mass growth rate curves from our model with $f_A = 0.23$ and $\alpha = 5/6$. These choice of values give a good match to the mass growth rate curves from simulations represented by the solid lines, which are obtained by smoothing the instantaneous values of $\Dot{m}$ represented by the grey points. The slopes are initially steeper as the cloud accelerates. As discussed in the Section~\ref{sect:analytics}, we find that $\alpha \sim 5/6$ seems to be an good fit to simulation data -- supporting the idea that both processes of cloud growth on the surface ($\alpha\sim 2/3$) and in a lengthening tail ($\alpha \sim 1$) are at play (or that the effective surface area scales in a fractal manner). 

As noted above, we also observe a `burn-in phase', where the mass growth is initially low because turbulence is developing around and behind the cloud due to instabilities, then ramps up quickly due to both turbulent onset and a rapid increase in surface area. Small sudden drops are associated with cold mass that exits the simulation box due to its fixed size, which are likely to occur at late times in our simulations. The computational cost of tracking cloud growth over longer periods of time increases significantly as the clouds keep growing in size and length which require increasingly larger boxes to contain. For the large 1\,kpc radius cloud, we were unable to run the simulation for a sufficient time to see the mass growth rate reach the same steady growth as convincingly as the smaller clouds, but nevertheless the mass growth is in line with model predictions for all growing clouds. 

\subsection{Scalings}
To verify our analytic scalings for $t_{\rm grow}$ in the subsonic and supersonic regimes, equations~\eqref{eq:tgrow_vgt} and \eqref{eq:tgrow_sat}, we vary each parameter to test the scalings explicitly. However, the parameters cannot be arbitrarily varied -- they are limited to the region of parameter space where the clouds survive. This is given by equation \eqref{eq:survival_criterion} and \eqref{eq:survive_supersonic} for subsonic and supersonic infall respectively.  
\begin{figure}
    \centering
	\includegraphics[width=\columnwidth]{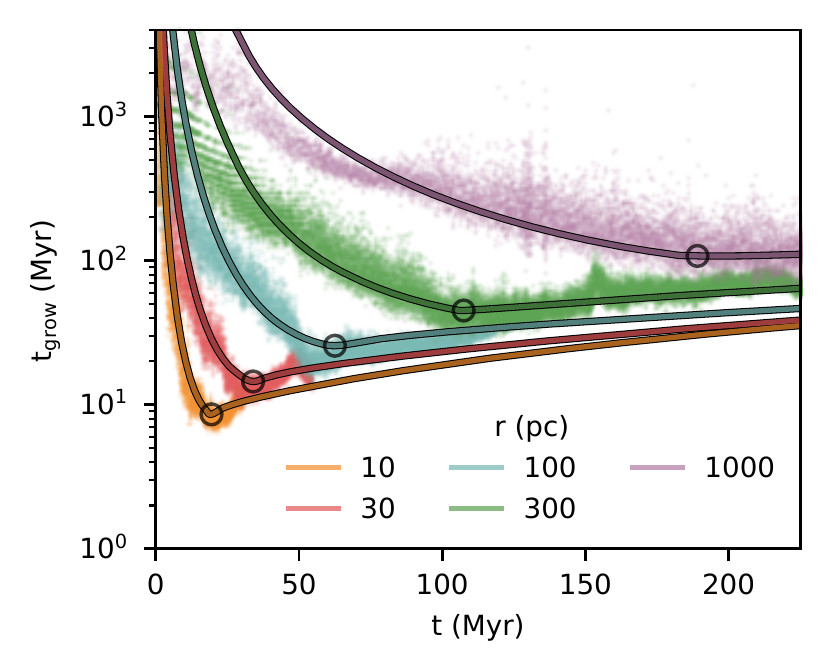}
	\includegraphics[width=\columnwidth]{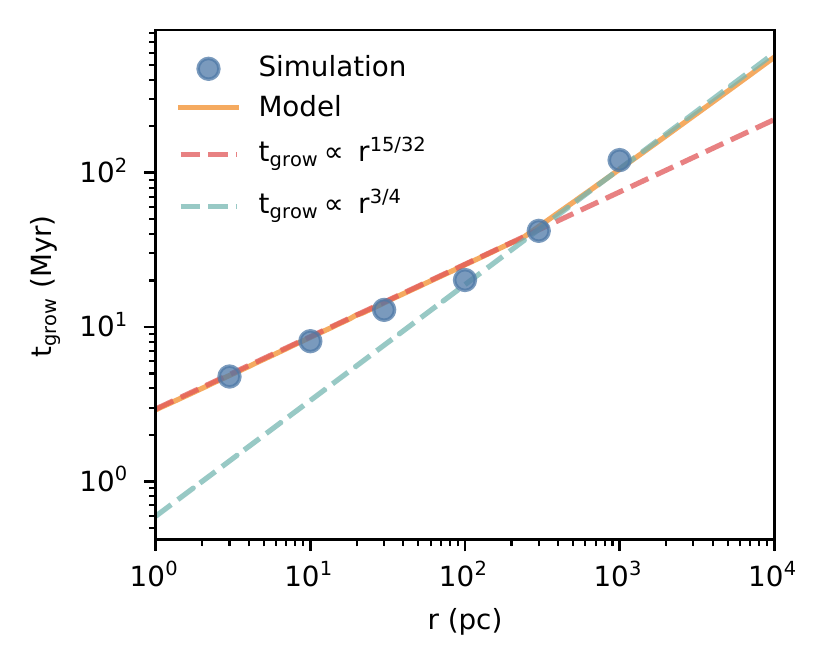}
	\caption{{\it Upper panel}: The growth time as a function of time for clouds of different sizes in the effective cooling regime ($\Lambda_0 = 100$). All clouds shown here are growing and survive. Solid lines show model predictions, while colored points represent simulation results. {\it Lower panel}: The growth time where turbulence is fully developed ($w_{\rm kh}(t)=1$) as a function of cloud size. Dashed lines show expected analytical scalings in the subsonic ($t_{\rm grow} \propto r^{15/32}$) and supersonic ($t_{\rm grow} \propto r^{3/4}$) regimes, while the solid orange line shows the model predictions. Both are in agreement.}
	\label{fig:04}
\end{figure}
\begin{figure}
    \centering
	\includegraphics[width=\columnwidth]{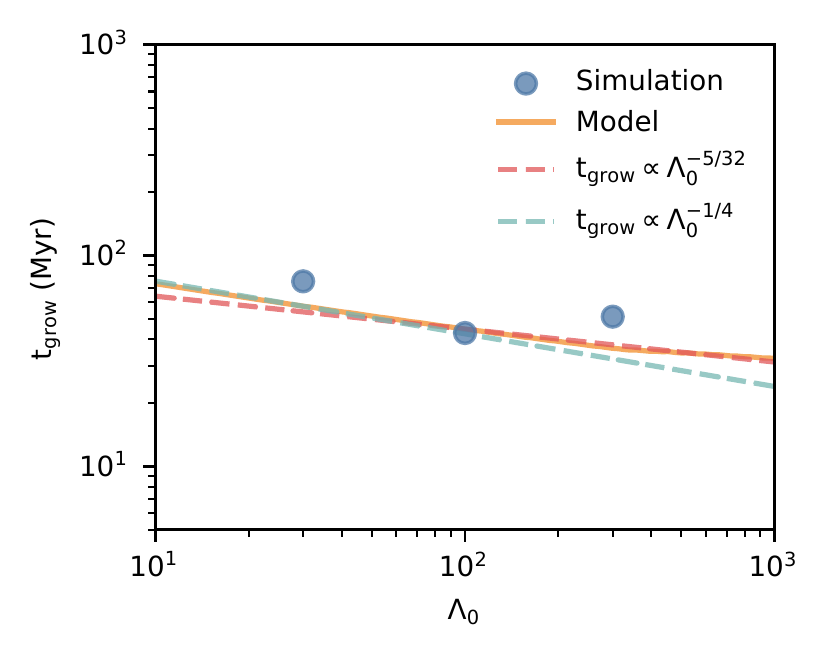}
	\caption{The growth time for different cooling strengths $\Lambda_0$, which modify the cooling time $t_{\rm cool} \propto \Lambda_0^{-1}$. Dashed lines show expected analytical scalings, while the solid orange line shows the model predictions. As expected, the dependence of $t_{\rm grow}$ on $t_{\rm cool}$ is weak.}
	\label{fig:04b}
\end{figure}

\subsubsection{Scaling With Cloud Size}
We first vary the initial cloud size $r$. The upper plot of Fig.~\ref{fig:04} shows $t_{\rm grow}$ as a function of time for the range of cloud sizes, while the lower plot shows the scaling of $t_{\rm grow}$ with $r$, measured at the times indicated by the black circles in the upper plot where the weight function in the model reaches unity, or in other words, turbulence and mixing has fully developed. In the upper plot, simulation results are represented by the small points colored by cloud size. Solid lines show model predictions. In the lower plot, the orange line represents the model predictions while the analytic scalings of $r^{15/32}$ and $r^{3/4}$ derived above (before and after saturation of turbulent velocities for subsonic and supersonic infall respectively) are plotted as dashed lines. The simulation results match the model and analytic scalings.

\subsubsection{Scaling With Cooling}
Next, we vary the cooling strength parameter $\Lambda_0$ by a factor of 3 above and below the fiducial value. Figure~\ref{fig:04b} shows the scaling of $t_{\rm grow}$ with $\Lambda_0 \propto 1/t_{\rm cool}$, along with the simulation and model results as before. The simulations are in agreement with the weak $t_{\rm cool}$ scaling. Despite this, as we will see later, survival is sensitive to cooling time rather than size, and hence it is difficult to probe the scaling to weaker cooling. Unfortunately, reducing the cooling strength further leads to cloud destruction. Higher cooling strengths require shorter timesteps and larger boxes, and are hence numerically challenging. While we vary the cooling strength explicitly here, strong cooling also corresponds to denser environments where higher densities lead to shorter cooling times. 
\subsubsection{Scaling With Gravity}
\begin{figure}
    \centering
	\includegraphics[width=\columnwidth]{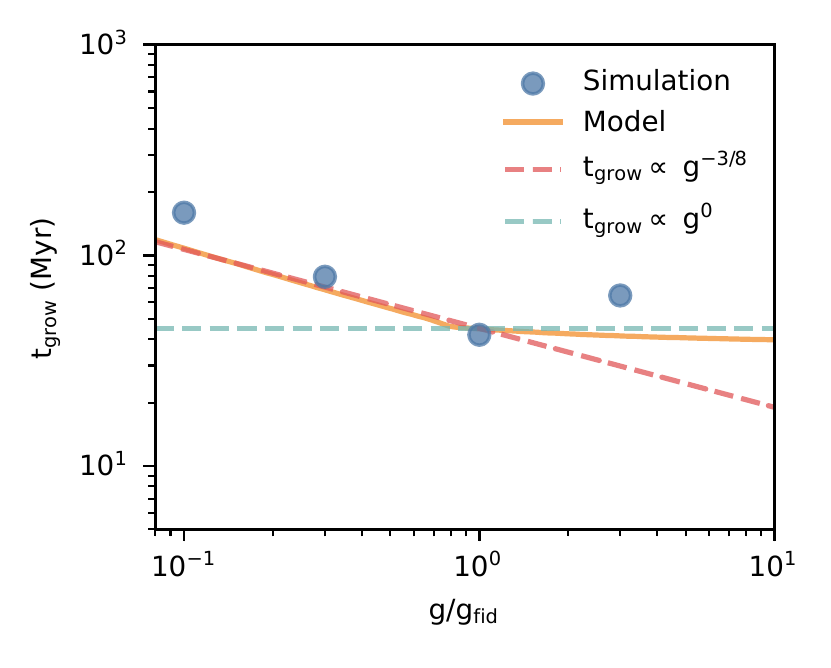}
	\caption{The growth time for different gravitational fields. Dashed lines show expected analytical scalings $t_{\rm grow} \propto g^{-3/8}, g^{0}$ for subsonic and supersonic infall respectively, while the solid orange line shows the model predictions.}
	\label{fig:04c}
\end{figure}
We also vary the gravitational strength $g$ from 0.1 to 3 times the fiducial value. Figure~\ref{fig:04c} shows the scaling of $t_{\rm grow}$ with $g$. As before, we also plot the model and the expected $g^{-3/8}$ and $g^0$ scalings, for subsonic and supersonic infall respectively. Simulation results are consistent with the model in both cases.

\subsubsection{Scaling With Density Contrast/Hot Gas Temperature}
\begin{figure}
    \centering
	\includegraphics[width=\columnwidth]{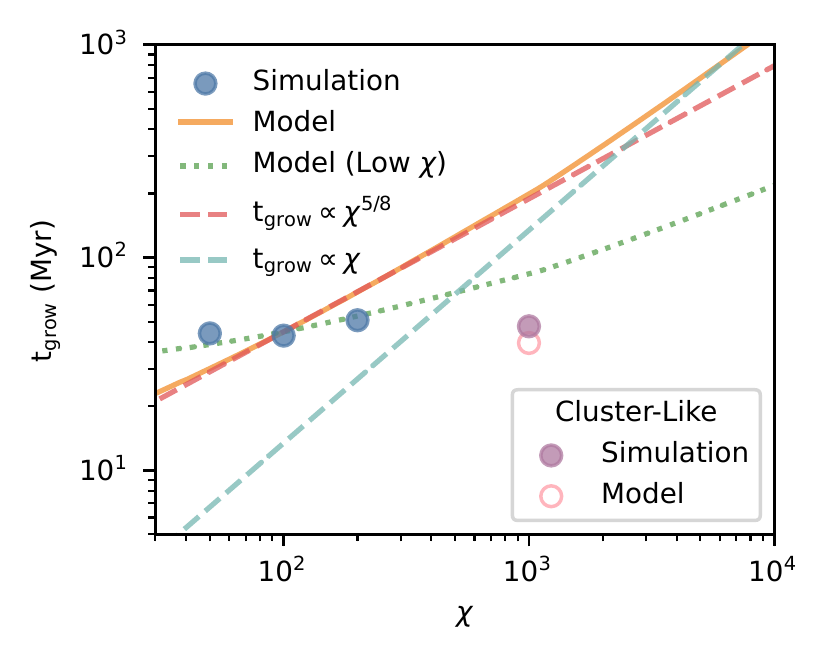}
	\caption{The growth time for different overdensities. Dashed lines show analytical scalings $t_{\rm grow} \propto \chi^{5/8},\chi$ for subsonic and supersonic infall respectively. At low overdensities ($\chi \lsim 100$), the simulations differ from the expected scalings, which we attribute to lower turbulent velocities in mixing layers. If this is taken into account, simulations and models (dotted green line) match. We also test one case at high overdensity $\chi \sim 1000$ for cluster-like parameters, where multiple parameters were varied. The model and simulations match well.}
	\label{fig:04d}
\end{figure}

Lastly we vary $\chi$ by changing the background temperature. Figure~\ref{fig:04d} shows the scaling of $t_{\rm grow}$ with $\chi$. Unlike the previous sections, we do not see the expected $\chi^{5/8}$ scaling. This can be understood by the scaling of the turbulent velocity $u'$ with $\chi$; in our derivation, we assumed $u'$ is independent of $\chi$. As seen in the middle panel of Fig.~12 of \citet{tan21}, this is true for $\chi \gtrsim 100$, but for $\chi \lesssim 100$, then $u^{\prime} \propto \sqrt{\chi}$. If we put in this scaling $u' \propto \sqrt{\chi}$, we see that the $\chi$ dependence of $t_{\rm grow}$ becomes weaker and better matches the simulation results. We expect the our predicted $t_{\rm grow} \propto \chi^{5/8}$ scaling to hold at higher $\chi$, but the simulations required to probe this regime in detail require very long boxes and are beyond the numerical scope of this work. We also plot a single simulation, along with the model expectation, where multiple parameters were varied, not just $\chi$, so as to sample a different region of parameter space with higher $\chi$. These are plotted as standalone points. For this particular simulation, the parameters we have used are $r=300$\,pc, $\chi=1000$, $g=4\times 10^{-8}$\,cm/s$^2$ and $n = 1$\,cm$^{-3}$. Cooling here is not boosted since we use a high density instead (i.e. $\Lambda_0=1$). We find that the growth time for this simulation remains in line with the model.

\subsection{Survival}
\label{sec:results-survival}
\begin{figure}
    \centering
	\includegraphics[width=\columnwidth]{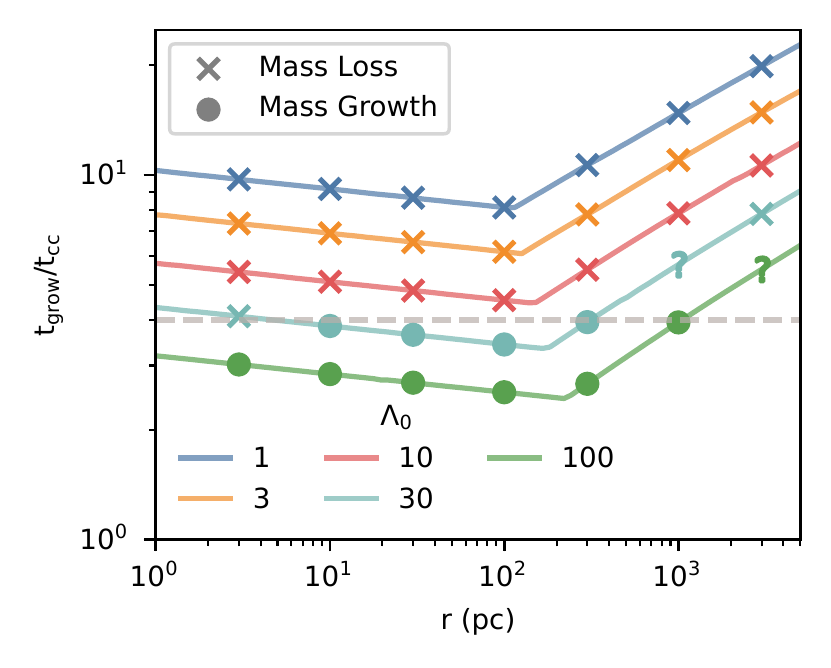}
	\caption{
	Overview of the fate of cold gas in the `constant background' case as a function of cloud size and different cooling strengths.
	Points denote whether clouds in the corresponding simulations are growing in mass or losing mass; question marks denote cases where the fate is uncertain. The breaks correspond to where the the turbulent velocity $u'$ saturates when the cloud velocity reaches the sound speed of the hot background $c_{\rm s,hot}$. This causes $t_{\rm grow}/t_{\rm cc}$ to increase with cloud size instead. In the simulations marked with `?', the final fate of the cold gas is unclear.}
	\label{fig:06}
\end{figure}

Since we are primarily interested in modeling clouds which are growing, it is useful to determine when we are in such a growth regime.
In Section~\ref{sec:survival}, we argued that this criterion is given by $t_{\rm grow} < f_{\rm S} t_{\rm cc}$, where $f_S$ is some constant factor of order unity.
We now test this by running a number of simulations to explore the parameter space, varying the initial cloud radius between 3 pc and 3 kpc, and the cooling time between the fiducial value and 100 times shorter.
Figure~\ref{fig:06} shows\footnote[7]{Question marks denote simulations where it is unclear what the fate of the cloud is.
For example, the cloud might break up, with one portion accelerating and getting destroyed, while leaving some much slower falling material behind it that possibly survives and grows. The cold material then hits the boundary of the box at the top or bottom and we cannot track further evolution. This seems to happen near our survival boundary, where the long term fate of the cloud can be sensitive to cloud dynamics. It also happens for the largest clouds.} the fate of simulated clouds for various cloud sizes and cooling times. Solid lines denote a contour of constant cooling strength, while the vertical axis shows the ratio of the growth time to the cloud crushing time $t_{\rm grow}/t_{\rm cc}$. These timescales are calculated by evaluating the model where our weight factor $w_{\rm kh}(t) = 1$. Physically, this is where turbulence has fully developed and $t_{\rm grow}$ stabilizes. Alternatively, evaluating $t_{\rm grow}/t_{\rm cc}$ at some time $\alpha t_{\rm cc}$ yields the same result, but can change the normalization of $t_{\rm grow}/t_{\rm cc}$ (this ratio gets larger as $\alpha$ gets smaller since  $w_{\rm kh}(t) < 1$). The implication here is that the threshold value of $f_{\rm S}$ is depends on when $t_{\rm grow}/t_{\rm cc}$ is evaluated.

In general, the results are in line with criterion $t_{\rm grow}/t_{\rm cc} \lsim f_{\rm S} \sim 4$ for survival, and the discussion in Section \ref{sec:survival}. Rather than being sensitive to cloud size, clouds get destroyed when cooling is weak, and only survive when cooling is strong enough. Cloud size does begin to play a role when $t_{\rm grow} > t_{\rm ff}$, so that infall velocities become supersonic, and $t_{\rm grow}/t_{\rm cc} \propto r^{1/2}$. As discussed in Section~\ref{sec:survival}, this happens when $r > r_{\rm sonic}$, (equation \eqref{eq:rsonic}); $r_{\rm sonic} \sim 200$pc in our models, where we see the change to a $t_{\rm grow}/t_{\rm cc} \propto r^{1/2}$ scaling. The low mass growth rates at high Mach number means that it is harder for clouds to fall supersonically and still survive; it is only possible in a limited size range $r_{\rm sonic} < r < r_{\rm SS}$ (where $r_{\rm SS}$ is given by equation \eqref{eq:rSS}).  
\begin{figure}
    \centering
	\includegraphics[width=\columnwidth]{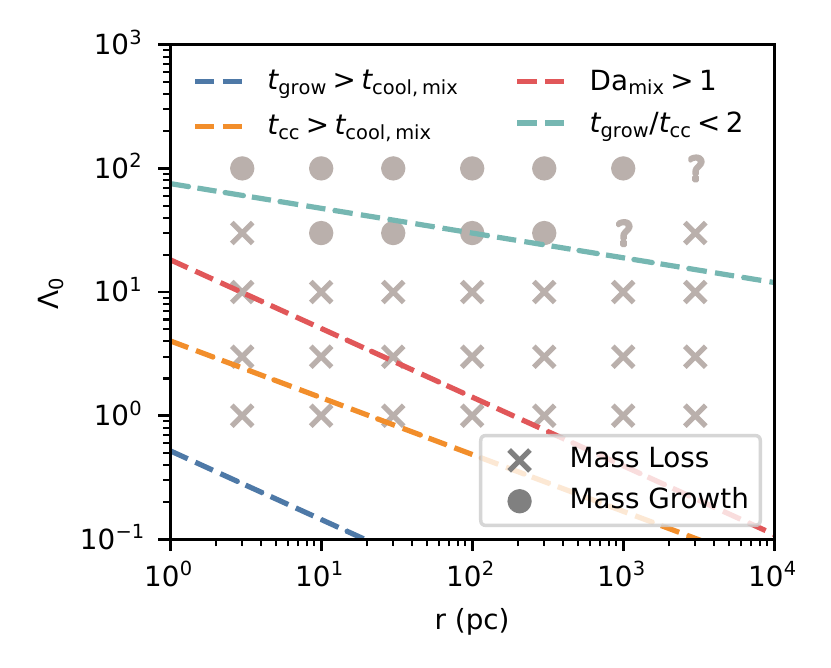}
	\caption{Comparison of various survival criteria (dashed lines) to the simulation results as a function of cloud radius and cooling strength. 
	The criteria are satisfied above the respective lines. 
	The symbols indicate whether a cloud grows or gets destroyed (as in Fig.~\ref{fig:06}).
	}
	\label{fig:A1}
\end{figure} 

To reinforce the point that $t_{\rm grow}/t_{\rm cc} < f_{\rm S}$ is a more stringent survival criteria than others, in Fig.~\ref{fig:A1} we show the boundaries in the $r-\Lambda_0$ plane for two other possible criteria: (i) $t_{\rm cool,mix} < t_{\rm cc}$, which is the criterion for cloud survival in a wind, (ii) ${\rm Da}_{\rm mix} \equiv L/(u^{\prime} t_{\rm cool,mix}) > 1$, which is the criterion for a multi-phase medium in the presence of turbulence and radiative cooling \citep{tan21}. The two criterion are closely related. In Fig.~\ref{fig:A1}, we see that clouds which satisfy these criterion are nonetheless destroyed, while the more restrictive criterion $t_{\rm grow}/t_{\rm cc} < f_{\rm S}$ straddles the boundary between destruction and survival. Note that for sufficiently small clouds, $t_{\rm grow} < t_{\rm cool,mix}$ (blue dashed line) instead of the other way round. However, this lies in the cloud destruction regime and thus is irrelevant. 

\subsection{Growth and Free-fall Timescales}
\begin{figure}
    \centering
	\includegraphics[width=\columnwidth]{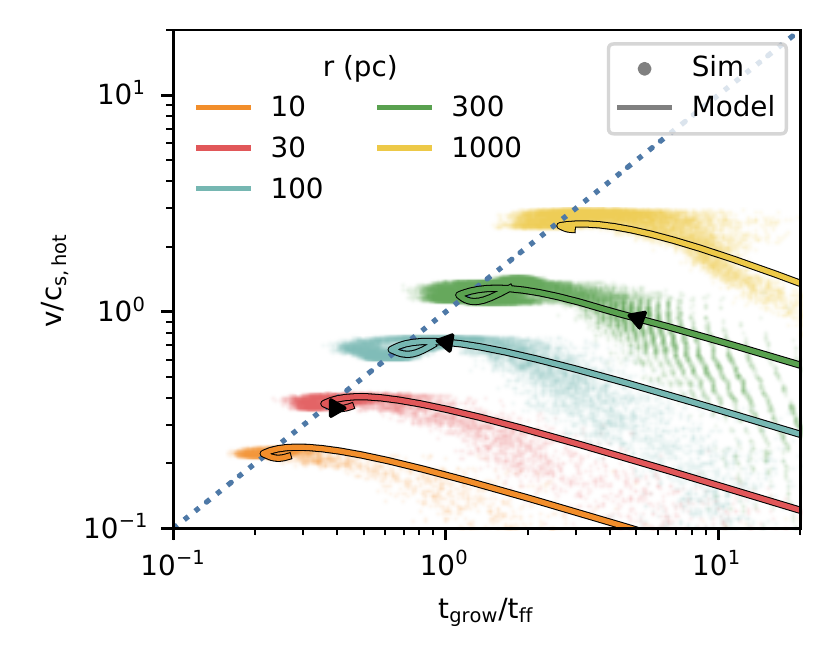}
	\caption{Evolution of the falling velocity of the cloud as a function of (evolving) $t_{\rm grow} / t_{\rm ff}$ for different cloud sizes. Black triangles indicate the direction of evolution at $t=t_{\rm ff}$. As the cloud accelerates, the growth time is decreasing until it stabilizes at the growth terminal velocity $v_{\rm T,grow}/c_{\rm s,hot}\sim t_{\rm grow}/t_{\rm ff}$.}
	\label{fig:linear_relation}
\end{figure}
In Section~\ref{sect:analytics}, we saw that if the drag force from mass accretion balances gravity such that $F_{\rm grav} \sim F_{\rm grow}$, then we expect that $t_{\rm grow}/t_{\rm ff} \sim  v_{\rm T,grow}/c_{\rm s,hot}$. We show that we do indeed see this in our simulations in Fig.~\ref{fig:linear_relation}. The blue dotted line shows the equality, while the colored points are simulation results for various cloud sizes over time. Solid lines show the model values for the same time range as the corresponding simulations. Initially, $t_{\rm grow}$ is large as turbulence develops, but once they reach the terminal velocity $v_{\rm T,grow} \sim g t_{\rm grow}$, falling clouds indeed obey the scaling $v_{\rm grow} \sim c_{\rm s,hot} (t_{\rm grow}/t_{\rm ff}$, as seen from the fact that the clouds evolve to the blue dotted line and stays there. 

\dobib

\section{Results : Stratified Background} \label{sect:results_stratified}
In our second set of simulations, we consider a more realistic setup of a cloud falling through an isothermal hydrostatic background. This means that $P,\rho \propto \exp(\frac{z}{H})$, where $z$ is the vertical height the cloud has fallen and $H$ is the scale height of the background medium. As mentioned in Section~\ref{sect:methods}, the density profile of the background is thus:
\begin{align}
    n(z) = n_0 \exp(\frac{z}{H}),
\end{align}
where $n_0 = 10^{-4}$\,cm$^{-3}$ is the initial background density, $z$ is the height the cloud has fallen and $H \equiv k_B T_{\rm hot}/g m_H = 2.8$\,kpc is the isothermal scale height (assuming the mean molecular weight $\mu = 1$). We define our origin where the cloud begins to fall, hence density increases rather than decreases exponentially with $z$. While the use of a constant gravitational acceleration $g$ is not in general a realistic assumption, this simplification helps in isolating the relevant physics. 

\subsection{Time Evolution}
We now present the time evolution of a simulation where the cloud comfortably survives, along with the model predictions for various quantities. Unlike the constant background setups, we do not artificially boost the cooling function in these simulations. Instead, the cooling time naturally varies with density and hence height. Fig.~\ref{fig:07} shows the evolution of these quantities over the course of a simulation with an initial cloud radius $r = 1$\,kpc and $g = g_{\rm fid} = 10^{-8}{\rm cm\,s}^{-2}$. As before, these are, from left to right and top to bottom - timescales, cloud velocity, distance fallen, and the total mass of cold gas. 

The upper left panel of Fig.~\ref{fig:07} shows the same timescales as in Fig.~\ref{fig:02}: The cooling time of the {\it mixed} gas $t_{\rm cool, mix}$, which decreases as the clouds falls, the free-fall time $t_{\rm ff} = c_{\rm s, hot}/g$, the cloud crushing time $t_{\rm cc} = \sqrt{\chi}r/v$, which uses the {\it initial} cloud radius $r$ and the instantaneous cloud velocity, and the instantaneous cloud growth time $t_{\rm grow} = m/\Dot{m}$, computed using the mass of cold gas. For the latter two timescales ($t_{\rm cc}$ and $t_{\rm grow}$), both model and simulation results are shown for comparison.  We have adjusted the value of $f_{\rm kh}$ in the weight term $w_{\rm kh}(t)$ to be 1 for the stratified background as that is more in line with simulation results. The suggests a more rapid onset of turbulence for clouds that are falling into a denser background (this parameter is of course, only a crude approximation of the relevant processes involved). The model performs well at matching the simulation results for both $t_{\rm cc}$ and  $t_{\rm grow}$, although marginally less so than for the constant background. This can be attributed to the cloud initially travelling through a region of parameter space where it is not in the growth regime.
Since our model does not include cloud destruction, this leads to a deviation of the simulation from the model. 
The velocity evolution of the cloud is shown in the upper right panel. 
The cloud initially accelerates ballistically, before the cooling drag force kicks in and slows the cloud down. Since the cooling drag force operates on a timescale $t_{\rm grow}$, the cloud remains ballistic until $t \sim t_{\rm grow}$. During this time, the cloud can reach velocities greater than the eventual terminal velocity $v_{\rm T,grow}=g t_{\rm grow}$. The subsequent deceleration due to cooling slows the cloud down such that the velocity turns around and starts to {\it decrease}. This has implications for cloud survival which we discuss further on. At late times the cloud velocity approaches a roughly constant value. We now delve into this further.

\subsection{Terminal Velocity}
\begin{figure*}
    \centering
	\includegraphics{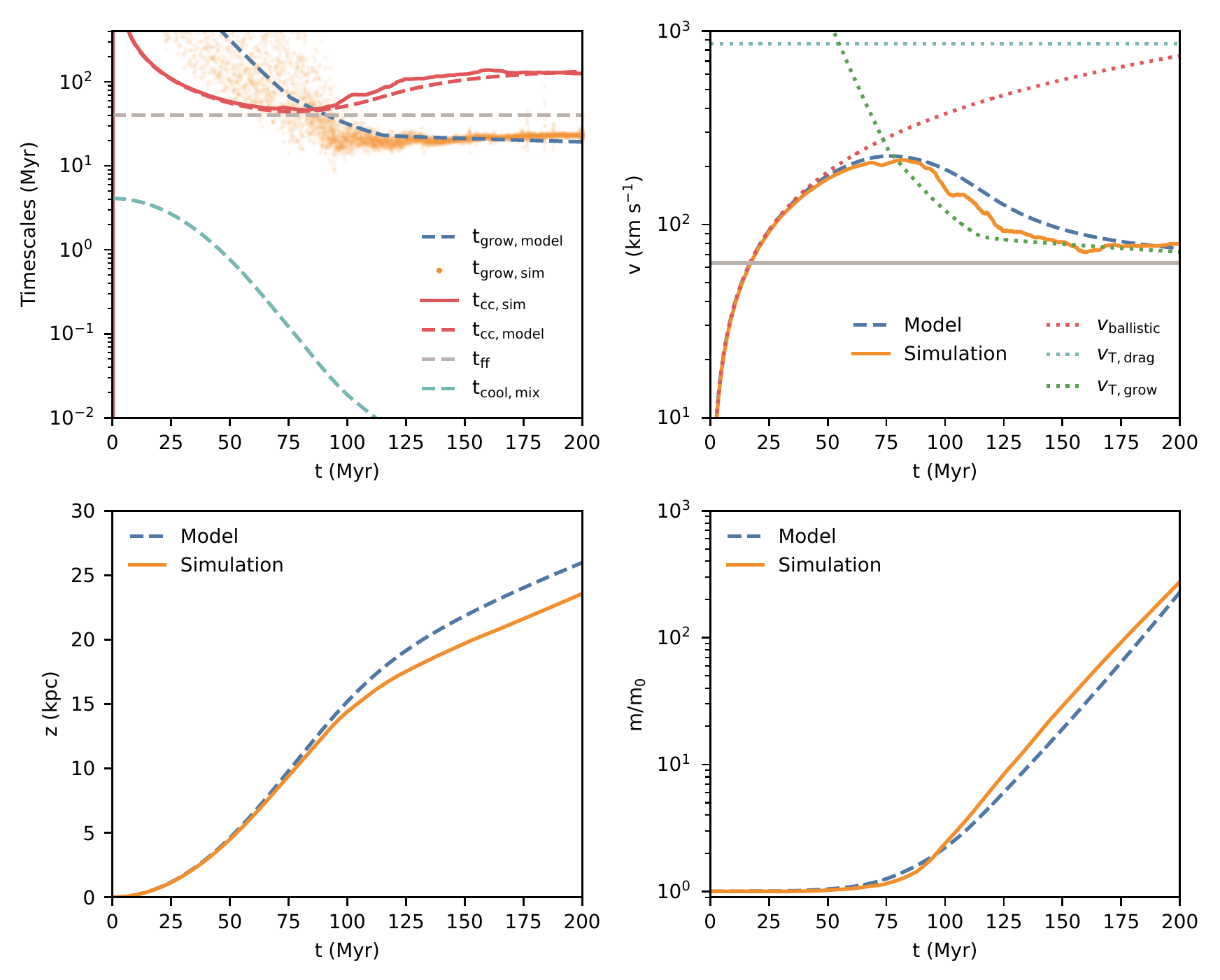}
	\caption{Time evolution of various quantities for a $r = 300$~pc cloud falling in a stratified background. From left to right, top to bottom, the panels compare the growth time $t_{\rm grow}$, the velocity $v$, the distance fallen $z$, and the cold gas mass $m$ of the cloud in the simulation versus the model. The upper panels also include comparison with other quantities of interest. Model predictions are in good agreement with simulations results.}
	\label{fig:07}
\end{figure*}

Previously, we argued that the terminal velocity should approach a value $v_{\rm T,grow} \approx g t_{\rm grow}$ (equation \eqref{eq:vtgrow}). Indeed, it does so, after some `overshoot' as described above. However, as apparent from equation \eqref{eq:tgrow_scaling}, $t_{\rm grow}$ itself is a function of parameters such as $t_{\rm cool}(t)$, $m(t)$, $\rho_{\rm h}(t)$ which change with time as the cloud falls through a stratified atmosphere. Thus, one might expect $t_{\rm grow}$ and consequently $v_{\rm T,grow}$ to vary with time as the hot plasma surrounding the cloud increases in density. Instead, what is surprising from Fig. \ref{fig:07} is that $t_{\rm grow}$ asymptotes to a {\it constant} value. Indeed, it does so quite early, before $v \rightarrow v_{\rm T,grow}$. How can we understand this? 

From equation \eqref{eq:tgrow_scaling}, and using $t_{\rm cool} \propto 1/n \propto \exp(\frac{-z}{H})$, we can write: 
\begin{align}
    t_{\rm grow}(t) \propto v(t)^{-3/5} \left( \frac{m(t)}{m_0} \right)^{1-\alpha} \exp(-(\frac{5}{4}-\alpha)\frac{z(t)}{H}).
    \label{eq:tgrow}
\end{align} 
as a time-dependent quantity. The rate at which $t_{\rm grow}$ changes is: 
\begin{equation}
    \frac{d {\rm ln} t_{\rm grow}}{dt} = \frac{\dot{t}_{\rm grow}}{t_{\rm grow}} = - \frac{3}{5} \frac{\dot{v}}{v} +  \frac{(1-\alpha)}{t_{\rm grow}} - (\frac{5}{4}-\alpha) \frac{v}{H}
    \label{eq:dln_tgrow}
\end{equation}
From equation \eqref{eq:vtgrow}, this can be contrasted with the rate at which $v$ evolves: 
\begin{equation}
    \frac{d {\rm ln} v}{dt} = \frac{\dot{v}}{v} = \frac{g}{v} - \frac{1}{t_{\rm grow}}
    \label{eq:dln_v}
\end{equation}
We can make two observations. Firstly, equation \eqref{eq:dln_tgrow} has terms of opposing sign. Thus, it is possible that $\dot{t}_{\rm grow} \rightarrow 0$, i.e. $t_{\rm grow} \approx$~const, rather than evolving with background quantities. Physically, this is because of a negative feedback loop. Suppose $t_{\rm grow}$ decreases as a cloud falls into denser surroundings. The subsequent increase in mass causes $t_{\rm grow}$ to increase (from equation \eqref{eq:tgrow}). The opposite is also true: if $t_{\rm grow}$ is large, the cloud will fall faster (due to weaker accretion drag) into denser regions, reducing $t_{\rm grow}$. Secondly, by comparing terms on the right-hand side of equations \eqref{eq:dln_tgrow} and \eqref{eq:dln_v}, the timescale on which $t_{\rm grow}$ equilibrates to its steady-state value is comparable to the timescale on which $v$ equilibrates to its steady-state value\footnote[9]{Indeed, because of `velocity overshoot', $t_{\rm grow}$ equilibrates first.}, $v_{\rm T,grow} = g t_{\rm grow}$. Thus, $\dot{v}, \dot{t}_{\rm grow} \rightarrow 0$ on similar timescales. From setting equations \eqref{eq:dln_tgrow} and \eqref{eq:dln_v} to zero, the steady-state value of $t_{\rm grow}$, and hence $v_{\rm T,grow}$, is given by: 
\begin{align}
    v_{\rm T,grow} = gt_{\rm grow} \approx \sqrt{\frac{1-\alpha}{\frac{5}{4}-\alpha}Hg} \approx \sqrt{\frac{2}{5}} c_{\rm s,hot} ,
    \label{eq:vT-stratified}
\end{align}
where in the last step we use $\alpha = 5/6$ and $g \approx c_s^2/H$ for an isothermal atmosphere. This velocity is shown by the grey line in Fig \ref{fig:07}. 
\begin{figure}
    \centering
	\includegraphics[width=\columnwidth]{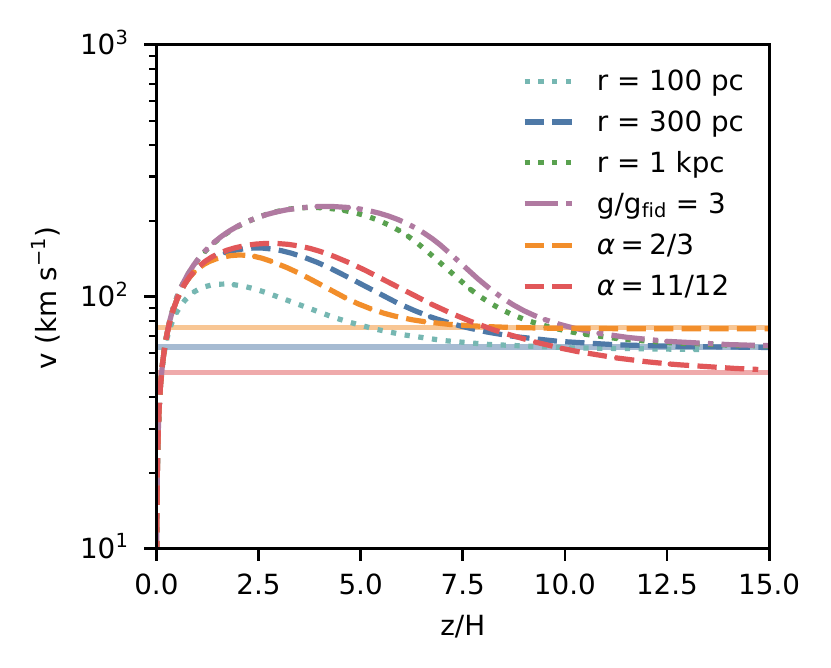}%
	\caption{Velocities in a stratified environment converge to a constant value that only depends on $\alpha$ (where area $A \propto M^{\alpha}$), independent of all other properties such as cloud size, or gravity. Curves show velocity evolution in our model (equations \eqref{eq:diff_eq1} -- \eqref{eq:diff_eq3}), while solid lines give the asymptotic velocities from equation \eqref{eq:vT-stratified}.}
	\label{fig:strat_model}
\end{figure}
This then has the remarkable implication that in an isothermal atmosphere with constant gravity, $f_{\rm sub-vir} = v_{\rm T}/c_{\rm s,hot} = t_{\rm grow}/t_{\rm ff}$ (equation \eqref{eqn:alpha}) of a cloud where accretion induced drag dominates is {\it independent} of all properties of the system except cloud geometry, specifically $\alpha$. For our measured value of $\alpha=5/6$ from infalling clouds with cometary tails, we predict $f_{\rm sub-vir} = [(1-\alpha)/(5/4-\alpha)]^{1/2} \approx 0.6$. In Fig. \ref{fig:strat_model}, we compare velocity evolution in our model (equations \eqref{eq:diff_eq1} -- \eqref{eq:diff_eq3}), to the asymptotic velocities from equation \eqref{eq:vT-stratified}, for different cloud sizes and gravitational fields. Equation \eqref{eq:vT-stratified}, which only depends on $\alpha$, correctly predicts the asymptotic velocity. Note, however, that reaching the asymptotic velocity requires falling through many scale heights, and a planar ${\rm g}\approx$~const isothermal atmosphere may not be realistic over such lengthscales. `Velocity overshoot' also implies that large clouds (which exhibit stronger overshoot) might be seen to fall faster than predicted. In systems with varying $g(r)$ and $T(r)$ (and thus non-constant scale heights), the result can be more complex, and the most straightforward way to arrive at predictions is to simply integrate the set of ODEs, equations \eqref{eq:diff_eq1} -- \eqref{eq:diff_eq3}. We will show an example in Section~\ref{sect:clusters}.  

\subsection{Scaling With Cloud Size and Gravity}
\begin{figure}
    \centering
	\includegraphics[width=\columnwidth]{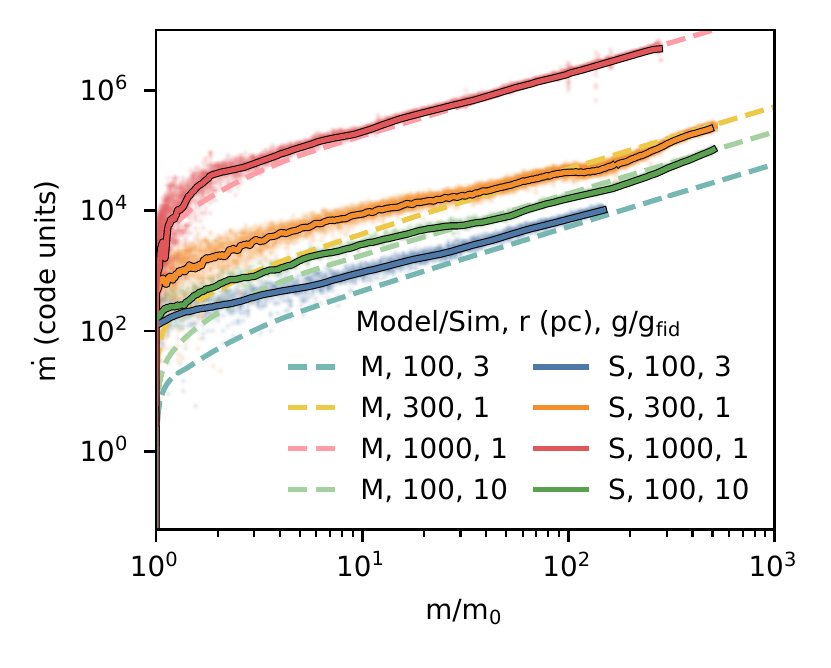}
	\caption{The mass growth rate as a function of cold gas mass for clouds of different initial sizes and different gravitational strengths. Curves are labelled by the initial parameters and whether they represent model solutions (M) or simulations (S), which are shown as dashed and solid lines respectively.}
	\label{fig:08}
\end{figure}

In Fig.~\ref{fig:08}, we compare the mass growth rates as a function of mass for simulations with varying initial cloud sizes and gravitational strengths to model predictions. Varying $g$ allow us to test the model for different scale heights. We can see that the model predictions are in good agreement with simulations results. In all cases, the simulations converge to the 1/$t_{\rm grow}$ slope predicted by the model. The divergence at early time is due to the fact that for this setup, the clouds  start in a destruction regime since cooling is relatively weak.

\subsection{Resolution Convergence}
\label{subsect:resolution}
\begin{figure}
    \centering
	\includegraphics[width=\columnwidth]{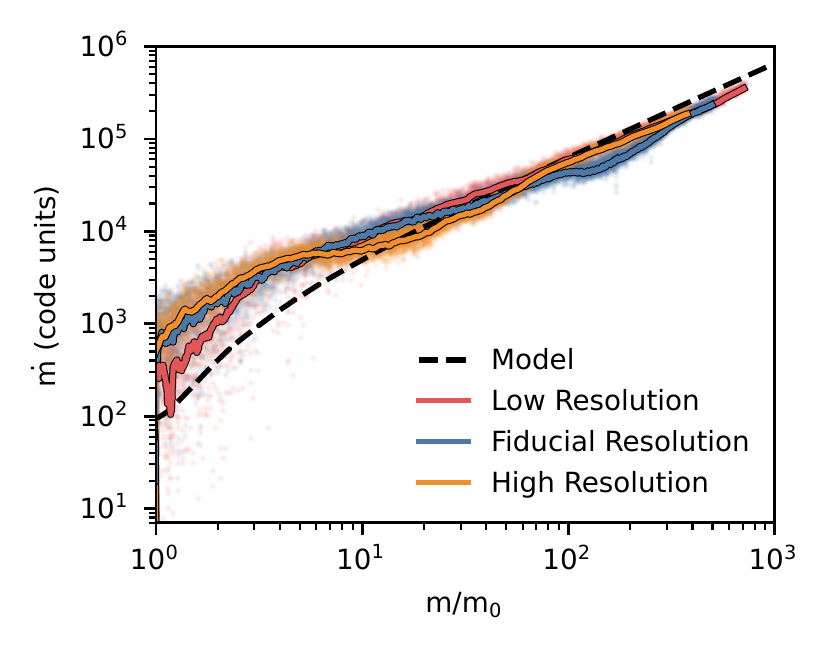}
	\caption{The mass growth rate as a function of cold gas mass for a $r=300$~pc cloud with $g=g_{\rm fid}$ at different resolutions ($8\times$ higher and lower mass resolution than in the fiducial run, respectively). The simulations are relatively well converged.}
	\label{fig:resolution}
\end{figure}

To test if our results for mass growth rates are converged. we run a $r=300$\,pc cloud with $g=g_{\rm fid}$ at various resolutions, varying the fiducial resolution by a factor of 2. Fig.~\ref{fig:resolution} shows that the three resolutions show little difference in mass growth rates and that the simulation appears to be converged, although the higher resolution simulation matches the model slightly better -- the cloud is disrupted less initially and reaches the model growth rate more rapidly. 

\subsection{Survival in a stratified background}
\begin{figure}
    \centering
	\includegraphics[width=\columnwidth]{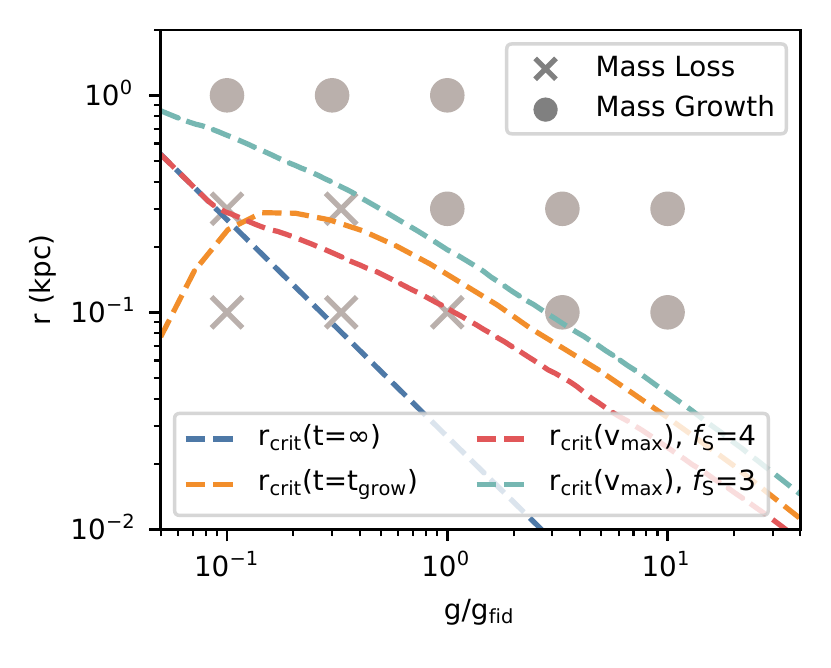}
	\caption{The fate of clouds of different size falling in stratified backgrounds with different gravitational strengths. Survival criterion evaluated at different times are shown. The best survival criterion is given by the teal curve, i.e. equation (\ref{eq:survive}) evaluated at the maximum velocity, for $f_S=3$.}
	\label{fig:survival_strat}
\end{figure}
\begin{figure*}
    \centering
	\includegraphics[width=\textwidth]{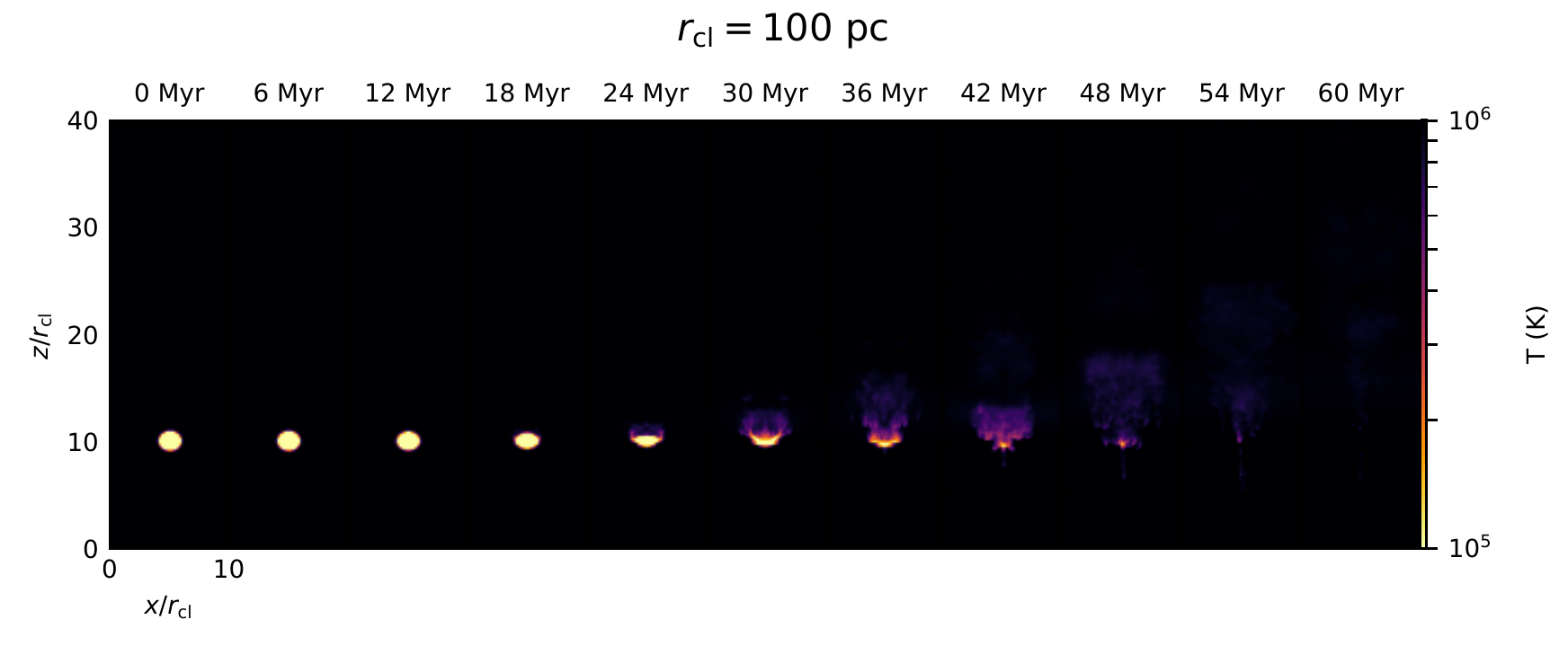}
	\includegraphics[width=\textwidth]{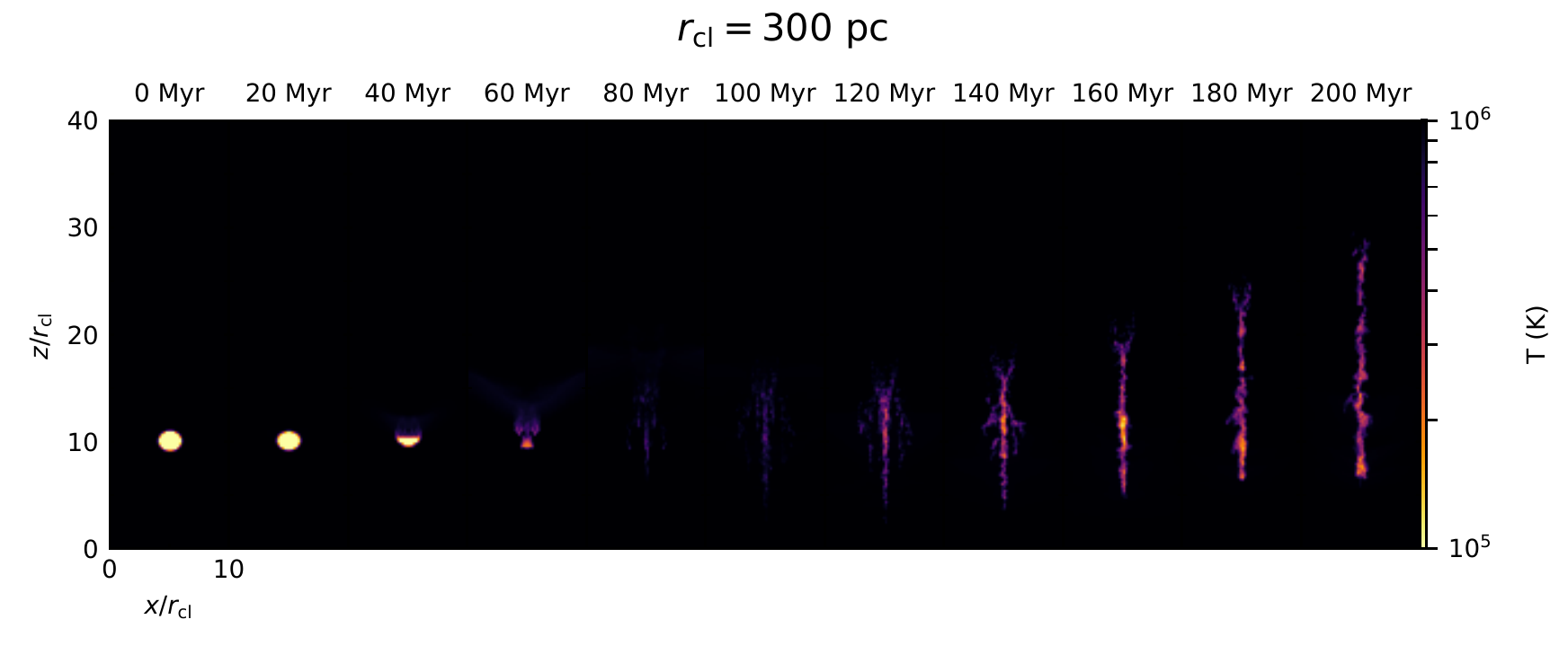}
	\caption{Snapshots of the projected density weighted temperature through the box (temperature here is hence just an indication of the amount of cold gas when projected along the y-axis) for a 100\,pc and a 300\,pc cloud at various points in their evolution. The former is disrupted completely while the latter reaches the survival zone and grows. $x$ and $z$ here simply reflect the size of the box along the respective axes normalized by cloud size.}
	\label{fig:snapshots}
\end{figure*}

For a cloud falling in a constant background we found that the survival criterion was given by a competition between the growth and destruction timescales of the cloud: $t_{\rm grow} < f_{\rm S} t_{\rm cc}$
where $f_{\rm S}$ is a constant factor. We wish to ascertain if the same condition applies to clouds falling in a stratified background.

In the case of a constant background, $t_{\rm grow}$ changes very little over time (once turbulence has developed), with only a very weak scaling with mass, and cooling is strong enough so $v$ approaches $gt_{\rm grow}$ without `overshooting', something we noted in Fig.~\ref{fig:07} above. For a stratified background, both these assumptions do not hold - $t_{\rm grow}$ changes continuously with background density, and an overshoot is often observed. 
Since our initial conditions are in the regime where clouds do not survive, surviving clouds are those that are able to survive long enough to enter the growth zone.

One ansatz would be to use the asymptotic value of $t_{\rm grow}$ and $v$ that we derived above in equation~\eqref{eq:vT-stratified} and evaluate the survival criteria there. This gives:
\begin{align}
    r > \frac{v^2_{\rm T,grow}}{g f_{\rm S} \sqrt{\chi}}
\end{align}
This condition is given by the blue dashed line in Fig.~\ref{fig:survival_strat}. Note that it is a {\it lower} bound on $r$, since $v_{\rm T,grow}$ is independent of $r$. It has the right qualitative behavior as a survival criterion, but does not seem to match the survival thresholds seen in the simulations.
Clouds have to fall many scale heights to reach the asymptotic velocity given by equation~\eqref{eq:vT-stratified}  -- often survival is determined much earlier. Indeed, the falling clouds often overshoot this asymptotic velocity as they initially fall ballistically, as seen in Fig.~\ref{fig:survival_strat}. We can estimate the time where gravity and cooling balance:
\begin{align}
    \Dot{m}v \sim \frac{m}{t_{\rm grow}}v \sim mg \frac{t}{t_{\rm grow}}
\end{align}
assuming the cloud is falling ballistically in this initial phase. Hence, $t \sim t_{\rm grow}$ is the time where the cloud is slowed from its ballistic free falling trajectory. If we evaluate equation~\eqref{eq:survive} at this time in the simulation, we can solve numerically for some $r_{\rm crit}$. Of course, this only makes sense if $v(t=t_{\rm grow}) > v_{\rm T,grow}$, i.e. there is an overshoot so $t_{\rm cc}$ is shorter. The larger the difference in the two velocities, the more likely the cloud is to be destroyed in this overshooting phase. In Fig.~\ref{fig:survival_strat}, we show this limit in the orange dashed line. We see that this matches the simulation results more closely for larger values of $g$, where the clouds accelerate to higher velocities. Ultimately, it is the {\it maximum} velocity that determines if a cloud survives. We thus show in the red and teal curves in Fig.~\ref{fig:survival_strat} the survival criterion evaluated at $v=v_{\rm max}$ from the model. The red curve use $f_{\rm S} = 4$ as in the previous section, while the teal curve has $f_{\rm S} = 3$, which seems to be a better match to the simulation results. It is unsurprising that we find a different value of  $f_{\rm S}$ here, since we are evaluating our quantities at a different time.

In Fig.~\ref{fig:snapshots}, we show the evolution of the 100 pc and the 300 pc cloud for $g=g_{\rm fid}$. The 100\,pc cloud does not survive and is disrupted completely, while the 300\,pc starts to get disrupted but survives long enough to reach the zone of growth and then grows. Note the tail growth in the surviving case. 
To summarize, we have looked at clouds that start outside the growth zone in a stratified medium, and find that in order to survive, the cloud has to make it to the growth zone. Since the cloud is accelerating ballistically before it reaches high enough pressures where cooling is efficient enough for it to grow and slow down, only large clouds can survive this infall.
We explore the implications of the survival conditions in this and the previous section on astrophysical systems of interest in the following section.

\dobib

\section{Discussion}  \label{sect:discussion}

\subsection{High Velocity Clouds}
\label{sect:HVC} 
3D simulations of clouds falling under gravity with mixing and cooling processes included have only been studied to a limited extent previously. \citet{heitsch09} concluded that clouds below $10^{4.5}$\,M$_{\odot}$ are disrupted within 10\,kpc. Notable differences in setup include a smaller box length along the tail direction and starting initially with colder clouds, as their temperature range extended down to 100\,K. \citet{heitsch22} focused on metallicity measurements, tracing original versus accreted cloud material. They found that most of the original cloud material does not survive and is instead replaced by accreted gas which mostly happens in the tail. \citet{gronnow22} observed cloud growth in MHD simulations but did not follow the clouds for many cloud crushing times. We have followed up by providing a model for the mass growth of such clouds based on the underlying process of turbulent mixing and cooling, so as to tackle the key questions of when HVCs can survive, how much mass they accrete, and how fast they travel. We then tested the model against a suite of numerical simulations. What then are the implications for HVCs?
\begin{figure}
    \centering
	\includegraphics[width=\columnwidth]{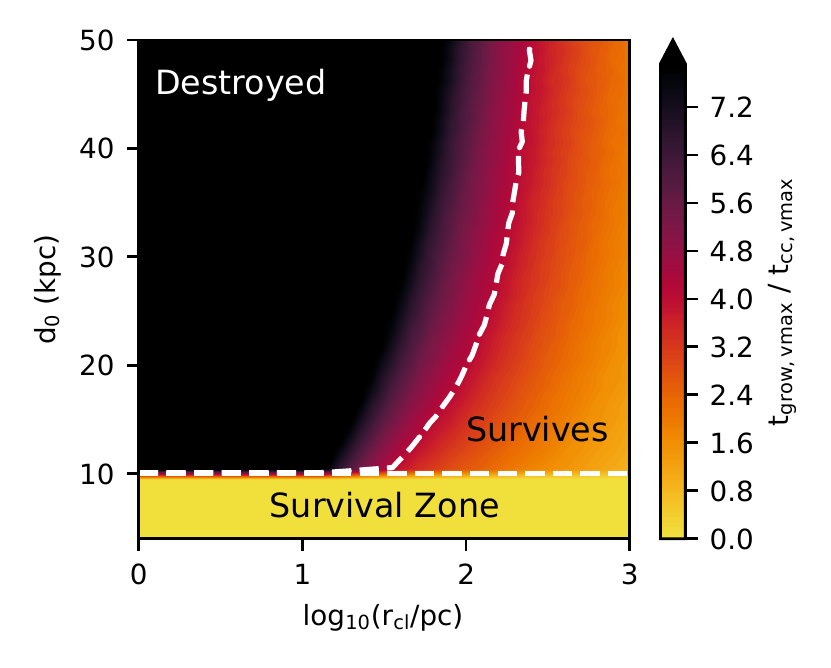}
	\caption{Survivability of HVCs in Milky Way conditions with a size $r_{\rm cl}$ and dropping height $d_0$. For clouds outside the survival region given by equation~\eqref{eq:safe_zone}, the color coding corresponds to the ratio $t_{\rm grow}/t_{\rm cc}$ evaluated at the maximum velocity along the cloud trajectory. The horizontal white dashed line shows where the survival criterion equation~\eqref{eq:survive} is satisfied for $f_{\rm S}=4$. Large clouds that fall in from large distances can still survive as they are not destroyed before reaching the survival region.}
	\label{fig:hvc_survival}
\end{figure}

In Fig.~\ref{fig:hvc_survival}, we show our estimates for cloud survival in a Milky Way like profile in the cloud size-initial height parameter space.
Specifically, we employ the profiles from \citet{salem15} who combine the density profile of \citet{miller2015} with a temperature profile mapped from a NFW halo \citep{navarro97}, which we also use to set the gravitational profile. In the region of interest, $T \sim 10^6$~K.
Figure~\ref{fig:hvc_survival} shows the ratio of the growth time and the cloud crushing time $t_{\rm grow}/t_{\rm cc}$ evaluated at the maximum velocity the cloud reaches along its trajectory.
We also show the threshold of survival (equation~\eqref{eq:survive}) at $\sim 4$ from the previous section. The analytic expectation (equation~\eqref{eq:safe_zone}) for where cooling is strong enough for clouds to survive regardless is demarcated by the white dashed line. Outside this region, larger clouds can survive falling from further out, simply from the fact that $t_{\rm cc}\propto r_{\rm cl}$.

More generally however, Fig.~\ref{fig:hvc_survival} shows that except for these larger ($\gtrsim 100\,$pc) clouds, HVCs in the Milky Way should only survive if they start at an initial height of $d_0\lesssim 10\,$kpc. Most HVC complexes detected do indeed fall within this regime -- with the notable detection of the ones associated with the LMC and its Leading Arm located at $\gtrsim 20\,$kpc \citep{richter2017}. 

While this prediction seems to explain the observed survival of most HVCs, we want to highlight that due to the mass transfer from the hot to the cold medium, most surviving clouds in the Milky Way in our model would fall at $v_{\rm GSR} \sim 70$~km/s (equation~\eqref{eq:vT-stratified}) and might thus have velocities $v_{\rm LSR}$ which are too low to be classified as HVCs. Such a population of intermediate to low velocity clouds is of course to be expected even from simply studying the velocity distribution of HVCs and ``filling in'' the gap at $v_{\rm LSR} \sim 0$, and has been the subject of several theoretical studies \citep[e.g.][]{Peek2007,Zheng2020} -- as well as observational attempts to locate them \citep[e.g.][]{Peek2009,Bish2021}. Thus far, there does not seem to be a firm conclusion on the existence of such a low-velocity population. Our work provides a theoretical foundation for the existence of such clouds and predicts an overabundance of them in the Milky Way halo at lower heights ($\lesssim 10\,$kpc).

An interesting example of a nearby HVC is the Smith Cloud \citep{smith63}, lying only 3 kpc below the galactic plane with a metallicity of $\sim 0.5$ M$_{\odot}$, and which is falling towards the galactic plane at velocity $v_z \sim 70$~km/s \citep{fox16}. A longstanding mystery has been explaining the survival of the Smith Cloud at its current location. 
A simple ballistic analysis suggests that the cloud might have already passed through the disk \citep[][]{lockman08} and should hence have been disrupted, in which case some mechanism is needed to explain its survival, such as the cloud being embedded in a dark matter sub-halo, which would shield the gas and extend its lifetime \citep{nichols09}. It is possible that the relative high metallicity and survival of the Smith Cloud can be potentially explained instead by accretion of ambient material driven by turbulent mixing and cooling. \citet{henley17} ran a wind tunnel simulation with the aim of reproducing a Smith cloud like setup, and found entrainment of the background gas largely in the tail of the cloud. \citet{galyardt16} ran simulations of the Smith Cloud with gravity and in a stratified background. They concluded that if the Smith Cloud was in a dark matter sub-halo, it would comprise gas accreted only after the sub-halo passed through the disk. Alternatively, if the Smith Cloud was not accompanied by such a sub-halo, then it must be on first approach, since the cloud would not survive its journey through the Galactic disk. Our model could naturally explain the survival of a Smith Cloud that was on first approach, as it fulfills the survival criterion Eq.~\eqref{eq:survive}, i.e., it falls within the `survival zone' of the Milky Way's halo. The trajectory in this case would be very different from the ballistic one since the accretion dynamically affects the cloud.

Since the terminal velocity is independent of the cloud size, one would expect no observable relationship between, for instance, cloud column density and infall velocity, although there may be significant scatter since this requires the cloud velocity to `turn around' and reach asymptotic terminal velocity. This is consistent with observations \citep{westmeier18}.

We have thus far considered clouds that are infalling from large distances and potentially feed the disk. In our model, HVCs and IVCs can continually grow in mass once they are near enough to the disk. It therefore also gives credence to the notion that fountain-driven accretion can supply the disk with fuel for star formation: cold gas thrown up into the halo `comes back with interest', by mixing with low metallicity halo gas which cools and increases the cold gas mass \citep{armillotta16,fraternali17}. Such low metallicity gas is required to satisfy constraints from disk stellar metallicities and chemical evolution models \citep{schonrich09,kubryk13}. The equations for mass transfer and velocity derived in this work can also be incorporated into semi-analytic `fountain flow' models and checked against observations.

\subsection{Clusters}
\label{sect:clusters} 

Galaxy clusters are amongst the largest virial systems in the universe and thus present opportune test beds for the comparison of observations and theoretical models of galactic properties and evolution. The hot intracluster medium (ICM) in such environments reaches temperatures in the range of $10^7$--$10^9$~K  which can be probed observationally via X-ray emission originating from the thermal bremsstrahlung radiation of this hot diffuse plasma \citep{sazarin86}.
However, the ICM does not exist simply in a single phase. Observations from measurements of carbon monoxide (CO) which traces cold molecular gas find an abundance in these central cluster galaxies, with molecular gas mass correlating with X-ray gas mass \citep{pulido18}. One theory for the origin of the cold molecular gas is that they develop from thermal instabilities triggered in the wakes of cooling updrafts of radio bubbles that rise and lift low entropy X-ray gas \citep{mcnamara16}. These form the cold filaments observed to trace the streamlines around and behind the bubbles, which should eventually decouple from the velocity structure of the hot flow and fall back towards the galaxy center \citep{russell19}.

A particularly interesting conundrum is the low observed velocities of the molecular gas measured by CO line emission in ALMA target systems \citep{mcnamara14,russell16,pulido18,russell19}. They are significantly smaller ($<100$~km~s$^{-1}$) than both stellar velocity dispersions ($200-300$~km~s$^{-1}$) and galaxy escape velocities ($\sim 1000$~km~s$^{-1}$), implying that the molecular gas is tightly bound to the galaxy and should be expected to be infalling. Even initially outflowing gas should at some point stall and fall back inwards. These low velocities are puzzling as models of free falling clouds in cluster potentials have estimated that they can be accelerated to hundreds of kilometers per second after falling just a few kpc \citep{lim08,russell16}. The large density contrast between the molecular gas and the hot background in the ICM means that ram pressure should do little to slow down these falling clouds, which would rapidly accelerate to high velocities. Small velocities would require the observed cold gas to have been falling gravitationally for only a short amount of time. While this could be explained if the infalling cold gas observed was mostly recently decoupled from the hot gas, there is no reason to suggest that this should be the case.  Furthermore, the rapid acceleration means we should see steep velocity gradients in these filaments. However, we observe shallower velocity gradients that are inconsistent with free-fall \citep{russell16,russell17}. Some observations find that free-fall models can match observations in outer filaments, but break down for inner regions \citep{lim08,vantyghem16}. One caveat here is that increasing the spatial resolution of observations can reveal more complex spatial and velocity structures \citep{lim08}. Lastly, if the molecular gas was free-falling, we would expect to generally detect higher velocities at smaller radii, but there is no evidence for this. A large influx of cold gas implies that circumnuclear disks should be more common in comparison with filaments, while the opposite is observed \citep{russell19}.

The conclusion then is that the picture of free falling clouds fails to explain a large number of observations with regard to these filaments, which suggest that the infalling cold gas has to be slowed by some alternative process other than ram pressure drag. One possibility which has been previously proposed is that magnetic stresses slow the clouds' descent, since it has been suggested that the cold filaments are significantly magnetically supported \citep{fabian2008}. However, the magnetic pressure that would be required to slow such a filament's infall along its length requires a strong non-radial magnetic field component with $\beta \sim 0.1$ \citep{russell16}.

Our results suggest an alternative explanation that naturally addresses the above issues. As noted above, observations of the prevalence of molecular gas are closely tied to systems with shorter cooling times. As shown in the previous section, the filamentary mass growth driven by turbulent mixing and cooling of these infalling cold filaments serve as a braking mechanism via accretion induced drag. This can significantly reduce the acceleration of the cold gas when the cooling time of mixed gas is short. To illustrate this point, we compare our model to the free fall model used in \citet{lim08} in their analysis of observed filaments in the cD galaxy NGC 1275 (Perseus A)  located in the Perseus cluster. For simplicity, we follow the approach of \citet{lim08} and adopt an analytic model of the mass density and gravitational potential of the form from \citet{hernquist90}. The mass density and gravitational potential as a function of radial distance are thus given by:
\begin{align}
    \rho(r) &= \frac{M}{2\pi}  \frac{a}{r}  \frac{1}{(r+a)^3} \\
    \phi(r) &= - \frac{GM}{r+a}
\end{align}
where $M$ is the total galactic mass, $r$ is the radial distance, and $a$ is a scale length. We also use the same values they deduce from luminosity observations of \citet{smith90} and an estimated mass to light ratio, with $M=8.3 \times 10^{11}$~M$_{\odot}$ and $a =  6.8$~kpc. We use the number density profile given in \citet{churazov03} for the Perseus cluster, which is mostly a constant $n = 4 \times 10^{-2}\ccm$ below 30\,kpc and adopt a constant temperature profile of $T = 10^7$\,K.
\begin{figure}
    \centering
	\includegraphics[width=\columnwidth]{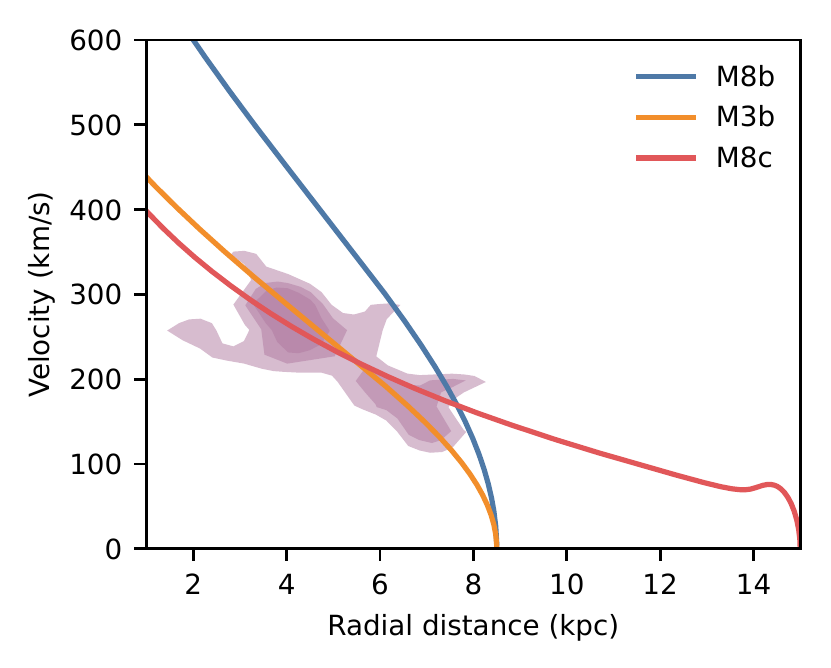}
	\caption{Observed velocity contours of the outer western filament in Per A from \citet{lim08} are shown in purple. Ballistic trajectories are shown by the blue and orange lines for galactic masses of $M=8.3 \times 10^{11}$~M$_{\odot}$ (as observed) and $M=3.4 \times 10^{11}$~M$_{\odot}$ (tuned to obtain the correct infall velocities) respectively. The red line shows the trajectory of a cloud in our model with a galactic mass of $M=8.3 \times 10^{11}$~M$_{\odot}$ but which is experiencing accretion drag. In the latter case, tuning of galactic mass is not necessary to explain observations.}
	\label{fig:cluster1}
\end{figure}

Figure~\ref{fig:cluster1} shows the observational contours of velocity as a function of radial distance from the center of Per A for the outer western filament as shown in Figure~10 of \citet{lim08}. In Fig.\ref{fig:cluster1}, we have also reproduced the free-fall trajectories used in \citet{lim08}, where they include one for galactic masses of $M=8.3 \times 10^{11}$\,M$_{\odot}$(M8b) and $M=3.4 \times 10^{11}$\,M$_{\odot}$(M3b), both starting from a radius of $8.5$\,kpc. The free-fall model that assumes the $M=8.3 \times 10^{11}$\,M$_{\odot}$ mass deduced from luminosity observations is unable to produce a good fit to observations, and hence the mass needs to be tuned to $M=3.4 \times 10^{11}$\,M$_{\odot}$ to fit a free-fall model to the observed contours. This tuning of mass and drop height is sensitive to both these factors, mainly due to the rapid acceleration by gravity in free-fall. In comparison, we show the same curve for $M=8.3 \times 10^{11}$~M$_{\odot}$ but using our model(M8c) (equations~\eqref{eq:diff_eq1}-\eqref{eq:diff_eq3}) which includes the braking effect due to growth from mixing and cooling. This shows the trajectory for a cloud where $r_{\rm cl}=50$~pc, assuming that $L/r \sim ~ 100$. We see that if the cloud initially falls from even a radial distance of 15\,kpc, it matches the observations well without changing the galaxy mass. Clouds can thus fall from a further distance out than observed. It should be noted that the conditions here are on the boundary of the survival criterion from equation~\eqref{eq:survival_criterion}, due to its strong scaling with $\chi$.
\begin{figure}
    \centering
	\includegraphics[width=\columnwidth]{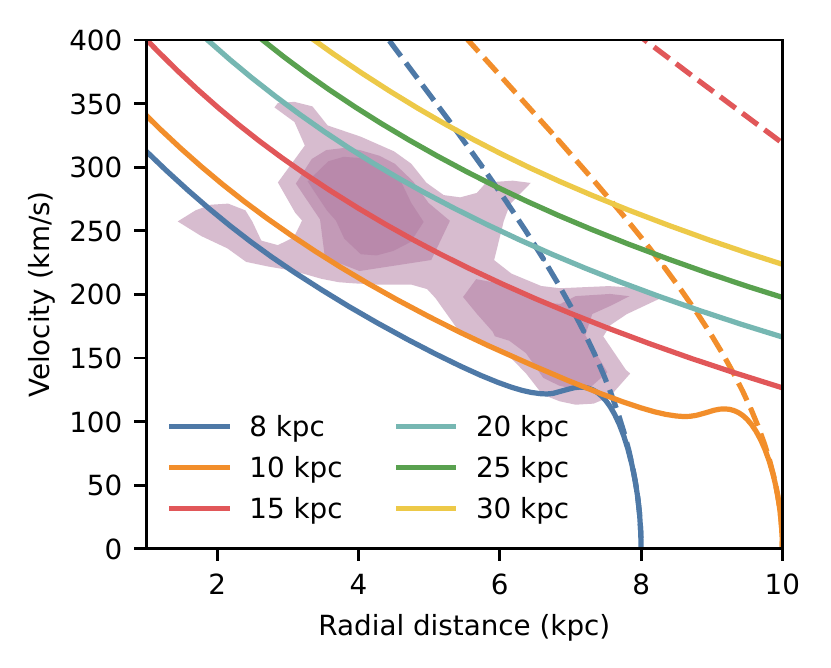}
	\caption{Observed velocity contours of the outer western filament in Per A from \citet{lim08} are shown in purple. Trajectories starting from different initial heights in our model are shown by the solid lines. Dashes lines show ballistic trajectories  with the same starting point. The velocities we predict are much less sensitive to the initial drop height compared to the ballistic model.}
	\label{fig:cluster2}
\end{figure}

In Fig.~\ref{fig:cluster2}, we show velocity trajectories for clouds dropped from various heights given by solid lines, with the 8\,kpc distance used as a lower bound. We find that even clouds that are dropped much further away do not accelerate as rapidly to high velocities as in the ballistic case. Ballistic trajectories for clouds dropped from the same heights are shown for comparison by the dashed lines, and can be seen to rapidly accelerate past observed velocities. On the other hand, the clouds in our model are slowed and stay within the range of observed velocities for much longer times. Hence, we are much less sensitive to the exact distance at which the cold gas first begins to fall.
Our results are consistent with the lower velocities and shallower velocity gradients observed relative to what would be expected from free-fall without requiring that the observed cold gas had only just recently cooled, or that magnetic drag from a strongly magnetized background must be present. In addition, the survival of cold gas and the filamentary morphology can also be naturally explained by cooling tails. 

\subsection{Other Implications}
We have found that it is more difficult for infalling cold material to survive, compared to their outflowing counterparts, which are eventually entrained and do not experience further shear forces thereafter. This conclusion has a range of wider implications which we will now touch on.

Assuming isobaric conditions, our survival criterion is equivalent to $t_{\rm cool,hot}/t_{\rm ff} \lsim 1$ (equation \eqref{eq:tcool_over_tff}), which is equivalent to the criterion for linear thermal instability in a plane parallel atmosphere. As previously remarked, this has the interesting implication that cold gas which forms via thermal instability should be able to survive infall, though this should be re-examined in a spherical potential, where the threshold for cold gas condensation changes, and $t_{\rm ff}$ (and gravitational acceleration) vary with radius. This is an interesting avenue for future work. 

Our results imply that clouds which grow in mass when they fall should undergo accretion-induced braking, a prediction which can be tested in larger scale simulations with more realistic set-ups. \citet{nelson20} find an abundance of cold clouds of sizes 1 kpc and smaller around the CGM of `red and dead' intermediate redshift elliptical galaxies in the TNG50 simulation. These clouds are mostly infalling, with the radial velocity distribution peaking at just one third of the virial velocity. They also find that the clouds are accreting and growing. They are long lived, surviving for cosmological timescales. This appears to be consistent with predictions from our model - that infalling cold clouds are growing and thus slowed to sub-virial velocities. It would be interesting to directly compare growth times $t_{\rm grow}$, and infall velocities, to see if the expectation $v_{\rm T} \sim g t_{\rm grow}$ is fulfilled.

Similarly, our results will affect the dynamics of cosmic cold streams feeding galaxies at high-$z$ \citep{dekel06,keres05}. Thus far, \citet{mandelker20} has found that the survival criterion for cold clouds seems to be able to translate relatively well to stream survival\footnote[10]{There is some controversy regarding the destruction timescale but for the relevant $\chi\sim100-1000$ the different possibilities agree \citep[cf. discussion in][]{bustard2021}}. However, in their studies they used a constant hot gas velocity -- similar to the outflowing cloud simulations -- which implies that their shear declines rapidly in the simulation. Since in reality cosmic streams are also accelerated by gravity, the situation for streams is likely closer to the setup studied here. This would imply that (a) an equivalently more stringent survival criterion would apply to streams, and (b) their terminal velocity is given by $\sim g t_{\rm grow}$. Indeed, unlike idealized simulations, cosmological simulations find that streams reach a roughly constant terminal velocity \citep{dekel09,goerdt15}; a result which has not been quantitatively explained. These implications directly affect the cold gas mass supplied towards the inner galaxies in dark matter halos.

Interestingly, coronal rain in our Sun is also observed to fall below free-fall velocities -- on average falling with only $\sim 1/3-1/2$  of the ballistic value \citep[see review by][]{Antolin2022FrASS...920116A}. While the temperatures and resulting overdensities are for coronal rain similar to what has mostly been considered here, the main difference is the strong magnetic field. Thus, most studies within the solar community have focused on magnetic fields as explanation of the slowdown and it has in fact been shown (using mostly one and two-dimensional simulations) that coronal rain can be efficiently decelerated due to a buildup of pressure in front of the cold cloud  \citep{Oliver2014a,Martinez-Gomez2020}. Clearly, the magnetic fields do play a major role here and affect the dynamics. However, it is noteworthy that mass transfer can also lead to significant slowdown. Plugging typical values found observationally ($n\sim 10^{11}\,{\rm cm}^{-3}$, $T\sim 2\times 10^4$, $r\sim 1\,$Mm, $\chi \sim 100$, $g=274\,{\rm m}\,{\rm s}^{-2}$; \citealp{Antolin2022FrASS...920116A}) into Eq.~\eqref{eq:tgrow_vgt} yields $v_{\rm term,drag}/v_{\rm term,grow}\sim 0.45$. Thus, the `accretion braking' process described in this work might be another important drag force at play; an interesting avenue for future work.

\subsection{Further Considerations}
While the model we have presented explores and captures the core physics at play, simplifications and assumptions have been made along the way. We discuss several considerations which could provide interesting avenues in order to expand and build on this model.

\subsubsection{Additional Physics}
There are a number of physical processes whose impact and importance we have not touched on in this work, but which could lead to complications and should be studied in future work. One such component is magnetic fields. Magnetic fields have been shown to significantly affect the morphology of clouds in both the wind tunnel and falling cloud setups, while their effect on mass growth is still uncertain \citep{gronnow17, gronnow18, gronke20-cloud, gronnow22, abruzzo22}.
For example, magnetic fields can suppress the KH instability, reducing mass entrainment rates \citep{ji19,gronnow22}, although mass growth rates in some full cloud simulations appear minimally impacted \citep{gronke20-cloud}. Another source of non-thermal physics that could be important to study in this context is cosmic rays \citep{huang22,armillotta22}. Self-gravity has been found to matter for compact HVCs \citep{sander21}. We have also not included explicit viscosity and thermal conduction (although we point out that for turbulent mixing layers the mass transfer is generally dominated by turbulent diffusion, \citealp{tan21}).

\subsubsection{Initial Cloud Morphology}
There is some uncertainty regarding an appropriate choice for the initial structure of the cloud. A spherical cloud is clearly an idealized choice. Instead of a uniform density sphere, smoothly varying density and temperature profiles connecting the two phases have been used for more realism \citep{heitsch09,kwak11,gritton14,sander21}. Furthermore, \citet{cooper09} found that fractal clouds were destroyed faster as compared to uniform spheres due to more rapid cloud breakup. \citet{schneider17} similarly found that an initially turbulent structure within the cloud would enhance fragmentation and ultimately facilitate cloud destruction. However, the above are all concerned with cloud destruction, where the clouds are in a regime where the cloud is ultimately destroyed over time ($t_{\rm cool,mix} > t_{\rm cc}$ for wind tunnel setups). The importance of the initial cloud structure can thus be understood in the context that it determines how the cloud is destroyed as it fragments and breaks up. However, if we are in the regime where one is concerned about cloud growth instead, then this dependence on the initial setup seems to matter less. \citet{gronke20-cloud} found that in the regime of cloud growth, there was little difference in either the mass growth or velocity evolution between an initially turbulent or uniform cloud. In fact, the turbulent case actually grew slightly faster, since it had a larger surface area at the start. Still, this suggests that the initial morphological evolution of the sphere does have some dependence on the choice of the initial structure of the cold gas cloud. In terms of numerical values, this creates some amount of uncertainty in our model, in particular with regards to the initial values of the cloud surface area and its initial evolution, which \citet{heitsch22} refers to as the `burn-in' phase. In our model, this uncertainty is folded in by calibrating a constant prefactor of order unity to the results from our simulations. It is possible that the precise value of this factor might vary depending on setup and initial cloud structure.

\subsubsection{Temperature Floor and Self Shielding}
In our simulations, we have assumed a temperature floor of $T \sim 10^4$~K. However, it would be useful to understand the phase structure of cold neutral gas that provides an additional layer of structure to the clouds \citep{girichidis2021,farber22} and how this might impact cloud growth and dynamics. This is especially for comparison with observations, which often detect warm gas surrounding cold cores. On a related note, we have assumed that all our clouds are optically thin. However, self-shielding could be important for the more massive clouds.

\subsubsection{Infall Conditions}
We have assumed our clouds fall directly towards the disk. However it is likely that most clouds will have some sort of rotational velocity component and hence fall inwards on some orbit trajectory. As mentioned in  \citet{heitsch09}, this component is more akin to the wind tunnel setups since net acceleration is reduced. We have also assumed a quiescent background - realistic environments are likely subject to large scale turbulence \citep[][]{gronke21}. This could affect mixing rates or significantly lengthen infall times and introduce a large stochastic variability in the infall velocity, much the same way a leaf falling to the ground in a windy environment follows a much longer trajectory. 
How this might affect cloud growth and dynamics is a natural follow up to this work. 

\subsubsection{Metallicity}
We have assumed solar metallicity everywhere in both phases. Depending on the origin of the cold cloud, it is possible that the metallicity of the original cloud and the background differ significantly. \citet{gritton14} and \citet{heitsch22} have showed that there is significant mixing of metals in such a case, with important implications for observables. 

\dobib

\section{Conclusions}  \label{sect:conclusions}
The growth and survival of infalling cold clouds has received considerably less attention compared to their outflowing counterparts. While the two appear to be similar problems at first glance, they have in fact a crucial difference between them, which is that infalling clouds continuously feel the force of gravity. This leads to very different dynamical evolution of the infalling clouds, and also a more stringent criterion for survival. Using 3D hydrodynamical simulations, we have studied the growth and survival of such clouds, considering both a constant background as a well as a more realistic stratified background. We have also developed a model for the dynamical evolution of these clouds based on turbulent mixing layer theory, and shown that they are able to predict the results of the simulations. These also agree well with analytical estimates. Our main findings are:

\begin{itemize}
  \item {\it Not a Wind Tunnel}:
  Infalling clouds do not correspond to wind tunnel setups, where the velocity shear is initially large and decreases as the cloud gets entrained. Instead, the velocity shear is initially small but increases as the cloud accelerates. This means that criteria such as $t_{\rm cool,mix} < t_{\rm cc}$ for survival are not applicable. 
  
  \item {\it Modelling Cloud Growth}:
  An important component determining how fast the cloud grows is the surface area of the cloud. We find that $A \propto m^{5/6}$. This is consistent with either a mix between surface and tail growth or a fractal surface area. Combining this with models of the inflow velocity allow us to model the growth time of the clouds, as given in equations~\eqref{eq:tgrow_scaling} and \eqref{eq:tgrow_scaling2}. We can hence evolve equations \eqref{eq:diff_eq1} -- \eqref{eq:diff_eq3} to model the evolution of cloud properties.
  
  \item  {\it Accretion Drag}:
  Clouds falling due to gravity can experience an alternative form of drag if they are growing via turbulent radiative mixing layers, since they are effectively accreting low momentum gas. This drag is dominant over the usual ram pressure drag as the clouds develop long tails along the direction of infall. This leads to much lower predicted infall velocities compared to models which only consider ballistic trajectories or ram pressure drag. In particular, the terminal velocity $v_{\rm T} \approx g t_{\rm grow}$, where $t_{\rm grow} = m/\dot{m}$ is given by equation~\eqref{eq:tgrow_vgt} for subsonic infall. 
  
  \item  {\it Relationship between Speed and Growth Rates}:
  The balance between gravity and growth results in $v_{\rm T}/c_{\rm s,hot} \sim t_{\rm grow}/t_{\rm ff}$. That is, the ratio of the terminal velocity and the virial velocity is
  also the ratio of the the growth time to the free-fall time. This is useful since infall velocities are measured observationally. The growth rate of the cloud can then be deduced.
  We expect sub-virial velocities ($v_{\rm T} < c_{\rm s,hot}$) to be indicative of considerable mass growth ($t_{\rm grow} < t_{\rm ff}$) in clouds. Observed sub-virial infall velocities are otherwise difficult to explain with existing models. In an isothermal atmosphere with constant gravity, we predict $v_{\rm T} \approx 0.6 c_{\rm s,hot}$, independent of all other properties of the system, although convergence to this asymptotic velocity can be slow. 
  
  \item {\it Criterion for Cloud Survival}:
  The criterion for clouds to survive and grow is  $t_{\rm grow} < 4 t_{\rm cc}$ (equation~\eqref{eq:survive}). The most important factor in determining cloud survival is the cooling time. We find that the ratio of $t_{\rm grow}/t_{\rm cc}$ is almost independent of cloud size (within a large practical range of parameter space). Hence, in order to survive and grow, clouds need only be within regions where densities/pressure are high enough such that cooling times are sufficiently short. For $\chi = 100$, this criterion can be written as 
  \begin{align}
     P > 3000 \,{\rm k}_B {\rm K}\,{\rm cm}^{-3}
    \left(\frac{g}{10^{-8}\,{\rm cm\,s}^{-2}}\right)^{4/5}
  \end{align}
  
  \item {\it Stratified Backgrounds and Cloud Size}:
  In stratified environments, clouds that start their infall beyond such survival `zones' can still survive provided they are not completely destroyed before reaching these zones. This favors larger clouds which have longer cloud crushing times. Larger clouds are hence more likely to be observed at distances where the above criterion is not satisfied. 
\end{itemize}

In summary, we have identified a new mechanism for the deceleration of clouds that has not been considered in existing models, with important bearings on cloud survival, growth, and dynamics. We have presented a model for cloud growth (equations \eqref{eq:diff_eq1} -- \eqref{eq:diff_eq3}), evolution (equations~\eqref{eq:tgrow_scaling} and \eqref{eq:tgrow_scaling2}), and survival (equation \eqref{eq:survive}) that agree well with simulations. These results can be applied to range of systems with infalling cold gas such as HVCs and clusters, and addresses important questions of survival, growth, and sub-virial velocities that have been highlighted by observations. Future work will refine this model with additional physics such as magnetic fields, cosmic rays and self-shielding, as well as allowing the gas to cool down to lower temperatures.

\section{Acknowledgements}
We thank Greg Bryan, Yakov Faerman, Drummond Fielding and Yong Zheng for helpful discussions, and the anonymous referee for a helpful and constructive report.
We have made use of the yt astrophysics analysis software suite \citep{yt}, matplotlib \citep{Hunter2007}, numpy \citep{van2011numpy}, and scipy \citep{2020SciPy-NMeth} whose communities we thank for continued development and support. 
We acknowledge support from NASA
grants NNX17AK58G, 19-ATP19-0205, HST-AR-15797.001-A,
NSF grant AST-1911198 and XSEDE grant TG-AST180036. 
MG thanks the Max Planck Society for support through the Max Planck Research Group.
This
research was supported in part by the National Science Foundation
under Grant No. NSF PHY-1748958 to KITP.

\section{Data Availability}
The data underlying this article will be shared on reasonable request
to the corresponding author.

\dobib

\newpage
\bibliography{references}

\begin{thebibliography}{}
\makeatletter
\relax
\def\mn@urlcharsother{\let\do\@makeother \do\$\do\&\do\#\do\^\do\_\do\%\do\~}
\def\mn@doi{\begingroup\mn@urlcharsother \@ifnextchar [ {\mn@doi@}
  {\mn@doi@[]}}
\def\mn@doi@[#1]#2{\def\@tempa{#1}\ifx\@tempa\@empty \href
  {http://dx.doi.org/#2} {doi:#2}\else \href {http://dx.doi.org/#2} {#1}\fi
  \endgroup}
\def\mn@eprint#1#2{\mn@eprint@#1:#2::\@nil}
\def\mn@eprint@arXiv#1{\href {http://arxiv.org/abs/#1} {{\tt arXiv:#1}}}
\def\mn@eprint@dblp#1{\href {http://dblp.uni-trier.de/rec/bibtex/#1.xml}
  {dblp:#1}}
\def\mn@eprint@#1:#2:#3:#4\@nil{\def\@tempa {#1}\def\@tempb {#2}\def\@tempc
  {#3}\ifx \@tempc \@empty \let \@tempc \@tempb \let \@tempb \@tempa \fi \ifx
  \@tempb \@empty \def\@tempb {arXiv}\fi \@ifundefined
  {mn@eprint@\@tempb}{\@tempb:\@tempc}{\expandafter \expandafter \csname
  mn@eprint@\@tempb\endcsname \expandafter{\@tempc}}}

\bibitem[\protect\citeauthoryear{{Abruzzo}, {Bryan}  \& {Fielding}}{{Abruzzo}
  et~al.}{2022a}]{abruzzo21}
{Abruzzo} M.~W.,  {Bryan} G.~L.,   {Fielding} D.~B.,  2022a, \mn@doi [\apj]
  {10.3847/1538-4357/ac3c48}, \href
  {https://ui.adsabs.harvard.edu/abs/2022ApJ...925..199A} {925, 199}

\bibitem[\protect\citeauthoryear{{Abruzzo}, {Bryan}  \& {Fielding}}{{Abruzzo}
  et~al.}{2022b}]{abruzzo22}
{Abruzzo} M.~W.,  {Bryan} G.~L.,   {Fielding} D.~B.,  2022b, \mn@doi [\apj]
  {10.3847/1538-4357/ac3c48}, \href
  {https://ui.adsabs.harvard.edu/abs/2022ApJ...925..199A} {925, 199}

\bibitem[\protect\citeauthoryear{{Afruni}, {Fraternali}  \&
  {Pezzulli}}{{Afruni} et~al.}{2019}]{afruni19}
{Afruni} A.,  {Fraternali} F.,   {Pezzulli} G.,  2019, \mn@doi [\aap]
  {10.1051/0004-6361/201835002}, \href
  {https://ui.adsabs.harvard.edu/abs/2019A&A...625A..11A} {625, A11}

\bibitem[\protect\citeauthoryear{{Antolin} \& {Froment}}{{Antolin} \&
  {Froment}}{2022}]{Antolin2022FrASS...920116A}
{Antolin} P.,  {Froment} C.,  2022, \mn@doi [Frontiers in Astronomy and Space
  Sciences] {10.3389/fspas.2022.820116}, \href
  {https://ui.adsabs.harvard.edu/abs/2022FrASS...920116A} {9, 820116}

\bibitem[\protect\citeauthoryear{Armillotta, Fraternali  \&
  Marinacci}{Armillotta et~al.}{2016}]{armillotta16}
Armillotta L.,  Fraternali F.,   Marinacci F.,  2016, \mn@doi [Monthly Notices
  of the Royal Astronomical Society] {10.1093/mnras/stw1930}, 462, 4157

\bibitem[\protect\citeauthoryear{{Armillotta}, {Fraternali}, {Werk},
  {Prochaska}  \& {Marinacci}}{{Armillotta} et~al.}{2017}]{armillotta17}
{Armillotta} L.,  {Fraternali} F.,  {Werk} J.~K.,  {Prochaska} J.~X.,
  {Marinacci} F.,  2017, \mn@doi [\mnras] {10.1093/mnras/stx1239}, \href
  {https://ui.adsabs.harvard.edu/abs/2017MNRAS.470..114A} {470, 114}

\bibitem[\protect\citeauthoryear{{Armillotta}, {Ostriker}  \&
  {Jiang}}{{Armillotta} et~al.}{2022}]{armillotta22}
{Armillotta} L.,  {Ostriker} E.~C.,   {Jiang} Y.-F.,  2022, \mn@doi [\apj]
  {10.3847/1538-4357/ac5fa9}, \href
  {https://ui.adsabs.harvard.edu/abs/2022ApJ...929..170A} {929, 170}

\bibitem[\protect\citeauthoryear{{Barenblatt} \& {Monin}}{{Barenblatt} \&
  {Monin}}{1983}]{barenblatt83}
{Barenblatt} G.~I.,  {Monin} A.~S.,  1983, \mn@doi [Proceedings of the National
  Academy of Science] {10.1073/pnas.80.11.3540}, \href
  {https://ui.adsabs.harvard.edu/abs/1983PNAS...80.3540B} {80, 3540}

\bibitem[\protect\citeauthoryear{{Begelman} \& {Fabian}}{{Begelman} \&
  {Fabian}}{1990}]{begelman90}
{Begelman} M.~C.,  {Fabian} A.~C.,  1990, \mnras, \href
  {http://adsabs.harvard.edu/abs/1990MNRAS.244P..26B} {244, 26P}

\bibitem[\protect\citeauthoryear{{Benjamin} \& {Danly}}{{Benjamin} \&
  {Danly}}{1997}]{benjamin&daly97}
{Benjamin} R.~A.,  {Danly} L.,  1997, \mn@doi [\apj] {10.1086/304078}, \href
  {https://ui.adsabs.harvard.edu/abs/1997ApJ...481..764B} {481, 764}

\bibitem[\protect\citeauthoryear{Bish, Werk, Peek, Zheng  \& Putman}{Bish
  et~al.}{2021}]{Bish2021}
Bish H.~V.,  Werk J.~K.,  Peek J.,  Zheng Y.,   Putman M.,  2021, \mn@doi [The
  Astrophysical Journal] {10.3847/1538-4357/abeb6b}, 912, 8

\bibitem[\protect\citeauthoryear{{Bregman}}{{Bregman}}{1980}]{bregman80}
{Bregman} J.~N.,  1980, \mn@doi [\apj] {10.1086/157776}, \href
  {https://ui.adsabs.harvard.edu/abs/1980ApJ...236..577B} {236, 577}

\bibitem[\protect\citeauthoryear{{Bustard} \& {Gronke}}{{Bustard} \&
  {Gronke}}{2022}]{bustard2021}
{Bustard} C.,  {Gronke} M.,  2022, \mn@doi [\apj] {10.3847/1538-4357/ac752b},
  \href {https://ui.adsabs.harvard.edu/abs/2022ApJ...933..120B} {933, 120}

\bibitem[\protect\citeauthoryear{{Chomiuk} \& {Povich}}{{Chomiuk} \&
  {Povich}}{2011}]{chomiuk11}
{Chomiuk} L.,  {Povich} M.~S.,  2011, \mn@doi [\aj]
  {10.1088/0004-6256/142/6/197}, \href
  {https://ui.adsabs.harvard.edu/abs/2011AJ....142..197C} {142, 197}

\bibitem[\protect\citeauthoryear{{Choudhury} \& {Sharma}}{{Choudhury} \&
  {Sharma}}{2016}]{choudhury16}
{Choudhury} P.~P.,  {Sharma} P.,  2016, \mn@doi [\mnras]
  {10.1093/mnras/stw152}, \href
  {https://ui.adsabs.harvard.edu/abs/2016MNRAS.457.2554C} {457, 2554}

\bibitem[\protect\citeauthoryear{{Churazov}, {Forman}, {Jones}  \&
  {B{\"o}hringer}}{{Churazov} et~al.}{2003}]{churazov03}
{Churazov} E.,  {Forman} W.,  {Jones} C.,   {B{\"o}hringer} H.,  2003, \mn@doi
  [\apj] {10.1086/374923}, \href
  {https://ui.adsabs.harvard.edu/abs/2003ApJ...590..225C} {590, 225}

\bibitem[\protect\citeauthoryear{{Cooper}, {Bicknell}, {Sutherland}  \&
  {Bland-Hawthorn}}{{Cooper} et~al.}{2009}]{cooper09}
{Cooper} J.~L.,  {Bicknell} G.~V.,  {Sutherland} R.~S.,   {Bland-Hawthorn} J.,
  2009, \mn@doi [\apj] {10.1088/0004-637X/703/1/330}, \href
  {https://ui.adsabs.harvard.edu/abs/2009ApJ...703..330C} {703, 330}

\bibitem[\protect\citeauthoryear{{Dekel} \& {Birnboim}}{{Dekel} \&
  {Birnboim}}{2006}]{dekel06}
{Dekel} A.,  {Birnboim} Y.,  2006, \mn@doi [\mnras]
  {10.1111/j.1365-2966.2006.10145.x}, \href
  {https://ui.adsabs.harvard.edu/abs/2006MNRAS.368....2D} {368, 2}

\bibitem[\protect\citeauthoryear{{Dekel} et~al.,}{{Dekel}
  et~al.}{2009}]{dekel09}
{Dekel} A.,  et~al., 2009, \mn@doi [\nat] {10.1038/nature07648}, \href
  {https://ui.adsabs.harvard.edu/abs/2009Natur.457..451D} {457, 451}

\bibitem[\protect\citeauthoryear{{Erb}}{{Erb}}{2008}]{erb08}
{Erb} D.~K.,  2008, \mn@doi [\apj] {10.1086/524727}, \href
  {https://ui.adsabs.harvard.edu/abs/2008ApJ...674..151E} {674, 151}

\bibitem[\protect\citeauthoryear{{Fabian}, {Johnstone}, {Sanders}, {Conselice},
  {Crawford}, {Gallagher}  \& {Zweibel}}{{Fabian} et~al.}{2008}]{fabian2008}
{Fabian} A.~C.,  {Johnstone} R.~M.,  {Sanders} J.~S.,  {Conselice} C.~J.,
  {Crawford} C.~S.,  {Gallagher} J.~S. I.,   {Zweibel} E.,  2008, \mn@doi
  [\nat] {10.1038/nature07169}, \href
  {https://ui.adsabs.harvard.edu/abs/2008Natur.454..968F} {454, 968}

\bibitem[\protect\citeauthoryear{{Farber} \& {Gronke}}{{Farber} \&
  {Gronke}}{2022}]{farber22}
{Farber} R.~J.,  {Gronke} M.,  2022, \mn@doi [\mnras] {10.1093/mnras/stab3412},
  \href {https://ui.adsabs.harvard.edu/abs/2022MNRAS.510..551F} {510, 551}

\bibitem[\protect\citeauthoryear{{Fielding}, {Ostriker}, {Bryan}  \&
  {Jermyn}}{{Fielding} et~al.}{2020}]{fielding20}
{Fielding} D.~B.,  {Ostriker} E.~C.,  {Bryan} G.~L.,   {Jermyn} A.~S.,  2020,
  \mn@doi [\apjl] {10.3847/2041-8213/ab8d2c}, \href
  {https://ui.adsabs.harvard.edu/abs/2020ApJ...894L..24F} {894, L24}

\bibitem[\protect\citeauthoryear{{Fox} et~al.,}{{Fox} et~al.}{2016}]{fox16}
{Fox} A.~J.,  et~al., 2016, \mn@doi [\apjl] {10.3847/2041-8205/816/1/L11},
  \href {https://ui.adsabs.harvard.edu/abs/2016ApJ...816L..11F} {816, L11}

\bibitem[\protect\citeauthoryear{{Fox}, {Richter}, {Ashley}, {Heckman},
  {Lehner}, {Werk}, {Bordoloi}  \& {Peeples}}{{Fox} et~al.}{2019a}]{fox19}
{Fox} A.~J.,  {Richter} P.,  {Ashley} T.,  {Heckman} T.~M.,  {Lehner} N.,
  {Werk} J.~K.,  {Bordoloi} R.,   {Peeples} M.~S.,  2019a, \mn@doi [\apj]
  {10.3847/1538-4357/ab40ad}, \href
  {https://ui.adsabs.harvard.edu/abs/2019ApJ...884...53F} {884, 53}

\bibitem[\protect\citeauthoryear{{Fox}, {Richter}, {Ashley}, {Heckman},
  {Lehner}, {Werk}, {Bordoloi}  \& {Peeples}}{{Fox} et~al.}{2019b}]{Fox2019}
{Fox} A.~J.,  {Richter} P.,  {Ashley} T.,  {Heckman} T.~M.,  {Lehner} N.,
  {Werk} J.~K.,  {Bordoloi} R.,   {Peeples} M.~S.,  2019b, \mn@doi [\apj]
  {10.3847/1538-4357/ab40ad}, \href
  {https://ui.adsabs.harvard.edu/abs/2019ApJ...884...53F} {884, 53}

\bibitem[\protect\citeauthoryear{{Fraternali}}{{Fraternali}}{2017}]{fraternali17}
{Fraternali} F.,  2017, in {Fox} A.,  {Dav{\'e}} R.,  eds,  Astrophysics and
  Space Science Library Vol. 430, Gas Accretion onto Galaxies. p.~323
  (\mn@eprint {arXiv} {1612.00477}), \mn@doi{10.1007/978-3-319-52512-9_14}

\bibitem[\protect\citeauthoryear{{Fraternali} \& {Binney}}{{Fraternali} \&
  {Binney}}{2006}]{2006MNRAS.366..449F}
{Fraternali} F.,  {Binney} J.~J.,  2006, \mn@doi [\mnras]
  {10.1111/j.1365-2966.2005.09816.x}, \href
  {https://ui.adsabs.harvard.edu/abs/2006MNRAS.366..449F} {366, 449}

\bibitem[\protect\citeauthoryear{{Fraternali} \& {Binney}}{{Fraternali} \&
  {Binney}}{2008}]{fraternali08}
{Fraternali} F.,  {Binney} J.~J.,  2008, \mn@doi [\mnras]
  {10.1111/j.1365-2966.2008.13071.x}, \href
  {https://ui.adsabs.harvard.edu/abs/2008MNRAS.386..935F} {386, 935}

\bibitem[\protect\citeauthoryear{{Fraternali}, {Marasco}, {Armillotta}  \&
  {Marinacci}}{{Fraternali} et~al.}{2015}]{fraternali15}
{Fraternali} F.,  {Marasco} A.,  {Armillotta} L.,   {Marinacci} F.,  2015,
  \mn@doi [\mnras] {10.1093/mnrasl/slu182}, \href
  {https://ui.adsabs.harvard.edu/abs/2015MNRAS.447L..70F} {447, L70}

\bibitem[\protect\citeauthoryear{{Galyardt} \& {Shelton}}{{Galyardt} \&
  {Shelton}}{2016}]{galyardt16}
{Galyardt} J.,  {Shelton} R.~L.,  2016, \mn@doi [\apjl]
  {10.3847/2041-8205/816/1/L18}, \href
  {https://ui.adsabs.harvard.edu/abs/2016ApJ...816L..18G} {816, L18}

\bibitem[\protect\citeauthoryear{{Girichidis}, {Naab}, {Walch}  \&
  {Berlok}}{{Girichidis} et~al.}{2021}]{girichidis2021}
{Girichidis} P.,  {Naab} T.,  {Walch} S.,   {Berlok} T.,  2021, \mn@doi
  [\mnras] {10.1093/mnras/stab1203}, \href
  {https://ui.adsabs.harvard.edu/abs/2021MNRAS.505.1083G} {505, 1083}

\bibitem[\protect\citeauthoryear{Gnat \& Sternberg}{Gnat \&
  Sternberg}{2007}]{Gnat2007}
Gnat O.,  Sternberg A.,  2007, \mn@doi [\apjs] {10.1086/509786}, 168, 213

\bibitem[\protect\citeauthoryear{{Goerdt} \& {Ceverino}}{{Goerdt} \&
  {Ceverino}}{2015}]{goerdt15}
{Goerdt} T.,  {Ceverino} D.,  2015, \mn@doi [\mnras] {10.1093/mnras/stv786},
  \href {https://ui.adsabs.harvard.edu/abs/2015MNRAS.450.3359G} {450, 3359}

\bibitem[\protect\citeauthoryear{{Gritton}, {Shelton}  \& {Kwak}}{{Gritton}
  et~al.}{2014}]{gritton14}
{Gritton} J.~A.,  {Shelton} R.~L.,   {Kwak} K.,  2014, \mn@doi [\apj]
  {10.1088/0004-637X/795/1/99}, \href
  {https://ui.adsabs.harvard.edu/abs/2014ApJ...795...99G} {795, 99}

\bibitem[\protect\citeauthoryear{{Gritton}, {Shelton}  \& {Galyardt}}{{Gritton}
  et~al.}{2017}]{gritton17}
{Gritton} J.~A.,  {Shelton} R.~L.,   {Galyardt} J.~E.,  2017, \mn@doi [\apj]
  {10.3847/1538-4357/aa756d}, \href
  {https://ui.adsabs.harvard.edu/abs/2017ApJ...842..102G} {842, 102}

\bibitem[\protect\citeauthoryear{{Gronke} \& {Oh}}{{Gronke} \&
  {Oh}}{2018}]{gronke18}
{Gronke} M.,  {Oh} S.~P.,  2018, \mn@doi [\mnras] {10.1093/mnrasl/sly131},
  \href {https://ui.adsabs.harvard.edu/abs/2018MNRAS.480L.111G} {480, L111}

\bibitem[\protect\citeauthoryear{{Gronke} \& {Oh}}{{Gronke} \&
  {Oh}}{2020a}]{gronke20-cloud}
{Gronke} M.,  {Oh} S.~P.,  2020a, \mn@doi [\mnras] {10.1093/mnras/stz3332},
  \href {https://ui.adsabs.harvard.edu/abs/2020MNRAS.492.1970G} {492, 1970}

\bibitem[\protect\citeauthoryear{{Gronke} \& {Oh}}{{Gronke} \&
  {Oh}}{2020b}]{gronke20-mist}
{Gronke} M.,  {Oh} S.~P.,  2020b, \mn@doi [\mnras] {10.1093/mnrasl/slaa033},
  \href {https://ui.adsabs.harvard.edu/abs/2020MNRAS.494L..27G} {494, L27}

\bibitem[\protect\citeauthoryear{{Gronke} \& {Oh}}{{Gronke} \&
  {Oh}}{2022}]{gronke22}
{Gronke} M.,  {Oh} S.~P.,  2022, arXiv e-prints, \href
  {https://ui.adsabs.harvard.edu/abs/2022arXiv220900732G} {p. arXiv:2209.00732}

\bibitem[\protect\citeauthoryear{{Gronke}, {Oh}, {Ji}  \& {Norman}}{{Gronke}
  et~al.}{2022}]{gronke21}
{Gronke} M.,  {Oh} S.~P.,  {Ji} S.,   {Norman} C.,  2022, \mn@doi [\mnras]
  {10.1093/mnras/stab3351}, \href
  {https://ui.adsabs.harvard.edu/abs/2022MNRAS.511..859G} {511, 859}

\bibitem[\protect\citeauthoryear{{Gr{\o}nnow}, {Tepper-Garc{\'\i}a},
  {Bland-Hawthorn}  \& {McClure-Griffiths}}{{Gr{\o}nnow}
  et~al.}{2017}]{gronnow17}
{Gr{\o}nnow} A.,  {Tepper-Garc{\'\i}a} T.,  {Bland-Hawthorn} J.,
  {McClure-Griffiths} N.~M.,  2017, \mn@doi [\apj] {10.3847/1538-4357/aa7ed2},
  \href {https://ui.adsabs.harvard.edu/abs/2017ApJ...845...69G} {845, 69}

\bibitem[\protect\citeauthoryear{{Gr{\o}nnow}, {Tepper-Garc{\'\i}a}  \&
  {Bland-Hawthorn}}{{Gr{\o}nnow} et~al.}{2018}]{gronnow18}
{Gr{\o}nnow} A.,  {Tepper-Garc{\'\i}a} T.,   {Bland-Hawthorn} J.,  2018,
  \mn@doi [\apj] {10.3847/1538-4357/aada0e}, \href
  {https://ui.adsabs.harvard.edu/abs/2018ApJ...865...64G} {865, 64}

\bibitem[\protect\citeauthoryear{{Gr{\o}nnow}, {Tepper-Garc{\'\i}a},
  {Bland-Hawthorn}  \& {Fraternali}}{{Gr{\o}nnow} et~al.}{2022}]{gronnow22}
{Gr{\o}nnow} A.,  {Tepper-Garc{\'\i}a} T.,  {Bland-Hawthorn} J.,   {Fraternali}
  F.,  2022, \mn@doi [\mnras] {10.1093/mnras/stab3452}, \href
  {https://ui.adsabs.harvard.edu/abs/2022MNRAS.509.5756G} {509, 5756}

\bibitem[\protect\citeauthoryear{{Heckman} \& {Thompson}}{{Heckman} \&
  {Thompson}}{2017}]{heckman17}
{Heckman} T.~M.,  {Thompson} T.~A.,  2017, arXiv e-prints, \href
  {https://ui.adsabs.harvard.edu/abs/2017arXiv170109062H} {p. arXiv:1701.09062}

\bibitem[\protect\citeauthoryear{{Heitsch} \& {Putman}}{{Heitsch} \&
  {Putman}}{2009}]{heitsch09}
{Heitsch} F.,  {Putman} M.~E.,  2009, \mn@doi [\apj]
  {10.1088/0004-637X/698/2/1485}, \href
  {https://ui.adsabs.harvard.edu/abs/2009ApJ...698.1485H} {698, 1485}

\bibitem[\protect\citeauthoryear{{Heitsch}, {Marchal}, {Miville-Desch{\^e}nes},
  {Shull}  \& {Fox}}{{Heitsch} et~al.}{2022}]{heitsch22}
{Heitsch} F.,  {Marchal} A.,  {Miville-Desch{\^e}nes} M.~A.,  {Shull} J.~M.,
  {Fox} A.~J.,  2022, \mn@doi [\mnras] {10.1093/mnras/stab3266}, \href
  {https://ui.adsabs.harvard.edu/abs/2022MNRAS.509.4515H} {509, 4515}

\bibitem[\protect\citeauthoryear{{Henley}, {Gritton}  \& {Shelton}}{{Henley}
  et~al.}{2017}]{henley17}
{Henley} D.~B.,  {Gritton} J.~A.,   {Shelton} R.~L.,  2017, \mn@doi [\apj]
  {10.3847/1538-4357/aa5df7}, \href
  {https://ui.adsabs.harvard.edu/abs/2017ApJ...837...82H} {837, 82}

\bibitem[\protect\citeauthoryear{{Hernquist}}{{Hernquist}}{1990}]{hernquist90}
{Hernquist} L.,  1990, \mn@doi [\apj] {10.1086/168845}, \href
  {https://ui.adsabs.harvard.edu/abs/1990ApJ...356..359H} {356, 359}

\bibitem[\protect\citeauthoryear{{Hillier} \& {Arregui}}{{Hillier} \&
  {Arregui}}{2019}]{2019ApJ...885..101H}
{Hillier} A.,  {Arregui} I.,  2019, \mn@doi [\apj] {10.3847/1538-4357/ab4795},
  \href {https://ui.adsabs.harvard.edu/abs/2019ApJ...885..101H} {885, 101}

\bibitem[\protect\citeauthoryear{{Hopkins}, {McClure-Griffiths}  \&
  {Gaensler}}{{Hopkins} et~al.}{2008}]{hopkins08}
{Hopkins} A.~M.,  {McClure-Griffiths} N.~M.,   {Gaensler} B.~M.,  2008, \mn@doi
  [\apjl] {10.1086/590494}, \href
  {https://ui.adsabs.harvard.edu/abs/2008ApJ...682L..13H} {682, L13}

\bibitem[\protect\citeauthoryear{{Huang}, {Chen}, {Johnson}  \&
  {Weiner}}{{Huang} et~al.}{2016}]{huang16}
{Huang} Y.-H.,  {Chen} H.-W.,  {Johnson} S.~D.,   {Weiner} B.~J.,  2016,
  \mn@doi [\mnras] {10.1093/mnras/stv2327}, \href
  {https://ui.adsabs.harvard.edu/abs/2016MNRAS.455.1713H} {455, 1713}

\bibitem[\protect\citeauthoryear{{Huang}, {Jiang}  \& {Davis}}{{Huang}
  et~al.}{2022}]{huang22}
{Huang} X.,  {Jiang} Y.-f.,   {Davis} S.~W.,  2022, \mn@doi [\apj]
  {10.3847/1538-4357/ac69dc}, \href
  {https://ui.adsabs.harvard.edu/abs/2022ApJ...931..140H} {931, 140}

\bibitem[\protect\citeauthoryear{Hunter}{Hunter}{2007}]{Hunter2007}
Hunter J.~D.,  2007, \mn@doi [Computing in Science \& Engineering]
  {10.1109/MCSE.2007.55}, 9, 90

\bibitem[\protect\citeauthoryear{{Ji}, {Oh}  \& {McCourt}}{{Ji}
  et~al.}{2018}]{ji18}
{Ji} S.,  {Oh} S.~P.,   {McCourt} M.,  2018, \mn@doi [\mnras]
  {10.1093/mnras/sty293}, \href
  {http://adsabs.harvard.edu/abs/2018MNRAS.476..852J} {476, 852}

\bibitem[\protect\citeauthoryear{{Ji}, {Oh}  \& {Masterson}}{{Ji}
  et~al.}{2019}]{ji19}
{Ji} S.,  {Oh} S.~P.,   {Masterson} P.,  2019, \mn@doi [\mnras]
  {10.1093/mnras/stz1248}, \href
  {https://ui.adsabs.harvard.edu/abs/2019MNRAS.487..737J} {487, 737}

\bibitem[\protect\citeauthoryear{{Joung}, {Bryan}  \& {Putman}}{{Joung}
  et~al.}{2012}]{joung12}
{Joung} M.~R.,  {Bryan} G.~L.,   {Putman} M.~E.,  2012, \mn@doi [\apj]
  {10.1088/0004-637X/745/2/148}, \href
  {https://ui.adsabs.harvard.edu/abs/2012ApJ...745..148J} {745, 148}

\bibitem[\protect\citeauthoryear{{Kere{\v{s}}}, {Katz}, {Weinberg}  \&
  {Dav{\'e}}}{{Kere{\v{s}}} et~al.}{2005}]{keres05}
{Kere{\v{s}}} D.,  {Katz} N.,  {Weinberg} D.~H.,   {Dav{\'e}} R.,  2005,
  \mn@doi [\mnras] {10.1111/j.1365-2966.2005.09451.x}, \href
  {https://ui.adsabs.harvard.edu/abs/2005MNRAS.363....2K} {363, 2}

\bibitem[\protect\citeauthoryear{{Klein}, {McKee}  \& {Colella}}{{Klein}
  et~al.}{1994}]{klein94}
{Klein} R.~I.,  {McKee} C.~F.,   {Colella} P.,  1994, \mn@doi [\apj]
  {10.1086/173554}, \href
  {https://ui.adsabs.harvard.edu/abs/1994ApJ...420..213K} {420, 213}

\bibitem[\protect\citeauthoryear{{Kubryk}, {Prantzos}  \&
  {Athanassoula}}{{Kubryk} et~al.}{2013}]{kubryk13}
{Kubryk} M.,  {Prantzos} N.,   {Athanassoula} E.,  2013, \mn@doi [\mnras]
  {10.1093/mnras/stt1667}, \href
  {https://ui.adsabs.harvard.edu/abs/2013MNRAS.436.1479K} {436, 1479}

\bibitem[\protect\citeauthoryear{{Kwak}, {Henley}  \& {Shelton}}{{Kwak}
  et~al.}{2011}]{kwak11}
{Kwak} K.,  {Henley} D.~B.,   {Shelton} R.~L.,  2011, \mn@doi [\apj]
  {10.1088/0004-637X/739/1/30}, \href
  {https://ui.adsabs.harvard.edu/abs/2011ApJ...739...30K} {739, 30}

\bibitem[\protect\citeauthoryear{{Li}, {Hopkins}, {Squire}  \& {Hummels}}{{Li}
  et~al.}{2020}]{li20}
{Li} Z.,  {Hopkins} P.~F.,  {Squire} J.,   {Hummels} C.,  2020, \mn@doi
  [\mnras] {10.1093/mnras/stz3567}, \href
  {https://ui.adsabs.harvard.edu/abs/2020MNRAS.492.1841L} {492, 1841}

\bibitem[\protect\citeauthoryear{{Lim}, {Ao}  \& {Dinh-V-Trung}}{{Lim}
  et~al.}{2008}]{lim08}
{Lim} J.,  {Ao} Y.,   {Dinh-V-Trung} 2008, \mn@doi [\apj] {10.1086/523664},
  \href {https://ui.adsabs.harvard.edu/abs/2008ApJ...672..252L} {672, 252}

\bibitem[\protect\citeauthoryear{{Lockman}, {Benjamin}, {Heroux}  \&
  {Langston}}{{Lockman} et~al.}{2008}]{lockman08}
{Lockman} F.~J.,  {Benjamin} R.~A.,  {Heroux} A.~J.,   {Langston} G.~I.,  2008,
  \mn@doi [\apjl] {10.1086/588838}, \href
  {https://ui.adsabs.harvard.edu/abs/2008ApJ...679L..21L} {679, L21}

\bibitem[\protect\citeauthoryear{{Maller} \& {Bullock}}{{Maller} \&
  {Bullock}}{2004}]{maller04}
{Maller} A.~H.,  {Bullock} J.~S.,  2004, \mn@doi [\mnras]
  {10.1111/j.1365-2966.2004.08349.x}, \href
  {http://adsabs.harvard.edu/abs/2004MNRAS.355..694M} {355, 694}

\bibitem[\protect\citeauthoryear{{Mandelker}, {Nagai}, {Aung}, {Dekel},
  {Birnboim}  \& {van den Bosch}}{{Mandelker} et~al.}{2020}]{mandelker20}
{Mandelker} N.,  {Nagai} D.,  {Aung} H.,  {Dekel} A.,  {Birnboim} Y.,   {van
  den Bosch} F.~C.,  2020, \mn@doi [\mnras] {10.1093/mnras/staa812}, \href
  {https://ui.adsabs.harvard.edu/abs/2020MNRAS.494.2641M} {494, 2641}

\bibitem[\protect\citeauthoryear{{Mart{\'i}nez-G{\'o}mez}, Oliver, Khomenko  \&
  Collados}{{Mart{\'i}nez-G{\'o}mez} et~al.}{2020}]{Martinez-Gomez2020}
{Mart{\'i}nez-G{\'o}mez} D.,  Oliver R.,  Khomenko E.,   Collados M.,  2020,
  \mn@doi [Astronomy and Astrophysics] {10.1051/0004-6361/201937078}, 634, 1

\bibitem[\protect\citeauthoryear{{McCourt}, {Sharma}, {Quataert}  \&
  {Parrish}}{{McCourt} et~al.}{2012}]{mccourt12}
{McCourt} M.,  {Sharma} P.,  {Quataert} E.,   {Parrish} I.~J.,  2012, \mn@doi
  [\mnras] {10.1111/j.1365-2966.2011.19972.x}, \href
  {http://adsabs.harvard.edu/abs/2012MNRAS.419.3319M} {419, 3319}

\bibitem[\protect\citeauthoryear{{McCourt}, {O'Leary}, {Madigan}  \&
  {Quataert}}{{McCourt} et~al.}{2015}]{mccourt15}
{McCourt} M.,  {O'Leary} R.~M.,  {Madigan} A.-M.,   {Quataert} E.,  2015,
  \mn@doi [\mnras] {10.1093/mnras/stv355}, \href
  {https://ui.adsabs.harvard.edu/abs/2015MNRAS.449....2M} {449, 2}

\bibitem[\protect\citeauthoryear{{McNamara} et~al.,}{{McNamara}
  et~al.}{2014}]{mcnamara14}
{McNamara} B.~R.,  et~al., 2014, \mn@doi [\apj] {10.1088/0004-637X/785/1/44},
  \href {https://ui.adsabs.harvard.edu/abs/2014ApJ...785...44M} {785, 44}

\bibitem[\protect\citeauthoryear{{McNamara}, {Russell}, {Nulsen}, {Hogan},
  {Fabian}, {Pulido}  \& {Edge}}{{McNamara} et~al.}{2016}]{mcnamara16}
{McNamara} B.~R.,  {Russell} H.~R.,  {Nulsen} P.~E.~J.,  {Hogan} M.~T.,
  {Fabian} A.~C.,  {Pulido} F.,   {Edge} A.~C.,  2016, \mn@doi [\apj]
  {10.3847/0004-637X/830/2/79}, \href
  {https://ui.adsabs.harvard.edu/abs/2016ApJ...830...79M} {830, 79}

\bibitem[\protect\citeauthoryear{{Miller} \& {Bregman}}{{Miller} \&
  {Bregman}}{2015}]{miller2015}
{Miller} M.~J.,  {Bregman} J.~N.,  2015, \mn@doi [\apj]
  {10.1088/0004-637X/800/1/14}, \href
  {https://ui.adsabs.harvard.edu/abs/2015ApJ...800...14M} {800, 14}

\bibitem[\protect\citeauthoryear{{Muller}, {Oort}  \& {Raimond}}{{Muller}
  et~al.}{1963}]{muller63}
{Muller} C.~A.,  {Oort} J.~H.,   {Raimond} E.,  1963, Academie des Sciences
  Paris Comptes Rendus, \href
  {https://ui.adsabs.harvard.edu/abs/1963CRAS..257.1661M} {257, 1661}

\bibitem[\protect\citeauthoryear{{Navarro}, {Frenk}  \& {White}}{{Navarro}
  et~al.}{1997}]{navarro97}
{Navarro} J.~F.,  {Frenk} C.~S.,   {White} S.~D.~M.,  1997, \apj, 490, 493
  (NFW)

\bibitem[\protect\citeauthoryear{{Nelson} et~al.,}{{Nelson}
  et~al.}{2020}]{nelson20}
{Nelson} D.,  et~al., 2020, \mn@doi [\mnras] {10.1093/mnras/staa2419}, \href
  {https://ui.adsabs.harvard.edu/abs/2020MNRAS.498.2391N} {498, 2391}

\bibitem[\protect\citeauthoryear{{Nichols} \& {Bland-Hawthorn}}{{Nichols} \&
  {Bland-Hawthorn}}{2009}]{nichols09}
{Nichols} M.,  {Bland-Hawthorn} J.,  2009, \mn@doi [\apj]
  {10.1088/0004-637X/707/2/1642}, \href
  {https://ui.adsabs.harvard.edu/abs/2009ApJ...707.1642N} {707, 1642}

\bibitem[\protect\citeauthoryear{Oliver, Soler, Terradas, Zaqarashvili  \&
  Khodachenko}{Oliver et~al.}{2014}]{Oliver2014a}
Oliver R.,  Soler R.,  Terradas J.,  Zaqarashvili T.~V.,   Khodachenko M.~L.,
  2014, \mn@doi [The Astrophysical Journal] {10.1088/0004-637X/784/1/21}, 784,
  21

\bibitem[\protect\citeauthoryear{{Peek}, {Putman}, {Sommer-Larsen}, {Heiles},
  {Stanimirovic}, {Douglas}, {Gibson}  \& {Korpela}}{{Peek}
  et~al.}{2007}]{Peek2007}
{Peek} J.~E.~G.,  {Putman} M.~E.,  {Sommer-Larsen} J.,  {Heiles} C.~E.,
  {Stanimirovic} S.,  {Douglas} K.,  {Gibson} S.,   {Korpela} E.,  2007, in
  American Astronomical Society Meeting Abstracts. p. 14.08

\bibitem[\protect\citeauthoryear{Peek, Heiles, Putman  \& Douglas}{Peek
  et~al.}{2009}]{Peek2009}
Peek J. E.~G.,  Heiles C.,  Putman M.~E.,   Douglas K.,  2009, \mn@doi [The
  Astrophysical Journal] {10.1088/0004-637X/692/1/827}, 692, 827

\bibitem[\protect\citeauthoryear{{P{\'e}roux} \& {Howk}}{{P{\'e}roux} \&
  {Howk}}{2020}]{2020ARA&A..58..363P}
{P{\'e}roux} C.,  {Howk} J.~C.,  2020, \mn@doi [\araa]
  {10.1146/annurev-astro-021820-120014}, \href
  {https://ui.adsabs.harvard.edu/abs/2020ARA&A..58..363P} {58, 363}

\bibitem[\protect\citeauthoryear{{Pulido} et~al.,}{{Pulido}
  et~al.}{2018}]{pulido18}
{Pulido} F.~A.,  et~al., 2018, \mn@doi [\apj] {10.3847/1538-4357/aaa54b}, \href
  {https://ui.adsabs.harvard.edu/abs/2018ApJ...853..177P} {853, 177}

\bibitem[\protect\citeauthoryear{{Putman} et~al.,}{{Putman}
  et~al.}{2009}]{putman09}
{Putman} M.~E.,  et~al., 2009, \mn@doi [\apj] {10.1088/0004-637X/703/2/1486},
  \href {https://ui.adsabs.harvard.edu/abs/2009ApJ...703.1486P} {703, 1486}

\bibitem[\protect\citeauthoryear{{Putman}, {Saul}  \& {Mets}}{{Putman}
  et~al.}{2011}]{putman11}
{Putman} M.~E.,  {Saul} D.~R.,   {Mets} E.,  2011, \mn@doi [\mnras]
  {10.1111/j.1365-2966.2011.19524.x}, \href
  {https://ui.adsabs.harvard.edu/abs/2011MNRAS.418.1575P} {418, 1575}

\bibitem[\protect\citeauthoryear{{Putman}, {Peek}  \& {Joung}}{{Putman}
  et~al.}{2012}]{putman12}
{Putman} M.~E.,  {Peek} J.~E.~G.,   {Joung} M.~R.,  2012, \mn@doi [\araa]
  {10.1146/annurev-astro-081811-125612}, \href
  {https://ui.adsabs.harvard.edu/abs/2012ARA&A..50..491P} {50, 491}

\bibitem[\protect\citeauthoryear{{Richter} et~al.,}{{Richter}
  et~al.}{2017}]{richter2017}
{Richter} P.,  et~al., 2017, \mn@doi [\aap] {10.1051/0004-6361/201630081},
  \href {https://ui.adsabs.harvard.edu/abs/2017A&A...607A..48R} {607, A48}

\bibitem[\protect\citeauthoryear{{Rubin}, {Prochaska}, {Koo}, {Phillips},
  {Martin}  \& {Winstrom}}{{Rubin} et~al.}{2014}]{rubin14}
{Rubin} K. H.~R.,  {Prochaska} J.~X.,  {Koo} D.~C.,  {Phillips} A.~C.,
  {Martin} C.~L.,   {Winstrom} L.~O.,  2014, \mn@doi [\apj]
  {10.1088/0004-637X/794/2/156}, \href
  {https://ui.adsabs.harvard.edu/abs/2014ApJ...794..156R} {794, 156}

\bibitem[\protect\citeauthoryear{{Russell} et~al.,}{{Russell}
  et~al.}{2016}]{russell16}
{Russell} H.~R.,  et~al., 2016, \mn@doi [\mnras] {10.1093/mnras/stw409}, \href
  {https://ui.adsabs.harvard.edu/abs/2016MNRAS.458.3134R} {458, 3134}

\bibitem[\protect\citeauthoryear{{Russell} et~al.,}{{Russell}
  et~al.}{2017}]{russell17}
{Russell} H.~R.,  et~al., 2017, \mn@doi [\apj] {10.3847/1538-4357/836/1/130},
  \href {https://ui.adsabs.harvard.edu/abs/2017ApJ...836..130R} {836, 130}

\bibitem[\protect\citeauthoryear{{Russell} et~al.,}{{Russell}
  et~al.}{2019}]{russell19}
{Russell} H.~R.,  et~al., 2019, \mn@doi [\mnras] {10.1093/mnras/stz2719}, \href
  {https://ui.adsabs.harvard.edu/abs/2019MNRAS.490.3025R} {490, 3025}

\bibitem[\protect\citeauthoryear{{Salem}, {Besla}, {Bryan}, {Putman}, {van der
  Marel}  \& {Tonnesen}}{{Salem} et~al.}{2015}]{salem15}
{Salem} M.,  {Besla} G.,  {Bryan} G.,  {Putman} M.,  {van der Marel} R.~P.,
  {Tonnesen} S.,  2015, \mn@doi [\apj] {10.1088/0004-637X/815/1/77}, \href
  {https://ui.adsabs.harvard.edu/abs/2015ApJ...815...77S} {815, 77}

\bibitem[\protect\citeauthoryear{{Sander} \& {Hensler}}{{Sander} \&
  {Hensler}}{2021}]{sander21}
{Sander} B.,  {Hensler} G.,  2021, \mn@doi [\mnras] {10.1093/mnras/staa3952},
  \href {https://ui.adsabs.harvard.edu/abs/2021MNRAS.501.5330S} {501, 5330}

\bibitem[\protect\citeauthoryear{{Sarazin}}{{Sarazin}}{1986}]{sazarin86}
{Sarazin} C.~L.,  1986, \mn@doi [Reviews of Modern Physics]
  {10.1103/RevModPhys.58.1}, \href
  {https://ui.adsabs.harvard.edu/abs/1986RvMP...58....1S} {58, 1}

\bibitem[\protect\citeauthoryear{{Scannapieco} \& {Br{\"u}ggen}}{{Scannapieco}
  \& {Br{\"u}ggen}}{2015}]{scannapieco15}
{Scannapieco} E.,  {Br{\"u}ggen} M.,  2015, \mn@doi [\apj]
  {10.1088/0004-637X/805/2/158}, \href
  {https://ui.adsabs.harvard.edu/abs/2015ApJ...805..158S} {805, 158}

\bibitem[\protect\citeauthoryear{{Schneider} \& {Robertson}}{{Schneider} \&
  {Robertson}}{2016}]{schneider16}
{Schneider} E.~E.,  {Robertson} B.~E.,  2016, preprint, \href
  {http://adsabs.harvard.edu/abs/2016arXiv160701788S} {} (\mn@eprint {arXiv}
  {1607.01788})

\bibitem[\protect\citeauthoryear{{Schneider} \& {Robertson}}{{Schneider} \&
  {Robertson}}{2017}]{schneider17}
{Schneider} E.~E.,  {Robertson} B.~E.,  2017, \mn@doi [\apj]
  {10.3847/1538-4357/834/2/144}, \href
  {https://ui.adsabs.harvard.edu/abs/2017ApJ...834..144S} {834, 144}

\bibitem[\protect\citeauthoryear{{Schneider}, {Robertson}  \&
  {Thompson}}{{Schneider} et~al.}{2018}]{schneider18}
{Schneider} E.~E.,  {Robertson} B.~E.,   {Thompson} T.~A.,  2018, \mn@doi
  [\apj] {10.3847/1538-4357/aacce1}, \href
  {https://ui.adsabs.harvard.edu/abs/2018ApJ...862...56S} {862, 56}

\bibitem[\protect\citeauthoryear{{Sch{\"o}nrich} \& {Binney}}{{Sch{\"o}nrich}
  \& {Binney}}{2009}]{schonrich09}
{Sch{\"o}nrich} R.,  {Binney} J.,  2009, \mn@doi [\mnras]
  {10.1111/j.1365-2966.2009.14750.x}, \href
  {https://ui.adsabs.harvard.edu/abs/2009MNRAS.396..203S} {396, 203}

\bibitem[\protect\citeauthoryear{{Shapiro} \& {Field}}{{Shapiro} \&
  {Field}}{1976}]{shapiro76}
{Shapiro} P.~R.,  {Field} G.~B.,  1976, \mn@doi [\apj] {10.1086/154332}, \href
  {https://ui.adsabs.harvard.edu/abs/1976ApJ...205..762S} {205, 762}

\bibitem[\protect\citeauthoryear{{Sharma}, {McCourt}, {Quataert}  \&
  {Parrish}}{{Sharma} et~al.}{2012}]{sharma12}
{Sharma} P.,  {McCourt} M.,  {Quataert} E.,   {Parrish} I.~J.,  2012, \mn@doi
  [\mnras] {10.1111/j.1365-2966.2011.20246.x}, \href
  {https://ui.adsabs.harvard.edu/abs/2012MNRAS.420.3174S} {420, 3174}

\bibitem[\protect\citeauthoryear{{Smith}}{{Smith}}{1963}]{smith63}
{Smith} F.~J.,  1963, \mn@doi [Plan. Space Sci.]
  {10.1016/0032-0633(63)90065-5}, 11, 1126

\bibitem[\protect\citeauthoryear{{Smith}, {Heckman}  \& {Illingworth}}{{Smith}
  et~al.}{1990}]{smith90}
{Smith} E.~P.,  {Heckman} T.~M.,   {Illingworth} G.~D.,  1990, \mn@doi [\apj]
  {10.1086/168848}, \href
  {https://ui.adsabs.harvard.edu/abs/1990ApJ...356..399S} {356, 399}

\bibitem[\protect\citeauthoryear{{Sparre}, {Pfrommer}  \& {Ehlert}}{{Sparre}
  et~al.}{2020}]{sparre20}
{Sparre} M.,  {Pfrommer} C.,   {Ehlert} K.,  2020, \mn@doi [\mnras]
  {10.1093/mnras/staa3177}, \href
  {https://ui.adsabs.harvard.edu/abs/2020MNRAS.499.4261S} {499, 4261}

\bibitem[\protect\citeauthoryear{{Steidel}, {Erb}, {Shapley}, {Pettini},
  {Reddy}, {Bogosavljevi{\'c}}, {Rudie}  \& {Rakic}}{{Steidel}
  et~al.}{2010}]{steidel10}
{Steidel} C.~C.,  {Erb} D.~K.,  {Shapley} A.~E.,  {Pettini} M.,  {Reddy} N.,
  {Bogosavljevi{\'c}} M.,  {Rudie} G.~C.,   {Rakic} O.,  2010, \mn@doi [\apj]
  {10.1088/0004-637X/717/1/289}, \href
  {https://ui.adsabs.harvard.edu/abs/2010ApJ...717..289S} {717, 289}

\bibitem[\protect\citeauthoryear{{Stone}, {Tomida}, {White}  \&
  {Felker}}{{Stone} et~al.}{2020}]{stone20}
{Stone} J.~M.,  {Tomida} K.,  {White} C.~J.,   {Felker} K.~G.,  2020, \mn@doi
  [\apjs] {10.3847/1538-4365/ab929b}, 249, 4

\bibitem[\protect\citeauthoryear{{Tan}, {Oh}  \& {Gronke}}{{Tan}
  et~al.}{2021}]{tan21}
{Tan} B.,  {Oh} S.~P.,   {Gronke} M.,  2021, \mn@doi [\mnras]
  {10.1093/mnras/stab053}, \href
  {https://ui.adsabs.harvard.edu/abs/2021MNRAS.502.3179T} {502, 3179}

\bibitem[\protect\citeauthoryear{{Thom}, {Peek}, {Putman}, {Heiles}, {Peek}  \&
  {Wilhelm}}{{Thom} et~al.}{2008}]{thom08}
{Thom} C.,  {Peek} J.~E.~G.,  {Putman} M.~E.,  {Heiles} C.,  {Peek} K.~M.~G.,
  {Wilhelm} R.,  2008, \mn@doi [\apj] {10.1086/589960}, \href
  {https://ui.adsabs.harvard.edu/abs/2008ApJ...684..364T} {684, 364}

\bibitem[\protect\citeauthoryear{{Thompson}, {Quataert}, {Zhang}  \&
  {Weinberg}}{{Thompson} et~al.}{2016}]{thompson16}
{Thompson} T.~A.,  {Quataert} E.,  {Zhang} D.,   {Weinberg} D.~H.,  2016,
  \mn@doi [\mnras] {10.1093/mnras/stv2428}, \href
  {https://ui.adsabs.harvard.edu/abs/2016MNRAS.455.1830T} {455, 1830}

\bibitem[\protect\citeauthoryear{{Townsend}}{{Townsend}}{2009}]{townsend}
{Townsend} R.~H.~D.,  2009, \mn@doi [\apjs] {10.1088/0067-0049/181/2/391}, 181,
  391

\bibitem[\protect\citeauthoryear{{Tripp}}{{Tripp}}{2022}]{tripp22}
{Tripp} T.~M.,  2022, \mn@doi [\mnras] {10.1093/mnras/stac044}, \href
  {https://ui.adsabs.harvard.edu/abs/2022MNRAS.511.1714T} {511, 1714}

\bibitem[\protect\citeauthoryear{{Turk}, {Smith}, {Oishi}, {Skory}, {Skillman},
  {Abel}  \& {Norman}}{{Turk} et~al.}{2011}]{yt}
{Turk} M.~J.,  {Smith} B.~D.,  {Oishi} J.~S.,  {Skory} S.,  {Skillman} S.~W.,
  {Abel} T.,   {Norman} M.~L.,  2011, \mn@doi [\apjs]
  {10.1088/0067-0049/192/1/9}, 192, 9

\bibitem[\protect\citeauthoryear{Van Der~Walt, Colbert  \& Varoquaux}{Van
  Der~Walt et~al.}{2011}]{van2011numpy}
Van Der~Walt S.,  Colbert S.~C.,   Varoquaux G.,  2011, Computing in Science \&
  Engineering, 13, 22

\bibitem[\protect\citeauthoryear{{Van Woerden}, Wakker, Schwarz  \& de
  Boer}{{Van Woerden} et~al.}{2004}]{vanwoerden}
{Van Woerden} H.,  Wakker B.~P.,  Schwarz U.~J.,   de Boer K.~S.,  eds, 2004,
  {High Velocity Clouds}  Astrophysics and Space Science Library Vol. 312,
  \mn@doi{10.1007/1-4020-2579-3.
}

\bibitem[\protect\citeauthoryear{{Vantyghem} et~al.,}{{Vantyghem}
  et~al.}{2016}]{vantyghem16}
{Vantyghem} A.~N.,  et~al., 2016, \mn@doi [\apj] {10.3847/0004-637X/832/2/148},
  \href {https://ui.adsabs.harvard.edu/abs/2016ApJ...832..148V} {832, 148}

\bibitem[\protect\citeauthoryear{{Veilleux}, {Cecil}  \&
  {Bland-Hawthorn}}{{Veilleux} et~al.}{2005}]{veilleux05}
{Veilleux} S.,  {Cecil} G.,   {Bland-Hawthorn} J.,  2005, \mn@doi [\araa]
  {10.1146/annurev.astro.43.072103.150610}, \href
  {https://ui.adsabs.harvard.edu/abs/2005ARA&A..43..769V} {43, 769}

\bibitem[\protect\citeauthoryear{{Virtanen} et~al.,}{{Virtanen}
  et~al.}{2020}]{2020SciPy-NMeth}
{Virtanen} P.,  et~al., 2020, \mn@doi [Nature Methods]
  {https://doi.org/10.1038/s41592-019-0686-2}, 17, 261

\bibitem[\protect\citeauthoryear{{Voit}, {Donahue}, {Zahedy}, {Chen}, {Werk},
  {Bryan}  \& {O'Shea}}{{Voit} et~al.}{2019}]{voit19}
{Voit} G.~M.,  {Donahue} M.,  {Zahedy} F.,  {Chen} H.-W.,  {Werk} J.,  {Bryan}
  G.~L.,   {O'Shea} B.~W.,  2019, \mn@doi [\apjl] {10.3847/2041-8213/ab2766},
  \href {https://ui.adsabs.harvard.edu/abs/2019ApJ...879L...1V} {879, L1}

\bibitem[\protect\citeauthoryear{{Wakker} \& {van Woerden}}{{Wakker} \& {van
  Woerden}}{1991}]{wakker91}
{Wakker} B.~P.,  {van Woerden} H.,  1991, \aap, \href
  {https://ui.adsabs.harvard.edu/abs/1991A&A...250..509W} {250, 509}

\bibitem[\protect\citeauthoryear{{Wakker}, {York}, {Wilhelm}, {Barentine},
  {Richter}, {Beers}, {Ivezi{\'c}}  \& {Howk}}{{Wakker}
  et~al.}{2008}]{wakker08}
{Wakker} B.~P.,  {York} D.~G.,  {Wilhelm} R.,  {Barentine} J.~C.,  {Richter}
  P.,  {Beers} T.~C.,  {Ivezi{\'c}} {\v{Z}}.,   {Howk} J.~C.,  2008, \mn@doi
  [\apj] {10.1086/523845}, \href
  {https://ui.adsabs.harvard.edu/abs/2008ApJ...672..298W} {672, 298}

\bibitem[\protect\citeauthoryear{{Westmeier}}{{Westmeier}}{2018}]{westmeier18}
{Westmeier} T.,  2018, \mn@doi [\mnras] {10.1093/mnras/stx2757}, \href
  {https://ui.adsabs.harvard.edu/abs/2018MNRAS.474..289W} {474, 289}

\bibitem[\protect\citeauthoryear{{Yang} \& {Ji}}{{Yang} \& {Ji}}{2022}]{yang22}
{Yang} Y.,  {Ji} S.,  2022, arXiv e-prints, \href
  {https://ui.adsabs.harvard.edu/abs/2022arXiv220515336Y} {p. arXiv:2205.15336}

\bibitem[\protect\citeauthoryear{{Zahedy}, {Chen}, {Johnson}, {Pierce},
  {Rauch}, {Huang}, {Weiner}  \& {Gauthier}}{{Zahedy} et~al.}{2019}]{zahedy19}
{Zahedy} F.~S.,  {Chen} H.-W.,  {Johnson} S.~D.,  {Pierce} R.~M.,  {Rauch} M.,
  {Huang} Y.-H.,  {Weiner} B.~J.,   {Gauthier} J.-R.,  2019, \mn@doi [\mnras]
  {10.1093/mnras/sty3482}, \href
  {https://ui.adsabs.harvard.edu/abs/2019MNRAS.484.2257Z} {484, 2257}

\bibitem[\protect\citeauthoryear{{Zhang}, {Thompson}, {Quataert}  \&
  {Murray}}{{Zhang} et~al.}{2017}]{zhang17}
{Zhang} D.,  {Thompson} T.~A.,  {Quataert} E.,   {Murray} N.,  2017, \mn@doi
  [\mnras] {10.1093/mnras/stx822}, \href
  {https://ui.adsabs.harvard.edu/abs/2017MNRAS.468.4801Z} {468, 4801}

\bibitem[\protect\citeauthoryear{{Zheng}, {Putman}, {Peek}  \& {Joung}}{{Zheng}
  et~al.}{2015}]{zheng15}
{Zheng} Y.,  {Putman} M.~E.,  {Peek} J.~E.~G.,   {Joung} M.~R.,  2015, \mn@doi
  [\apj] {10.1088/0004-637X/807/1/103}, \href
  {https://ui.adsabs.harvard.edu/abs/2015ApJ...807..103Z} {807, 103}

\bibitem[\protect\citeauthoryear{{Zheng} et~al.,}{{Zheng}
  et~al.}{2020}]{Zheng2020}
{Zheng} Y.,  et~al., 2020, \mn@doi [\apj] {10.3847/1538-4357/ab960a}, \href
  {https://ui.adsabs.harvard.edu/abs/2020ApJ...896..143Z} {896, 143}

\makeatother
\end{thebibliography}
\bsp
\label{lastpage}
\end{document}